\documentclass[12pt,preprint,natbib,iop]{emulateapj}	%Use this for all other purposes (emulateapj)
\usepackage{graphicx,epsfig,hyperref}	%Comment out hyperref for submission

\newcommand{\ran}{\rangle}
\newcommand{\for}[1]{\begin{equation} #1 \end{equation}}
\newcommand{\st}[1]{_{\rm{#1}}}
\newcommand{\sh}[1]{^{\rm{#1}}}
\newcommand{\gl}[1]{(\ref{#1})}

\slugcomment{Accepted for publication in the Astrophysical Journal (January 16 2011)}

\shorttitle{The star formation history of mass-selected galaxies in the COSMOS field}
\shortauthors{A. Karim et al.}

\begin{document}

\title{The star formation history of mass-selected galaxies in the COSMOS field} 

%AUTHORS

\author{A. Karim\altaffilmark{1,12}, E. Schinnerer\altaffilmark{1}, A. Mart\'\i nez-Sansigre\altaffilmark{2,3,4}, M. T. Sargent\altaffilmark{1}, A. van der Wel\altaffilmark{1}, 
H.-W. Rix\altaffilmark{1}, O. Ilbert\altaffilmark{5}, V. Smol{\v{c}}i{\'{c}}\altaffilmark{6,7}, C. Carilli\altaffilmark{8}, M. Pannella\altaffilmark{8}, A. M. Koekemoer\altaffilmark{9}, 
E. F. Bell\altaffilmark{10} and M. Salvato\altaffilmark{11}}

%AFFILIATIONS
\altaffiltext{1}{Max-Planck-Institut f\"ur Astronomie, K\"onigstuhl 17, D-69117 Heidelberg, Germany [email: karim@mpia.de]}
\altaffiltext{2}{Institute of Cosmology and Gravitation, University of Portsmouth, Dennis Sciama Building, Burnaby Road, Portsmouth, PO1 3FX, United Kingdom}
\altaffiltext{3}{Astrophysics, Department of Physics, University of Oxford, Keble Road, Oxford OX1 3RH, United Kingdom}
\altaffiltext{4}{SEP{\it net}, South-East Physics network}
\altaffiltext{5}{Laboratoire d'Astrophysique de Marseille, BP 8, Traverse du Siphon, 13376 Marseille Cedex 12, France}
\altaffiltext{6}{ESO ALMA COFUND Fellow, European Southern Observatory, Karl-Schwarzschild-Strasse 2, 
85748 Garching b. M\"unchen, Germany}
\altaffiltext{7}{Argelander Institut for Astronomy, Auf dem H\"ugel 71, Bonn, 53121, Germany}
\altaffiltext{8}{National Radio Astronomy Observatory, P.O. Box 0, Socorro, NM 87801-0387, USA}
\altaffiltext{9}{Space Telescope Science Institute, 3700 San Martin Drive, Baltimore, MD 21218, USA}
\altaffiltext{10}{Department of Astronomy, University of Michigan, 500 Church Street, Ann Arbor, MI 48109, USA}
\altaffiltext{11}{Max-Planck-Institut f\"ur extraterrestrische Physik, Giessenbachstrasse 1, D-85748 Garching, Germany}
\altaffiltext{12}{International Max Planck Research School for Astronomy and Cosmic Physics at the University of Heidelberg}

%ABSTRACT

\begin{abstract}
We explore the redshift evolution of the specific star formation rate (SSFR) for galaxies of different stellar mass by
drawing on a deep 3.6~$\mu$m-selected sample of $>10^5$ galaxies in the 2~deg$^2$ COSMOS field. The
average star formation rate (SFR) for sub-sets of these galaxies is estimated with stacked 1.4~GHz radio
continuum emission.
We separately consider the total sample and a subset of galaxies that shows evidence for substantive recent star
formation in the rest-frame optical spectral energy distributions. At redshifts $0.2 < z < 3$ both populations show a
strong and mass-independent decrease in their SSFR towards the present epoch. It is best described by a power-
law $(1+z)^n$, where $n \sim 4.3$ for all galaxies and $n \sim 3.5$ for star forming (SF) sources. The decrease
appears to have started at $z>2$, at least for high-mass ($M_* \gtrsim 4 \times 10^{10}~M_{\odot}$) systems
where our conclusions are most robust. Our data show that there is a tight correlation with power-law dependence,
$\rm{SSFR} \propto {M_*}^{\beta}$, between SSFR and stellar mass at all epochs. The relation tends to
flatten below $M_* \approx  10^{10}~M_{\odot}$ if quiescent galaxies are included; if they are excluded from the
analysis a shallow index $\beta_{\rm{SFG}} \approx -0.4$ fits the correlation. On average, higher mass objects
always have lower SSFRs, also among SF galaxies. At $z > 1.5$ there is tentative evidence for an upper
threshold in SSFR that an average galaxy cannot exceed, possibly due to gravitationally limited molecular gas
accretion. It is suggested by a flattening of the SSFR-$M_*$ relation (also for SF sources), but affects massive
($> 10^{10}~M_{\odot}$) galaxies only at the highest redshifts. Since $z=1.5$ there thus is no direct evidence
that galaxies of higher mass experience a more rapid waning of their SSFR than lower mass SF systems. In this
sense, the data rule out any strong 'downsizing' in the SSFR.
We combine our results with recent measurements of the galaxy (stellar) mass function in order to determine
the characteristic mass of a SF galaxy: we find that since $z\sim 3$ the majority of all new stars were always
formed in galaxies of  $M_*  = 10^{10.6\pm 0.4}~ M_{\odot}$. In this sense, too, there is no 'downsizing'.
Finally, our analysis constitutes the most extensive SFR density determination with a single technique out to
$z=3$. Recent Herschel results are consistent with our results, but rely on far smaller samples.
\end{abstract}

%KEYWORDS


%TEXTBODY

\section{Introduction}
Over the last years multi-waveband surveys of various wide fields have lead to estimates of star formation rates (hereafter SFRs) and stellar masses for large numbers of galaxies out to high redshifts. Both quantities are crucial for understanding galaxy evolution. On the one hand an evolution of the observed number density of galaxies as a function of stellar mass, i.e. the mass function, reveals how the stars are distributed among galaxies at different cosmic epochs. If, on the other hand, an increase in stellar mass of any population of galaxies can solely be explained by the rate at which new stars are formed within these systems or if other mechanisms are dominant can only be discussed if the corresponding SFRs themselves are known.

A number of studies (\citealp[e.g.][]{LILL96, MADA96, CHAR01, LEFL05, SMOL09A, DUNN09, RODI09, GRUP10, BOUW10, RUJO10} and for a compilation \citealp{HOPK04} and \citealp{HOPK06}) revealed that the star formation rate density (hereafter SFRD), i.e. the SFR per unit comoving volume, rapidly declines over the last $\sim 10~$Gyr following the purported maximum of star formation activity in the universe. The question of whether the stellar mass content of galaxies could be a major driver for this decline has gained significant interest after the discovery of a tight correlation of SFR and stellar mass for star forming (hereafter SF) galaxies with an intrinsic scatter of only about 0.3~dex \citep[e.g.][]{BRIN04, NOES07A,ELBA07}. This relation was studied in the local universe \citep[]{BRIN04, SALI07} suggesting an apparent bimodality in the SFR-$M_*$ plane if all galaxies are taken into account. It was also found to exist for SF galaxies at $z \lesssim 1.2$ \citep[e.g.][]{NOES07A, ELBA07, BELL07, WALC08} and further out to $z \approx 2.5$ \citep[]{DADD07,PANN09}.\footnote{It needs to be mentioned that at $z \sim 2$ \citet{DERB06} only found a weak correlation between SFR and stellar mass. However, their galaxy sample selection at ultraviolet wavelengths preferentially traces SFR rather than stellar mass, thus potentially biasing their results towards a flatter SFR-$M_*$ relation.} Consequently the stellar mass normalized SFR (hereafter specific SFR or SSFR), i.e. the SFR at a given epoch divided by the stellar mass the galaxy possesses at the same cosmic epoch, shows a tight (anti-)correlation. 

By studying the SSFR galaxies of different stellar masses can be directly compared. The SSFR itself defines a typical timescale that can be interpreted as a current efficiency of star formation within a galaxy compared to its past average star formation activity. The compilation of the studies mentioned \citep[e.g.][]{PANN09, GONZ09, DUTT09}, not using a common tracer for star formation nor selection technique for separating the SF galaxy fraction and data originating from various wide fields, suggests a steep evolution of the normalization of the SSFR-M$_*$ relation\footnote{In the following we will refer to this relation for SF galaxies as the SSFR-sequence.} for {\emph{SF} galaxies. 
Studies, covering a broad dynamical range in stellar mass, have been carried out for {\emph{all}} galaxies and confirmed the SSFR, as a function of redshift, to be even more rapidly increasing from $z=0$ to $z \approx 1$ \citep[e.g.][]{FEUL05A, ZHEN07A, DAME09A, DAME10} as well as throughout an even wider range in redshift \citep[e.g.][]{FEUL05B, PERE08, DUNN09, DAME09B}. It has been claimed that the steepness of the SSFR-increase with redshift might be a challenge for a cold dark matter concordance model ($\Lambda$CDM) suggested by comparisons to predictions from semi-analytical models (SAMs) (\citealp[see][]{SANT09,DAME09B, FIRM09} and, \citealp[e.g.,][for theoretical results based on a SAM]{GUOW08}). 

It was recently discussed by \citet{STAR09} and \citet{GONZ09}, at least for moderately massive SF galaxies $(M_{\ast} \sim 5 \times 10^{9}~M_{\odot})$, that the rapidly evolving SSFR might turn constant in the early universe. Their data show constant SSFRs up to the highest redshift ranges ($z \approx 7-8$) probed so far. This significant deviation of the SSFR-evolution from a power-law ($\rm{SSFR} \propto (1+z)^n$), fitting well the data below $z\approx 2$, could be a hint for different physical mechanisms regulating star formation in the early universe \citep{GONZ09}. However, this deviation could also be a result of observational data significantly underestimating the SSFRs at these high redshifts \citep{DUTT09} caused by selection biases \footnote{Note the very small number of galaxies currently studied in the extreme high redshift regime. Also note the highly discrepant SSFR-estimates presented by \citet{YABE09} and \citet{SCHA10} at the most extreme redshifts as summarized by \citet{BOUC09} in their Fig. 13.}. Recent theoretical models propose an enhanced merger rate \citep{KHOC10} at high $z$ in order to account for the purported constancy of the SSFR. This is in contrast to pure steady cold-mode gas accretion above a limiting dark matter halo mass (the so-called 'mass floor' of $M\st{DM} \sim 10^{11}~M_{\odot}$) \citep{BOUC09} reproducing well the observed slope of the SSFR-sequence at all $z < 2$. 

It was generally found that at $z < 2$ {\emph{all}} galaxies show a significant (negative) slope of the SSFR-M$_*$ relation leading to lower SSFRs in more massive galaxies. {\emph{Star forming}} galaxies also seem to show this behavior but the trend tends to be significantly weaker especially at $z > 1$ where, based on the sBzK selection technique, the slope was found to be practically vanishing \citep[]{DADD07,PANN09}. It therefore is an ongoing debate if this phenomenon of a decreasing slope of the SFR-M$_*$ relation for SF galaxies with redshift is real or just an artifact \citep[for an introduction and a summary of the conflicting observational results see e.g.][]{FONT09}. 
This effect is commonly interpreted as star formation efficiency being shifted from higher mass objects in the cosmic past to lower mass objects in the present and sometimes referred to as 'cosmic downsizing' \citep{COWI96}. 
Most recently, based on first Herschel/PACS far-infrared data, even the opposite effect, the so-called SSFR-upsizing at $z \gtrsim 1.5$, has been proposed \citep{RODI10}.\footnote{This trend is weakly supported by the earlier findings of \citet{OLIV10}.}    

More measurements are needed to understand the relation of SFR and stellar mass and its evolution with redshift.
This holds especially true for the population of SF galaxies. An accurate measurement of the (S)SFR-sequence at all epochs is key for a better understanding of galaxy evolution. As it was claimed \citep[e.g.][]{NOES07A} a tight correlation of SFR and stellar mass disfavors star formation histories (SFHs) of individual normal galaxies that are mainly driven by stochastic processes, such as mergers. Quite contrarily it favors smooth SFHs in such a way that the SFH at any cosmic epoch of a galaxy is solely determined by its stellar mass content measured at the corresponding redshift unless the galaxy becomes subject to quenching of star formation. In this sense the SFR-M$_*$ relation at a given redshift is regarded an isochrone for galaxy evolution in the same manner the Hertzsprung-Russel-Diagram is an isochrone for the evolution of a stellar population at a given age.\footnote{The (S)SFR-mass relation is therefore also sometimes referred to as 'the galaxy main sequence' \citep{NOES07A} that is, for an individual galaxy of stellar mass M$_*$, connected by evolutionary tracks \citep[e.g. the so-called tau-model discussed in][]{NOES07B} at distinct cosmic epochs \citep[see also][for a summary]{NOES09}.}  
It should be mentioned, however, that \citet{COWI08} disagree with this conclusion which underlines the importance of future studies that use a sufficiently deep direct SFR tracer to study the intrinsic dispersion of the SSFRs .\footnote{\citet{COWI08} cannot confirm the low level of intrinsic dispersion in the SSFR-$M_*$ plane found by \citet{NOES07A} and they discuss other hints they find supporting SFHs to be rather dominated by episodic bursts. We emphasize that the larger dispersion of SSFRs might be caused by the relatively broad bins in redshift used by \citet{COWI08} given the steep increase with redshift of SSFRs at $z<1.5$ while studying {\emph{all}} massive galaxies.} 

Several tracers across the electromagnetic spectrum are used to estimate the star formation rate of a normal galaxy\footnote{'Normal' galaxies are defined as systems that do not host an active galactic nucleus.}.
While rest-frame ultraviolet (UV) light originates mainly from massive stars and thus directly traces young stellar populations it will be strongly attenuated by dust. The absorbed UV emission is thermally reprocessed by heating the dust which in turn reemits at infrared (IR) wavelengths. Star formation also leads to emission in the radio continuum since charged cosmic particles are accelerated in shocks within the remnants of supernovae (SNR) leading to non-thermal synchrotron radiation (\citealp[e.g.][]{BELL78A} and \citealp[e.g.][for observations of individual SNRs]{MUXL94}). Thermal free-free emission ({\it{Bremsstrahlung}}) in general contributes only weakly to the 1.4~GHz signal \citep[see e.g.][]{COND92} but might become dominant in low-mass systems where the synchrotron emission was empirically found to be strongly suppressed \citep{BELL03A}. Also empirically the phenomenon of radio emission triggered by star formation results in its well known strong correlation with the far-IR output of a given SF galaxy \citep[e.g.][]{HELO85, COND92, YUNR01, BELL03A} that appears to persist out to high ($z > 2$) redshifts in a non-evolving fashion \citep[e.g.][]{SARG10A, SARG10B}.

A major advantage of radio emission as a tracer for star formation is its obvious independence of any correction for dust attenuation. Due to well known underlying physical processes the spectral energy distribution of a normal galaxy in the low ($\lesssim 5$)~GHz regime shows a $F_{\nu} \propto \nu^{\alpha\st{rc}}$ shape \citep[e.g.][]{BELL78A}. While $\alpha\st{rc} = -0.8$ is found to be a typical value for the radio spectral index (\citealp[e.g.][for a summary]{COND92, BELL03A} but also \citealp[e.g.][for early results]{SCHE68, BELL78B}) no further spectral features are expected in this frequency range thus leading to a robust K-correction up to high ($z \lesssim 3$) redshifts.\footnote{Unless radiative losses, e.g. inverse Compton scattering against the cosmic microwave background, steepen the spectral index to values $\sim$-1.3.} Both advantages directly confront the rather uncertain dust attenuation coefficient for UV light and the presence of polycyclic aromatic hydrocarbon (PAH) emission features redshifted (at $z \gtrsim 0.8$) into the 24~$\mu$m band commonly used as an estimator for the total infrared (TIR) emission. Also the combination of UV and mid-IR emission tracing star formation is limited since it is typically tested in moderately SF systems at low redshift \citep[for a summary see][]{CALZ09} which might not resemble high redshift galaxies with higher SFRs and larger dust content. Finally, even at a resolution of $\sim 5''$ achieved by current UV and IR telescopes blending of sources becomes a severe issue for the faint end of the sources \citep[see, e.g., ][]{ZHEN07B}. Current radio interferometers such as the (E)VLA and (e)Merlin achieve resolutions of $\lesssim 2''$ that are needed to unambiguously identify optical counterparts. This unambiguity is particularly important in a stacking experiment as otherwise flux density from nearby sources might contribute to the emission of an individual object. A drawback of using radio emission to trace star formation is the generally low sensitivity to the normal galaxy population even in the deepest radio surveys to-date which usually limits the analysis to a stacking approach. Therefore, current radio surveys allow one to study average SFR properties while they cannot shed light on the intrinsic dispersion of individual sources. This situation will improve with future EVLA surveys.   

Studying the stellar-mass
dependence of the SFH requires a mass-complete sample in order to prevent inferred evolutionary trends
from being mimicked by sample incompleteness.
Early type galaxies containing predominantly older stellar populations
and showing therefore a prominent $4000~\AA$ break \citep[see e.g.][]{GORG99} are likely to be excluded in optical surveys above $z \sim 1$ even
at deep limiting magnitudes as the break is redshifted into the selection band. Optical selection, thus, 
potentially limits any study of a stellar mass-complete sample to the bright (i.e. high-mass) end or is effectively rather a selection by
unobscured SFR than by stellar mass if the full sample is considered for the analysis.

Channel 1 of the IRAC instrument onboard the Spitzer Space Telescope provides us with the $3.6~\mu \rm{m}$ waveband that samples the rest-frame $K$-band at $z\sim 0.5$ to the rest-frame $z$-band at $z \sim 3$. It is therefore ideal in probing mainly
the light from old low-mass stars while not being severely affected by dust. For the analysis presented here, hence, a deep and rich
($\sim 100,000$ sources at $z \le 3$) $3.6~\mu \rm{m}$ galaxy sample in combination
with accurate photometric redshifts and stellar-mass estimates has been used \citep{ILBE09B}. With a sky coverage of $2~\rm{deg}^2$ the Cosmic Evolution Survey\footnote{http://cosmos.astro.caltech.edu} (COSMOS) provides the largest cosmological deep field to-date \citep[see][for an overview]{SCOV07B}. The uniquely large COSMOS $3.6~\mu \rm{m}$ galaxy sample offers uniform high-quality pan-chromatic data for {\emph{all}} sources enabling us to study the SSFR in small bins in both stellar mass and redshift. 
Additionally the evolution of the stellar mass-functions has been studied already based on the same sample and its SF sub-population \citep{ILBE09B}. As it was argued \citep[e.g. in][]{DADD09} the combination of the individual evolutions of the mass function and the (S)SFR-sequence might be the most important observational constraints for understanding the stellar mass built-up on cosmic scales jointly resulting in a potentially peaking and declining SFRD.

This paper is organized as follows. In Sec. \ref{sec:data} we present our principle and ancillary COSMOS data sets and the selection of our sample. Sec \ref{sec:stack} contains a detailed
description of our stacking algorithm and the derivation of average SFRs from the 1.4~GHz image stacks. Additional methodological considerations pertaining to both sample selection and flux density estimation by image stacking are to be found in the Appendices. Readers who wish to directly proceed to our results and their interpretations can find
those regarding the relation of SSFR and stellar mass in Sec. \ref{sec:ssfr}. Our measurements of the CSFH and a simple model that reproduces these observations are discussed in Sec. \ref{sec:csfh}. Both Sections (\ref{sec:ssfr} and \ref{sec:csfh}) contain a detailed discussion of how our results relate to the recent literature. We summarize our findings in Sec. \ref{sec:summ}. 
  
Throughout this paper all observed magnitudes are given in the AB system. We assume a standard cosmology with $H_{0}=70$~(km/s)/Mpc, $\Omega_M=0.3$ and $\Omega_{\Lambda}=0.7$ consistent with the latest WMAP results \citep{KOMA09} as well as a radio spectral index of $\alpha\st{rc} = -0.8$ in the notation given above if not explicitly stated otherwise. A \citet{CHAB03} initial mass function (IMF) is used for all stellar mass and SFR calculations in this article. Results from previous studies in the literature have been converted accordingly.\footnote{Logarithmic masses and SFRs based on a \citet{SALP55} IMF, a \citet{KROU01} IMF and a \citet{BALD03} IMF are converted to the Chabrier scale by adding -0.24~dex, 0~dex and 0.02~dex, respectively.} 

\section{The pan-chromatic COSMOS data used}
\label{sec:data}
In order to study the redshift evolution of galaxies in general, and the evolution of their SFRs in particular, a complete
and large sample of normal galaxies is needed as it not only provides representative but also statistically significant insights.

The large area of $2~\rm{deg}^2$ covered by the COSMOS survey, fully imaged at optical wavelengths by the Hubble space telescope (HST) \citep{SCOV07A, KOEK07}, is necessary to minimize the effect of cosmic variance. Deep UV GALEX \citep{ZAMO07} to ground-based optical and near-infrared (NIR) \citep{TANI07, CAPA07} imaging of the equatorial field\footnote{The COSMOS field is centered at $\rm{RA} =\rm10:00:28.6$ and $\rm{Dec} = \rm{+02:12:21.0}$ (J2000)} yielded
accurate photometric data products for $\sim 1 \times 10^6$ galaxies down to 26.5th magnitude in the $i$-band \citep{ILBE09A, CAPA07}. Thanks to extensive spectroscopic efforts at optical wavelentghs using VLT/VIMOS and Magellan/IMACS \citep{LILL07, TRUM07} the estimation of photometric redshifts for all these sources could be accurately calibrated. Ongoing deep Keck/DEIMOS campaigns (PIs Scoville, Capak, Salvato, Sanders and Karteltepe) extent the spectroscopically observed wavelength regime to the NIR which is critical to improve the photometric calibration for faint sources at high redshifts. 
In addition to observations of the whole or parts of the COSMOS field in the X-ray \citep{HASI07, ELVI09} and millimeter \citep{BERT07, SCOT08}, imaging by Spitzer in the mid- to far-IR \citep{SAND07} as well as interferometric radio data \citep{SCHI04,  SCHI07,  SCHI10} covering the full $2~\rm{deg}^2$ have been obtained.
      
\subsection{VLA-COSMOS radio data}

Radio observations of the full (2~$\rm{deg}^2$) COSMOS field were carried out with the Very Large Array (VLA) at 1.4~GHz (20~cm) in several campaigns between 2004 and 2006. The entire field was observed in A- and C-configuration \citep{SCHI07} where the 23 individual pointings were arranged in a hexagonal pattern. Additional observations of the central seven pointings in the more compact A-configuration \citep{SCHI10} were obtained in order to achieve a higher 1.4~GHz sensitivity in the area overlapping with the COSMOS MAMBO millimeter observations \citep{BERT07}. In both cases the data reduction was done using standard procedures from the Astronomical Imaging Processing System (AIPS) \citep[see][for details]{SCHI07}.
At a resolution of $1.5'' \times 1.4''$ the final map has a mean rms of $\sim 8$~$\mu$Jy/beam in the central $30'\times 30'$ and $\sim 12$~$\mu$Jy/beam over the full area, respectively. 
Using the SAD algorithm within AIPS, a total of 2,865 sources were identified at more than $5\sigma$ significance in the final VLA-COSMOS mosaic \citep{SCHI10}.  
As the outermost parts of the map are not covered by multiple pointings the noise increases rapidly towards the edges. In this study we therefore exclude these peripheral 
regions resulting in a final useable area of 1.72~$\rm{deg}^2$. 

\subsection{A 3.6~$\mu$m selected galaxy sample within the COSMOS photometric (redshift) catalogs}
\label{sec:irsamp}
Deep Spitzer IRAC data mapping the entire COSMOS field in all four channels have been obtained during the S-COSMOS observations \citep{SAND07}.
The data reduction yielding images and associated uncertainty maps for all the four channels is described in \citet{ILBE09B} (I10 hereafter). 
For the $3.6~\mu \rm{m}$ channel a source catalog has been obtained by O. Ilbert and M. Salvato (private communication) using the SExtractor package \citep{BERT96}. Given the point spread function (PSF) of 1.7'' a Mexican hat filtering of the $3.6~\mu \rm{m}$ image within SExtractor was used in 
order to assure careful deblending of the sources. 

The resulting sample of $3.6~\mu$m sources down to a limiting magnitude of $m_{\rm{AB}}(3.6~\mu \rm{m}) = 23.9$ in the 2.3~deg$^2$ field, not considering the masked areas around bright sources ($K_s < 12$), areas of poor image quality and the field boundaries, consists of 306,000 sources.\footnote{As a stacking analysis depends on the input sample prior masked areas consequently reduce further the effective area for this study. All space densities reported in this work are therefore computed for an effective field size of 1.49~deg$^2$.} 

As detailed in I10 photometric redshifts (hereafter photo-$z$'s) were assigned to all 3.6~$\mu$m detected sources. The vast majority of sources is also detected at optical wavelengths and therefore contained in the COSMOS photo-z catalog\footnote{This optically deep sample has a limiting magnitude of 26.2 in the $i^+$ selection band (see Tab. 1 in \citet{SALV09}).} \citep{ILBE09A} so that in general photometric information from 31 narrow-, intermediate and broad-band FUV-to-mid-IR filterbands was available.\footnote{As described in detail by I10 all photo-z's used in our study were obtained using a $\chi^2$ template-fitting procedure implemented in the code {\it{Le Phare}} \citep{ARNO02, ILBE06} and a library of 21 templates. Additional stellar templates were used to reject stars (i.e. sources with a lower $\chi^2$ values for the stellar compared to the galaxy templates) from the final galaxy sample.} Within the remaining 4~\% (i.e. a total of 8507) of the 3.6~$\mu$m sources 2714 are also contained in the COSMOS $K-$band selected galaxy sample \citep{MCCR10} and are also regarded as real sources. I10 assigned photo-z's to these extremely faint objects using the available NIR-to-IRAC photometry. 

The quality of the photo-$z$'s was estimated (for details see I10) by using spectroscopic redshifts for a total of 4,148 sources at $m_{AB}(i^+) < 22.5$ from the zCOSMOS survey \citep{LILL09}.
At a rate of $< 1~ \%$ of outliers the accuracy was found to be $\sigma_{(z_{\rm{phot}} - z_{\rm{spec}})/(1+z_{\rm{spec}})} = 0.0075$ down to the magnitude limit of the spectroscopic sample. For all objects within the $3.6~\mu \rm{m}$ selected catalog -- regardless of $i$-band magnitude the accuracy was derived by using
the $1\sigma$ uncertainty on the photo-z's from the probability distribution function which yields a conservative estimate of the photo-$z$ uncertainty as detailed in \citet{ILBE09A}. At $1.25 < z < 2$ the relative photo-z uncertainty is 0.08 and thus higher by a factor of four compared to the median value for the full ($m_{\rm{AB}}(3.6~\mu \rm{m}) \ge 23.9$) sample.\footnote{For a color-selected sub-set of galaxies for which spectroscopic redshifts from the zCOSMOS-faint survey (Lilly et al., in prep.) were available the photo-$z$ accuracy was directly tested at $1.5<z<3$. This yields an accuracy of $\sigma_{\Delta z/(1+z)}=0.04$ with 10\% of catastrophic failures.} We account for this when binning the data in redshift by choosing increasing bin widths with increasing redshift.\footnote{It should be mentioned, however, that the projected-pair analysis by \citet{QUAD09} independently shows that photo-$z$'s from data sets with broad- and intermediate band photometry like the COSMOS catalog are not expected to have very different photo-$z$ errors at $z>1.5$ than at lower redshifts.} It is worth noting that the photo-z accuracy is degraded at magnitudes fainter than $m_{AB}(i^+) = 25.5$ (See Fig. 12 in \citet{ILBE09A}). Our choice of lower stellar mass limits (see Sec. \ref{sec:comp}) and our stellar mass binning-scheme (see Sec. \ref{sec:comp}) automatically ensures a low fraction ($< 15~\%$) of these optically very faint objects within the lowest mass-bin above our mass limit at any redshift. The fraction of such faint objects effectively vanishes towards higher masses as also pointed out by I10.\footnote{I10 use comparable mass limits and their Fig. 8 the strong decline of the fraction of optically faint objects with mass at all $z$.} 

\subsection{Estimation of stellar masses}
Stellar masses for all objects within the $3.6~\mu$m selected parent sample have been computed by I10. Here, we briefly summarize the method and the important findings.
For the estimation of stellar masses based on a Chabrier IMF stellar population synthesis models generated with the package provided by \citet{BRUZ03} (BC03) have been used.
Furthermore an exponentially declining SFH and a \citet{CALZ00} dust extinction law have been assumed. Spitzer MIPS $24 \mu \rm{m}$ flux densities \citep[from][]{LEFL09} have been 
included in the SED template fitting as an additional constraint on the stellar mass. Systematic uncertainties on the stellar masses, caused by the use of photo-z's, the choice of the dust extinction law and library of stellar population synthesis models, have been investigated. No systematic effect due to the use of photo-z's is apparent. 
Stellar masses derived from the BC03 templates are systematically higher by 0.13-0.15 dex compared to the newer Charlot \& Bruzual (2007) versions \citep{BRUZ07} that have an improved treatment of thermally pulsing asymptotical giant branch (TP-AGB) stars. As BC03 models are commonly used in the literature, both studies, I10 and this work, are based on BC03 mass estimates. 

\subsection{Spectral classification}
\label{sec:class}
A number of studies suggest the existence of a bimodality in the SSFR-M$_*$ plane \citep[e.g.][]{SALI07, ELBA07, SANT09, RODI10} leading to
a tight SSFR-sequence to be in place only for SF galaxies. Therefore a deselection of quiescent, i.e. non SF,
objects is needed.  

Following I10 we classify galaxies with a best-fit BC03 template that has an intrinsic (i.e. dust unextincted) rest-frame color redder than $(\rm{NUV}-r^+)\st{temp} = 3.5$ as quiescent.
Several authors \citep[e.g.][]{WYDE07, MART07A, ARNO07} suggest this color to be an excellent indicator for the recent over past average SFR as it directly traces the ratio of young (light-weighted average age of $\sim 10^8$~yr) and old ($\ge 10^9$~yr) stellar populations. Seeking for a color bimodality that discriminates galaxies with currently high from those with low star formation activity the $\rm{NUV}-r$ color appears therefore to be superior to purely optical rest-frame colors such as $U-V$ \citep[e.g.][]{BELL04}. 

Using a dust uncorrected $\rm{NUV}-r^+$ versus $r^+ -J$ rest-frame color-color diagram\footnote{Here the absolute magnitudes were inferred from the observed magnitudes not accounting for dust reddening.} I10 showed that in the range $0 \le z \le 2$ for $(\rm{NUV}-r^+)\st{temp} > 3.5$ quiescent galaxies are well separated from the parent sample without severe contamination by dust-obscured SF galaxies. This quiescent population is therefore comparable to the one classified by \citet{WILL09} based on a $U-V$ versus $V-J$ rest-frame color-color diagram. 

Furthermore our quiescent population shows a clear separation from the parent sample with respect to galaxy morphology. I10 visually classified a subset of 1,500 isolated and bright galaxies from the $3.6~\mu$m parent sample using HST/ACS images and found the quiescent population among those to be clearly dominated by elliptical (E/S0) systems. A further cut ($(\rm{NUV}-r^+)\st{temp} < 1.2$) was shown to efficiently separate late type spiral and irregular galaxies from early type spirals as well as the remaining tail of elliptical systems. As any such color cut effectively is a cut in star formation activity we discuss the spectral pre-classification of SF systems in more detail in Appendix \ref{sec:app_sfg}. 

\subsection{AGN contamination}
\label{sec:agncont}
A major concern arising in the context of using radio emission to trace star formation is contaminating flux from active galactic nuclei (AGN). For some galaxies the total radio signal might even be dominated by an AGN. For our study, ideally, we should therefore remove all galaxies hosting an AGN from our sample. 

Cross-matching the most recent XMM-COSMOS photo-z catalog \citep{SALV09, BRUS10} with the $3.6~\mu$m selected parent sample delivered a total of 1,711 (i.e. $\sim 1~ \%$) X-ray detected objects. Most of these sources exhibit best-fit composite AGN/galaxy SEDs\footnote{Based on the \citet{SALV09} classification that uses an enhanced set of AGN/galaxy templates in order to fit the FUV-to-mid-IR SED and that includes further priors  (e.g. variability information) in the fitting procedure while delivering accurate photometric redshifts for all these sources.} while a minor fraction is well fitted by an SED showing no AGN contribution. However, here all X-ray detections are treated as potential AGN contaminants and thus removed from our sample.\footnote{Note that \citet{HICK09} and \citet{GRIF10} yield strong evidence that X-ray and radio selected AGN are mutually distinct populations such that it is actually questionable to remove X-ray selected objects from our samples. We confirmed that our results do not change significantly when including those objects and urge caution to remove more objects if deeper X-ray data compared to the XMM imaging used here is at hand.}   

Studies of the radio luminosity function \citep[e.g.][]{SADL02, COND02} agree that radio-AGN contribute half of the radio light in the local universe at radio luminosities slightly below $L_{1.4~\rm{GHz}} \sim 10^{23}$~W/Hz and outnumber SF galaxies above $\sim 2 \times 10^{23}$~W/Hz. Detailed multi-wavelength studies \citep{HICK09, GRIF10} yield that radio-AGN are hosted by red galaxies. The evolution out to $z \approx 1.3$ of the radio-AGN fraction for luminous (i.e. $L_{1.4~\rm{GHz}} > 4 \times 10^{23}$~W/Hz) radio-AGN as a function of stellar mass has been presented by \citet{SMOL09C} who selected a parent sample of red galaxies with rest-frame $\rm{U}-\rm{B}$ colors in a range close to our quiescent galaxy fraction. The derived AGN-fractions at a given stellar mass within the red galaxy population are therefore applicable to our sample.  

According to \citet{SMOL09C} (see their Fig. 11) the luminous radio-AGN fraction at $0.7 < z < 1.3$ is well below $25~\%$ at all $\log{(M_* [M_{\odot}])} < 11.5$ where it drops quickly to $\sim 1~\%$ at $\log{(M_* [M_{\odot}])} = 11$ and continuously to lower levels as stellar mass decreases. At masses lower than $\log{(M_* [M_{\odot}])} = 11$ the radio-AGN fractions are subject to non-negligible evolution between $0 < z < 1$ while the fractions at higher masses increase only mildly. However, given that the radio-AGN fractions are well below 1~\% in the former (i.e. low) mass range out to $z \sim 1.3$ it is unlikely that they rise above 10~\% at $z \gg 1$. The evolution of the radio-AGN fraction at the high-mass end is much slower but the fractions are high already in the local universe. We therefore set an arbitrary but reasonable threshold and exclude all quiescent objects above $\log{(M_* [M_{\odot}])} = 11.6$, where the expected radio-AGN fraction exceeds 50~\%, from our stacking analysis. As the radio-AGN fraction sharply drops below this limit the remainder of our full galaxy sample should be generally free from radio-AGN contamination. Within the highest mass bin probed here ($M_* > 10^{11}~M_{\odot}$; see Fig. \ref{fig:allbins}), however, the average fraction of radio AGN among the quiescent galaxies could still be $\sim 25~\%$ at $z > 1$. This fraction appears high but among the entire galaxy population (quiescent and SF sources) the percentage drops to at most $10~\%$ within our highest mass-bin at $z \sim 1$. As shown by I10 globally, but in particular at $M_* > 10^{11}~M_{\odot}$ the fraction of quiescent galaxies among the entire sample decreases strongly towards higher redshifts \citep[see also][]{TAYL09}.\footnote{The global stellar mass density of quiescent galaxies at $z=1.5$ is about an order of magnitude lower than the SF one.} An upper bound of $10~\%$ to the potential fraction of radio-AGN within our highest mass-bin hence is a well justified number at $z > 1$.

\begin{figure*} 
\centering
\includegraphics[angle=90,width=0.9\textwidth]{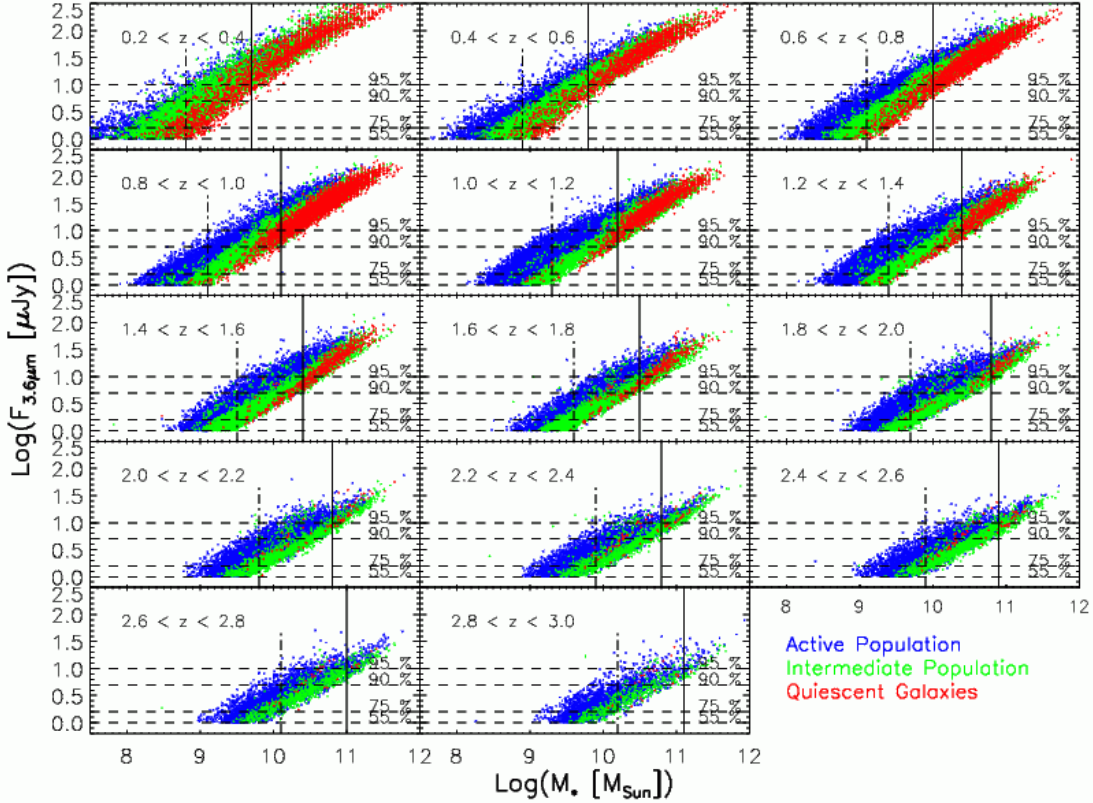}
\caption{\noindent Observed 3.6~$\mu$m flux versus stellar mass from SED fits. The 12 panels show photometric redshift bins, as $0.2 \le z_{\rm{phot}} \le 3$ indicated in the upper left part of each panel. Flux densities relate to AB-magnitudes via $m_{\rm{AB}}(3.6~\mu \rm{m}) = -2.5 \log_{10}(F_{3.6~\mu \rm{m}} \; [\mu \rm{Jy}]) +23.9$ (23.9 is the magnitude limit of the catalog). Blue points denote highly active SF systems, with intrinsic rest-frame template colors $(\rm{NUV}-r^+)\st{temp} < 1.2$; red points denote quiescent (low star formation activity) galaxies with $(\rm{NUV}-r^+)\st{temp} > 3.5$. Green points are objects of intermediate intrinsic rest-frame color (and hence star formation activity). Horizontal dashed lines mark the levels of the detection completeness, estimated through Monte Carlo simulations of artificial sources (see Sec. \ref{sec:comp}). The vertical dashed-dotted line in each panel denotes the lower mass limit, to which the sample of SF systems (i.e. the union of all blue and green points) is representative of the underlying SF population and the SFR is not affected by the intrinsic catalog incompleteness. The solid vertical line in each panel denotes the mass limit to which the {\emph{entire}} sample is regarded as representative.}
\label{fig:compaq} 
\end{figure*} 

Due to prominent spectral features we regard the SED-fits for quiescent objects as most trustworthy such that also the SED-derived SFRs are expected to be accurate for individual objects. These SFRs therefore serve as a prior for revealing potential radio-AGN among the radio-detections in our sample. Hence we correlated our sample with the latest version of the VLA-COSMOS catalog \citep{SCHI10} and excluded those objects showing radio-derived SFRs more than twice as large as the SED-derived values. We find that the overall number of objects excluded in each sample to be stacked is negligible. The same holds for very luminous ($L_{1.4~\rm{GHz}} > 10^{25}$~W/Hz) radio sources among the radio-detections that are most likely high-power radio-AGN. We therefore excluded also these objects relying on individual photo-z's in order to estimate the radio luminosity. The total fraction of galaxies among all objects in a given bin that we exclude by these two criteria amounts -- on average -- to less than 0.3~\% such that only a fraction of radio detections is rejected. We stress the smallness of this percentage as the advantage of our radio-approach is its insensitivity to dust obscuration which might be challenged by relying on individual optical best-fit SEDs as we partially do when removing some of the radio-detected objects. It should be noted that the high-power radio-AGN candidates are exclusively hosted by red galaxies within our sample. Hence, X-ray detected sources are the only objects that have been removed from our SF samples. 

As the radio-based SFR-results presented in this paper (see Sec. \ref{sec:ssfr}) are based on a median stacking approach (see Sec. \ref{sec:stack}) a minor fraction of contaminating outliers such as AGN is even tolerable. We conclude that contamination of the stacked radio flux densities caused by AGN emission at radio frequencies is not a siginifcant source of uncertainty in the context of this study and that our conclusions would not change if we included the radio-AGN candidates in our analysis.          

\subsection{Completeness considerations}
\label{sec:comp}
In the following we will discuss the completeness of our (sub-)samples. It is important to distinguish between two kinds of effects. While the full $3.6~\mu$m-selected source catalog (1) is subject to a flux density-dependent level of detection incompleteness we are interested in (2) how representative for the underlying population a given subset of galaxies is at a given mass. Our lower mass limits hence need to be chosen such that the objects at hand remain sufficiently representative.

I10 evaluated the efficiency of the source extraction procedure (and hence the detection completeness) with Monte Carlo simulations of mock point-sources inserted into the $3.6~\mu \rm{m}$ mosaic. At the flux density cut of $1~\mu \rm{Jy}$ ($m_{\rm{AB}}(3.6~\mu \rm{m}) = 23.9$) the catalog was found to be 55~\% complete; 90~\% completeness is reached at $F_{3.6~\mu \rm{m}} \approx 5~\mu$Jy ($m_{\rm{AB}}(3.6~\mu \rm{m})=22.15$).
This rather shallow decline in detection completeness towards the magnitude limit is due to source confusion.

Fig. \ref{fig:compaq} shows the distribution of 3.6~$\mu$m flux density with stellar mass in narrow redshift slices for our source catalog, color coded by the spectral type of the galaxies (see Sec. \ref{sec:class}). The Monte Carlo detection completeness levels of the catalog are indicated by horizontal dashed black lines starting from the flux density limit at the bottom to the 95~\% completeness limit at the top in each panel. Each sub-population shows a clear correlation between $3.6~\mu$m flux density and stellar mass, and the quiescent population residing at the high-mass end at all flux densities. While SF sources (the union of all blue and green data points) span the entire range of 3.6~$\mu$m flux densities at all redshifts, hardly any quiescent objects with low flux densities are observed at intermediate and high redshifts. We consequently find fewer and fewer low-mass quiescent objects as redshift increases. This is certainly the combined effect of a general absence of such sources at higher redshifts plus the loss of these objects at low flux densities due to the global detection incompleteness of our catalog.

Detection incompleteness affects all sources in at a given $3.6~\mu$m flux density, regardless of their spectral type. However, the different distribution of quiescent and SF sources with respect to $3.6~\mu$m flux density necessitates that a different lower mass limit (`{\it representativeness limit}', hereafter) be adopted, depending on whether we consider the redshift evolution of SF galaxies or that of the entire galaxy population. We now discuss how the limiting mass is set for these two samples:
\begin{itemize}
\item In the case of the {\it entire galaxy population}, it is important to be working with a sample in which the fractional contribution of quiescent and SF sources reflects the true population fractions as closely as possible. The probability that this is the case becomes larger, the better the underlying population is sampled; i.e. it rises with increasing detection completeness. We therefore require an intrinsic catalog completeness of 90~\% (corresponding $m\st{AB}(3.6~\mu \rm{m})=22.15$) at all masses considered. This is an arbitrary but reasonable threshold as the intrinsic catalog completeness rises rapidly towards higher flux densities.\\
In order to evaluate the actual mass representativeness limit we need to define yet another type of completeness level, which we shall refer to as {\it statistical completeness}. By applying the analytical scheme described in detail in Appendix \ref{sec:app_comp} we ensure that the statistical completeness of our sample always reaches at least 95~\%. This value sets the actual level of representativeness of a given sub-sample. In the following we will also present results for sub-samples below the evaluated mass-limits which will be indicated separately. Those results represent strict upper limits in (S)SFR.
\item For studying the {\it SF population} we need not be as conservative because we are dealing with a {\it single} sub-population that is subject to less internal variation of SF activity as a (bimodal) sample including both quiescent and SF systems. We thus consider sources down to the limiting flux density of the 3.6~$\mu$m catalog when we compute the mass limits at a given redshift. Since this implies that at low stellar masses the flux distribution is sharply cut due to the magnitude limit of our catalog, we still need to use the scheme presented in Appendix \ref{sec:app_comp} to identify stellar mass limits that provide a representative flux density distribution for SF galaxies.
As visible in all panels of Fig. \ref{fig:compaq}, the lowest mass bin always contains objects over the full range of detection completeness, from 55~\% to 100\%. One might expect -- and the SED fits confirm this -- that among galaxies of a given mass, those with the fainter fluxes have lower SSFRs. Failure to include them (due to detection incompleteness) would thus yield average radio-derived SSFRs that are biased towards higher values. We wish to emphasize, however, that our choice of the statistical completeness level ensures that this bias is small above our mass limit and that our samples hence are `representative' in the sense that they can be expected to render a meaningful measurement of, e.g., the average SSFR of the underlying population.
\end{itemize}

\begin{deluxetable}{ccc}
\tablecaption{\label{tab:comp}Stellar mass limits for all/SF galaxies}
%\tablenum{1}
\tablehead{\colhead{} & \colhead{All galaxies} & \colhead{SF systems} \\ \colhead{$z$} & \colhead{$\log(M_{\ast}~[M_{\odot}])\st{lim}$} & \colhead{$\log(M_{\ast}~[M_{\odot}])\st{lim}$}} 
\startdata
0.3 & 9.7 & 8.8\\
0.5 & 9.8 & 8.9\\
0.7 & 10.0 & 9.1\\
0.9 & 10.1 & 9.1\\
1.1 & 10.2 & 9.3\\
1.3 & 10.4 & 9.4\\
1.5 & 10.4 & 9.5\\
1.7 & 10.5 & 9.6\\
1.9 & 10.8 & 9.7\\
2.1 & 10.8 & 9.8\\
2.3 & 10.8 & 9.9\\
2.5 & 10.9 & 9.9\\
2.7 & 11.0 & 10.1\\
2.9 & 11.1 & 10.2\\
\enddata
\tablecomments{The lower stellar mass limits above which our samples are regarded representative. Those limits are as shown in Fig. \ref{fig:compaq} and \ref{fig:gaussall} and have been derived based on the scheme that is detailed in Appendix \ref{sec:app_comp}.}
\end{deluxetable}

The stellar mass representativeness limits for the whole sample and the SF systems are marked in Fig. \ref{fig:compaq} as vertical lines for each redshift bin in the range $0.2 < z\st{phot} < 3$ and listed in Tab. \ref{tab:comp}. Note that they increase with redshift. As a consequence, our results will be based on fewer mass bins at high redshift and the aforementioned bias in the lowest mass bin may therefore have a larger impact on fitting trends. Very conservatively speaking, our results for SF objects presented in the following should generally be regarded as most robust at $z \lesssim 1.5$ while evolutionary trends inferred at the high mass end are robust out to our redshift limit of $z=3$. We will also show results for SF galaxies obtained at masses lower than the individual mass limits and treat them as not entirely representative. Such measurements will be indicated with different symbols in our plots and we will discuss any further implications in Sec. \ref{sec:others}.

The final sample of galaxies with $m_{\rm{AB}}(3.6~\mu \rm{m}) = 23.9$ and $z_{\rm{phot}} < 3$ consists of 165,213 sources over an effective area of $\sim 1.5~$deg$^2$. Fig. 3 in I10 shows the redshift distribution with a median of $z_{\rm{phot}} \sim 1.1$. After adopting a lower redshift limit of $z_{\rm{phot}}=0.2$ in order to account for the small local volume sampled by our effective area and our binning scheme 113,610 sources\footnote{This number already considers the upper limiting mass for quiescent galaxies as discussed in Sec. \ref{sec:agncont} and excludes further 328 sources (i.e. 0.3~\%) classified as radio-AGN.} (90,957 SF galaxies) enter our analysis. This is by far the largest galaxy sample used for studying the dependence between SFR and stellar mass throughout cosmic time. Fig. \ref{fig:allbins} shows the adopted binning scheme and the number of galaxies contained in each stellar mass and photo-z bin.        

\begin{figure*}
\vspace{-3cm} 
\includegraphics[angle=90,width=\textwidth]{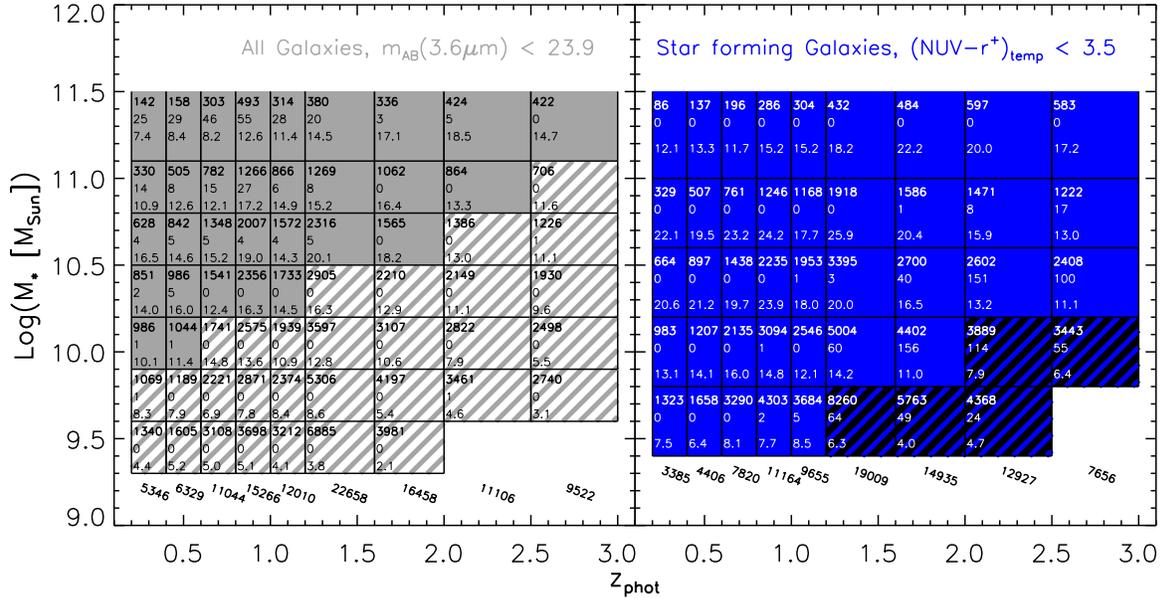}
\caption{\noindent Binning scheme in stellar mass and photometric redshift for the entire (left) and the SF (right) sample. Hatched bins lie below the corresponding 
limits denoted in Fig. \ref{fig:compaq} and are hence regarded representative of the underlying galaxy population.
The top number in each box is the total number of galaxies used in the radio stack; the bottom number shows the signal to
noise ratio achieved in the radio stack. In the left panel, the middle number is the amount of potential radio-AGN (not detected in the X-ray) that has been excluded from the stack. In the right panel this number gives the amount of optically very faint sources only detected redwards from the K-band. No radio-AGN candidate has been found among the radio-detected sources in the SF sample and only X-ray detected objects have been removed. The total number of galaxies per redshift bin is given below the panels.}
\label{fig:allbins} 
\end{figure*}
     
\section{Method and implementation of radio image stacking}
\label{sec:stack}
The bulk of objects in our $3.6~\mu \rm{m}$ selected sample is not individually detected in the 1.4~GHz continuum. An estimation of the SFR based on the radio flux density for {\it{every}} object in the sample is therefore impossible. On the other hand, studying only radio-detected galaxies in this sample yields effectively a selection by SFR and not by stellar mass since only radio-bright, i.e. highly active star forming, normal galaxies remain.\footnote{The currently deepest radio surveys \citep[e.g.][with rms$_{1.4~\rm{GHz}} \sim 3~\mu\rm{Jy}$]{OWEN08} individually detect galaxies with $\rm{SFRs} \gtrsim 50$~$M_{\sun}/ \rm{yr}$ at a redshift of $z=1$.} By co-adding postage stamp cutout images of the 1.4~GHz map at the positions of sources in the sample it is possible to estimate the typical radio properties for a specific galaxy population. Usually referred to as stacking, this technique has proven to be a powerful tool to estimate the typical flux density of galaxies with a given property, not only in the radio \citep[e.g.][]{WHIT07, CARI08, DUNN09, PANN09, GARN09, BOUR10, MESS10} but also in the mid-IR \citep[e.g.][]{ZHEN06, ZHEN07A, ZHEN07B, MART07B, BOUR10}, far-IR \citep[e.g.][]{NLEE10, RODI10, BOUR10} as well as sub-mm \citep[e.g.][]{GREV09, ALEJ09}. The list can be extended to other wavebands always requiring a galaxy sample representative for the underlying population.

\subsection{Median stacking and error estimates}
\label{sec:medst}
Our stacking algorithm uses cutouts with sizes of $40'' \times 40''$, centered on the position of the optical counterpart. Since the COSMOS astrometric reference system was provided by the VLA-COSMOS observations the positional accuracy between radio and optical sources should be well within the errors of both datasets. As detailed in \citet{SCHI07} the relative and absolute astrometry of the VLA data are 130 and $<55$~mas respectively. In other words the average distribution of radio flux follows the one at optical wavelengths and the central pixel in any stacked image was always the brightest one. Averaging over pixels located at the same position in each stamp hence is an astrometrically well-defined problem.

It can be approached by computing either the mean or the median of the mentioned set of pixels. The resulting stamp then shows the spatial distribution of the average radio emission for the sample studied. For an input sample of $N$ galaxies its background noise level should correspond to $\sim 1/\sqrt{N}$ of the noise measured in a single radio stamp.\footnote{Our image stacking implementation automatically monitors the decrease of the background noise level. For all results presented here it was verified that this decrease follows a $\sim 1/\sqrt{N}$ law.}
Any sample of galaxies in a given bin of redshift and stellar mass likely contains also a fraction of sources with radio detections. Even if this fraction is small, the mean is sensitive to the large excess in 1.4~GHz flux density compared to the average radio emission of the individual non-detections. On the other hand, setting a threshold and excluding radio detections from the stack artificially changes the sample and the results, hence, depend on the threshold applied. 

\begin{figure*} 
\includegraphics[width=\textwidth]{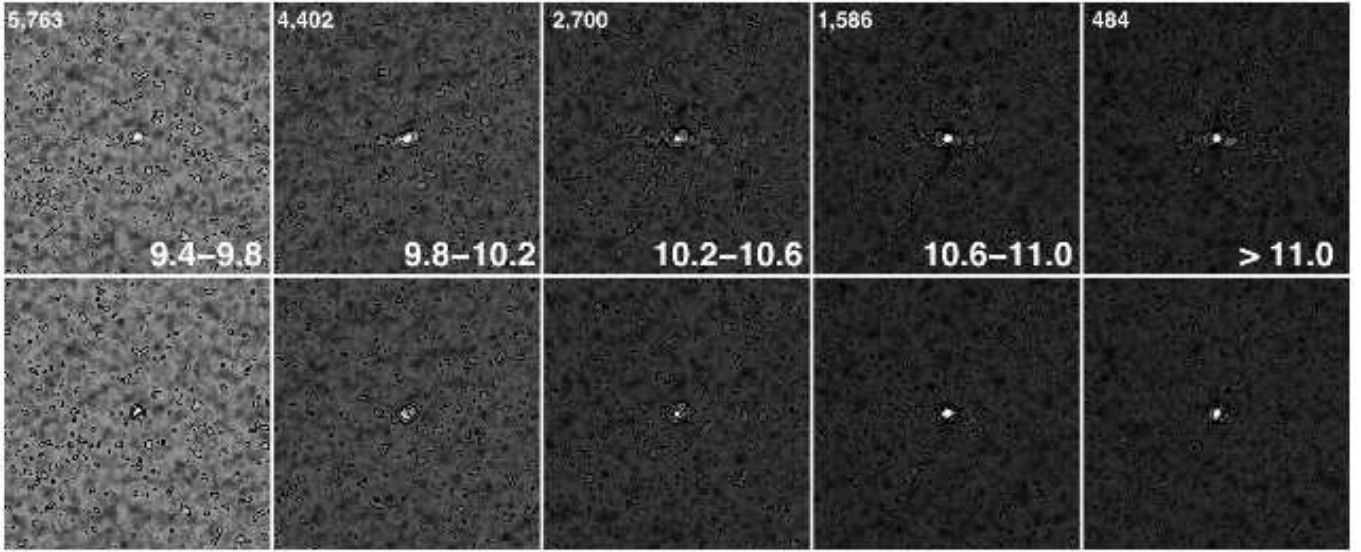}
\caption{\noindent Examples for $40'' \times 40''$ (i.e. ($115 \times 115$)~pixels) 1.4~GHz postage stamp images obtained via median stacking of star forming galaxies (see Sec. \ref{sec:class}) in
the redshift bin between $1.6 < z < 2$. The number of galaxies for which individual radio cutout images from the VLA-COSMOS map (resolution of $1.5'' \times 1.4''$) have been co-added is given at the upper left of each stamp in the top row while the number at the lower right denotes the bin extent in $\log(M_*~[M_{\odot}])$. Due to the high signal-to-noise ratios (SNRs) achieved, generally, a clear (dirty) beam pattern is visible. The bottom row shows the corresponding CLEANed stamps (see Sec. \ref{sec:im2flux}, for details). Contour levels are at 2, 4, 5~$\sigma\st{bg}$ and followed by steps of 5~$\sigma\st{bg}$ (The individual SNRs are given in Fig. \ref{fig:allbins} and flux densities measured as well as the background noise levels reached are listed in Tab. \ref{tab:fullsfg}).}
\label{fig:stamps} 
\end{figure*}

In addition, foreground objects and other extended radio bright features (e.g. lobes from radio galaxies) need to be handled with care and might be a source of contamination affecting the noise in the final stamp but also potentially the signal itself. It is therefore beneficial to exclude stamps showing these features from a mean stack. Typically, significantly less than $1~\%$ of objects in a sample are rejected and the effect of this artificial cut on the sample thus is negligible. However, by resorting to the median, the stacking technique becomes more robust against outliers allowing the use of the entire input sample.\footnote{We applied the different stacking techniques discussed above to some of our sub-samples. We found the median flux densities obtained to be within $\lesssim 7~\%$ of those obtained when using a mean stacking technique that excludes radio-stamps including extended foreground features. For the mean stack we co-added objects in a given sub-sample that are not individually detected in the radio imaging and the flux density of the detected sources has been added to the flux density obtained from the stack in a noise-weighted fashion. This ensures that those objects that are not individually radio-detected -- i.e. the bulk of our sources -- are most strongly weighted}. Non-uniform noise properties within the radio map can also be addressed by applying a weighted scheme to compute the median (see Appendix \ref{sec:app_stat}). While it is often argued that there is no straight-forward way of interpreting the sample median compared to the sample mean, 
\citet{WHIT07} showed that the median is a well-defined estimator of the mean of the underlying population in the presence of a dominant noise background.

Although, strictly speaking, these arguments only apply to the case of pure point sources, the condition of a dominant noise background is given in our study. One has to be aware of the fact that there is in principle no possibility to access the intrinsic distribution of radio peak fluxes of the underlying population as a whole. The observed distribution merely is the intrinsic one as smeared out by the gaussian noise background. However, it still contains information that needs to be used in order to find proper confidence limits for any statistic applied. Based on the above arguments, we expect the broadened distribution to be not only shifted but also skewed towards positive flux density values. As a result, the uncertainty for the obtained peak flux density is poorly estimated by the background noise in the final stamp. Using a bootstrapping technique (see Appendix \ref{sec:app_boot}) allows us to obtain more realistic, asymmetric error bars for our measured peak flux densities.

\subsection{Integrated flux densities, luminosities and SFRs from stacked radio images}
\label{sec:im2flux}
So far we considered only the average peak flux density which, to first order, would not require to stack individual cutouts but only their central pixel. However, the typical galaxy of a given sample might exhibit extended radio emission. In that case the peak flux density is no longer equivalent to the total source flux but underestimates the typical radio flux density and hence all other quantities derived from it.

The effect of bandwidth smearing (BWS), chromatic aberration caused by the finite bandwidth used during the VLA-COSMOS observations leads to a spatial broadening of a source even if it is intrinsically point-like. Within a single pointing the BWS increases with increasing radial distance from the pointing center and the effect is analytically well determined \citep[e.g.][]{BOND08}. For a mosaic like the VLA-COSMOS map that consists of many overlapping pointings the effect becomes analytically unpredictable due to the varying uncertainties introduced by the calibration and observing conditions. 

For all our samples we constructed median co-added cutout images (Fig. \ref{fig:stamps}) and determined 
accurate RMS-noise estimates (hereafter $\sigma_{\rm{Stack}}$) for the image stacks as described in Sec. \ref{sec:medst}. These $(115 \times 115)$~pixel$^2$ dirty maps were processed within AIPS.\footnote{Note that only bright ($> 45$~$\mu$Jy) radio sources have been CLEANed in the individual pointings prior to the assembly of the final mosaic. Hence, a stack of fainter sources will display a clear beam pattern as seen in Fig. \ref{fig:stamps} which must be deconvolved.} 

We used the task PADIM to make the stacked images equal in size to a $(512 \times 512)$~pixel$^2$ image of the VLA-COSMOS synthesized (dirty) beam by filling the outer image frame with additional pixels of constant value
The task APCLN with a circular CLEAN box of radius of seven pixels (i.e. $2.45''$) around the central component was then used to CLEAN each dirty map down to a flux density threshold of $2.5 \times \sigma_{\rm{Stack}}$\footnote{This is a conservative threshold. We confirmed that this choice does not lead to systematic biases by CLEANing individual stacked images down to $1 \times \sigma_{\rm{Stack}}$. Integrated flux densities obtained from both approaches do not differ by more than 3~\% and do not lead to mass-dependent effects. The mentioned fluctuations are well within the error margins.}.

Integrated flux densities, as well as source dimensions and position angles after deconvolution with the CLEAN beam were obtained by fitting a single-component Gaussian elliptical model to the CLEAN image within a quadratic box of $(15 \times 15)$~pixel$^2$ around the central pixel using the task JMFIT.
Errors on the integrated flux densities have been estimated according to \citet{HOPK03} and rely on the combined information on the best-fit source model and the bootstrapping results from the image stacking:
\for{\label{eq:errftot}\frac{\sigma_{\rm{Total}}}{\langle F_{\rm{Total}} \rangle}=\sqrt{\left( \frac{\sigma_{\rm{data}}}{\langle F_{\rm{Total}} \rangle} \right)^2 + \left( \frac{\sigma_{\rm{fit}}}{\langle F_{\rm{Total}}\rangle} \right)^2 },}
where (\citealp{WIND84, COND97} and also the explanations in \citealp{SCHI04, SCHI10}) 
\begin{eqnarray} \label{eq:errftotd}
\frac{\sigma_{\rm{data}}}{\langle F_{\rm{Total}} \rangle} &=& \sqrt{\left(\frac{S}{N}\right)^{-2} + \left( \frac{1}{100} \right)^2}\\ \label{eq:errftotf}
\frac{\sigma_{\rm{fit}}}{\langle F_{\rm{Total}} \rangle} &=& \sqrt{\frac{2}{\rho_S} + \left( \frac{\theta\st{B} \theta\st{b}}{\theta\st{M} \, \theta\st{m}} \right) \left( \frac{2}{\rho^2_{\psi}} + \frac{2}{\rho^2_{\phi}} \right)}.
\end{eqnarray}
$\theta\st{M}=1.5''$ is the major axis and $\theta\st{m}=1.4''$ the minor axis of the beam while $\theta\st{M}$ and $\theta\st{m}$ are the major and minor axis of the measured (hence convolved) flux density distribution. In order to include the bootstrapping error estimates we set $S/N=\langle F_{\rm{peak}} \rangle/\sigma_{\rm{bs}}$, i.e. the ratio of the peak flux density in the stacked dirty map and the 68~\% confidence interval resulting from the bootstrapping. The same applies to the parameter-dependent estimators of the fit entering equation \gl{eq:errftotf} that are given by:
\for{\rho^2_X = \frac{\theta\st{M} \, \theta\st{m}}{4 \, \theta\st{B} \theta\st{b}} \left( 1 + \frac{\theta\st{B}}{\theta\st{M}} \right)^{a} \left( 1 + \frac{\theta\st{b}}{\theta\st{m}} \right)^{b} \left(\frac{S}{N} \right)^2}
and $a = b = 1.5$ for $ \rho_F$, $a = 2.5$ and $b = 0.5$ for $ \rho\st{M}$ as well as $a =0.5$ and $b = 2.5$ for $ \rho\st{m}$.

For a given sub-sample centered at a given median redshift $\langle z\st{phot} \rangle$ the average (median stacking based) integrated flux density $\langle F_{\rm{Total}} \rangle$ observed at 1.4~GHz can be directly converted into a rest-frame 1.4~GHz luminosity using a K-correction that depends on the radio spectral index $\alpha\st{rc}$ (here $\alpha\st{rc}=-0.8$, \citealp[e.g.][]{COND92}):
\begin{eqnarray}
\label{eq:flux2lum}
\nonumber
\langle L_{1.4~\rm{GHz}} \rangle ~[\rm{W/Hz}] &=& 9.52 \times 10^{12} \, \langle F_{\rm{Total}} \rangle~[\mu \rm{Jy}] \\
&\times& \bigg(D_L~[\rm{Mpc}]\bigg)^2 \, 4 \pi \bigg(1+\langle z\st{phot} \rangle \bigg)^{1+\alpha\st{rc}}
\end{eqnarray}
with $D_L$ the luminosity distance at this median photo-z of all objects inside the bin. 

\begin{figure} 
\includegraphics[width=0.5\textwidth]{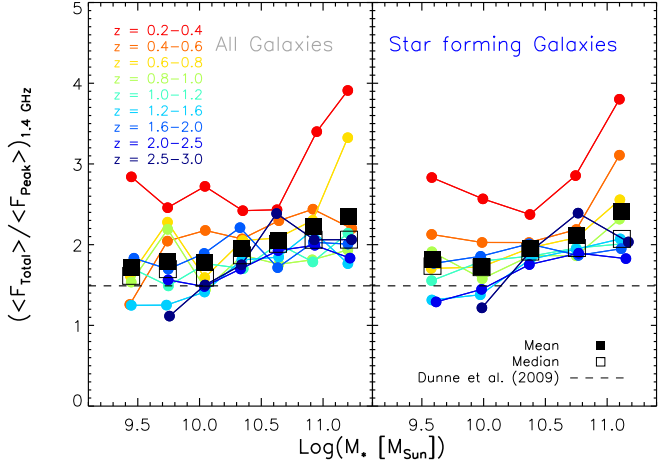}
\caption{\noindent Ratio of integrated to peak flux density at 1.4~GHz of the stacked radio images for different sample sub-sets. The left panel shows results for the entire sample, the right panel the sub-set of blue (SF) galaxies. The data is color-coded by redshift and the dashed black line depicts the uniform correction factor used in the radio stacking study by \citet{DUNN09}. It is evident that the extent of radio emission is not uniform across our samples because our radio imaging has a higher angular resolution compared to the VLA map used by \citet{DUNN09}. All measured data points are listed in Tab. \ref{tab:fullall} and \ref{tab:fullsfg}.}
\label{fig:pktotall} 
\end{figure}

It was pointed out by \citet{DUNN09} that the median redshift might not be appropriate for estimating the radio luminosity if the peak of the radio flux density distribution does not coincide with the median of the photo-$z$ distribution. They overcame this problem by deriving (and subsequently effectively stacking) luminosities according to Eq. \gl{eq:flux2lum} for all objects relying both on the individual photo-$z$'s and peak flux density measurements at the pixel corresponding to the position in the input catalog. At $z>0.2$ they afterwards applied a common (i.e. redshift independent) factor
to the median of all obtained luminosities to correct for the difference between peak and total flux density as well as for the effect of BWS. A similar approach was recently also used by \citet{BOUR10}.
The method by \citet{DUNN09} is justified given their data as they find for $z > 0.2$ that the ratio of total to peak flux density does not change significantly, in particular not as a function of K-band magnitude.
However, our data does not yield such a uniform behavior with respect to mass in the correction factor as Fig. \ref{fig:pktotall} shows.  
Indeed, if we were to state an average peak to total flux density conversion it would be a function of mass. An explanation for the discrepancy of our findings compared to \citet{DUNN09} can be found in the use of higher resolved A-array data in our case compared to the B-array data constituting their radio continuum imaging. Hence, both results are correct given the respective data used and show that higher resolved radio data needs to be treated differently. 
The spread in conversion factors within our sub-samples is large and lower redshift objects show a significantly larger $\langle F_{\rm{Total}}\rangle/\langle F_{\rm{Peak}}\rangle$ ratio\footnote{Note that a larger conversion factor is equivalent to a larger source extent. Since it is unlikely that the varying number counts in our sub-samples are responsible for mass- or redshift-dependent source sizes we infer that higher mass objects are intrinsically more extended at all redshifts compared to their lower mass siblings. The larger correction factors at lower $z$ can be explained by the increasing angular diameter distance towards higher $z$.} (see Fig. \ref{fig:pktotall}). Moreover, further variations might arise depending on the galaxy population studied. Hence, if high-resolution data is used, results are more robust when first total flux densities are individually derived for any radio stacking experiment before computing radio luminosities. As it is apparent from the \citet{DUNN09} results their method should be considered, however, if stacking is used to infer the average radio luminosity of an entire galaxy population with a broad redshift distribution ($\Delta z \gtrsim 1$). As our broadest bins in redshift have $\Delta z = 0.5$ -- and this only at $z \gg 1$ where they span a much smaller range in time -- it is indeed more accurate to rely on our approach given our radio imaging.  

\begin{deluxetable*}{ccccccccc}
\thispagestyle{empty}
\vspace{-2.5cm}
%\rotate
\tabletypesize{\tiny}
\tablecaption{\label{tab:fullall}Radio stacking results for {\emph{the entire mass-selected sample}}}
%\tablenum{2}
\tablehead{\colhead{$\Delta \log(M_{*})$} & \colhead{$ \langle \log(M_{*}) \rangle$} &  \colhead{$\Delta z_{\rm{phot}}$} & \colhead{$ \langle z\st{phot} \rangle$} & \colhead{$ \langle F\st{Peak}\rangle $} & \colhead{$ \langle F\st{Total}\rangle $} & \colhead{rms} & \colhead{$  \langle L_{1.4~\rm{GHz}}\rangle / L_c$} & \colhead{  $\langle$SFR$\rangle$} \\ 
\colhead{$M_{*}~[M_{\odot}]$} & \colhead{$M_{*}~[M_{\odot}]$} & \colhead{} & \colhead{} & \colhead{$[\mu$Jy/beam]} & \colhead{$[\mu$Jy]} & \colhead{$[\mu$Jy/beam]}& \colhead{$L_c=6.4 \times 10^{21}$~W/Hz} & \colhead{$[M_{\odot}/$yr]}} 
\startdata
9.3-9.6 & 9.45\tablenotemark{$\dagger$} & 0.2-0.4 & 0.27 & $1.7^{+0.3}_{-0.4}$ & $4.9^{+1.4}_{-2.2}$ & $0.399^{+0.015}_{-0.002}$ & $0.2^{+0.0}_{-0.1}$ & $0.6^{+0.1}_{-0.2}$\\
 & 9.44\tablenotemark{$\dagger$} & 0.4-0.6 & 0.49 & $2.0^{+0.4}_{-0.3}$ & $2.5^{+1.1}_{-1.0}$ & $0.388^{+0.006}_{-0.006}$ & $0.3^{+0.1}_{-0.1}$ & $0.9^{+0.3}_{-0.3}$\\
 & 9.44\tablenotemark{$\dagger$} & 0.6-0.8 & 0.69 & $1.4^{+0.2}_{-0.3}$ & $2.2^{+1.2}_{-1.9}$ & $0.276^{+0.004}_{-0.006}$ & $0.6^{+0.4}_{-0.6}$ & $1.5^{+0.6}_{-1.0}$\\
 & 9.45\tablenotemark{$\dagger$} & 0.8-1.0 & 0.89 & $1.3^{+0.2}_{-0.3}$ & $1.9^{+0.5}_{-0.9}$ & $0.246^{+0.007}_{-0.013}$ & $1.0^{+0.3}_{-0.5}$ & $2.1^{+0.5}_{-1.0}$\\
 & 9.44\tablenotemark{$\dagger$} & 1.0-1.2 & 1.11 & $1.1^{+0.3}_{-0.2}$ & $1.9^{+1.6}_{-1.2}$ & $0.275^{+0.005}_{-0.010}$ & $1.7^{+1.4}_{-1.1}$ & $3.5^{+2.8}_{-2.3}$\\
 & 9.45\tablenotemark{$\dagger$} & 1.2-1.6 & 1.40 & $0.7^{+0.2}_{-0.1}$ & $0.9^{+0.5}_{-0.4}$ & $0.179^{+0.001}_{-0.003}$ & $1.4^{+0.9}_{-0.6}$ & $2.8^{+1.8}_{-1.2}$\\
 & 9.47\tablenotemark{$\dagger$} & 1.6-2.0 & 1.78 & $0.5^{+0.2}_{-0.3}$ & $0.9^{+0.8}_{-1.0}$ & $0.243^{+0.004}_{-0.006}$ & $2.6^{+2.3}_{-2.8}$ & $5.2^{+4.6}_{-5.7}$\\ \hline
9.6-9.9 & 9.74\tablenotemark{$\dagger$} & 0.2-0.4 & 0.27 & $3.9^{+0.6}_{-0.2}$ & $9.6^{+2.5}_{-1.0}$ & $0.469^{+0.017}_{-0.033}$ & $0.3^{+0.1}_{-0.0}$ & $0.9^{+0.2}_{-0.1}$\\
 & 9.74\tablenotemark{$\dagger$} & 0.4-0.6 & 0.49 & $3.6^{+0.3}_{-0.4}$ & $7.4^{+1.3}_{-1.9}$ & $0.455^{+0.010}_{-0.004}$ & $1.0^{+0.2}_{-0.2}$ & $2.0^{+0.3}_{-0.4}$\\
 & 9.74\tablenotemark{$\dagger$} & 0.6-0.8 & 0.69 & $2.4^{+0.2}_{-0.3}$ & $5.4^{+0.9}_{-1.3}$ & $0.341^{+0.028}_{-0.002}$ & $1.6^{+0.3}_{-0.4}$ & $3.2^{+0.6}_{-0.8}$\\
 & 9.74\tablenotemark{$\dagger$} & 0.8-1.0 & 0.89 & $2.1^{+0.3}_{-0.2}$ & $4.5^{+1.4}_{-0.9}$ & $0.265^{+0.002}_{-0.010}$ & $2.4^{+0.8}_{-0.5}$ & $4.9^{+1.6}_{-0.9}$\\
 & 9.75\tablenotemark{$\dagger$} & 1.0-1.2 & 1.10 & $2.7^{+0.3}_{-0.3}$ & $4.0^{+1.4}_{-1.3}$ & $0.320^{+0.022}_{-0.005}$ & $3.5^{+1.3}_{-1.2}$ & $7.2^{+2.6}_{-2.4}$\\
 & 9.73\tablenotemark{$\dagger$} & 1.2-1.6 & 1.41 & $1.9^{+0.2}_{-0.3}$ & $2.3^{+1.1}_{-1.4}$ & $0.216^{+0.006}_{-0.007}$ & $3.7^{+1.8}_{-2.3}$ & $7.6^{+3.6}_{-4.7}$\\
 & 9.75\tablenotemark{$\dagger$} & 1.6-2.0 & 1.82 & $1.3^{+0.2}_{-0.2}$ & $2.1^{+1.0}_{-1.0}$ & $0.232^{+0.003}_{-0.015}$ & $6.2^{+3.0}_{-3.1}$ & $12.7^{+6.0}_{-6.2}$\\
 & 9.75\tablenotemark{$\dagger$} & 2.0-2.5 & 2.21 & $1.3^{+0.2}_{-0.3}$ & $2.0^{+0.7}_{-0.8}$ & $0.271^{+0.015}_{-0.008}$ & $8.9^{+3.0}_{-3.8}$ & $18.2^{+6.1}_{-7.6}$\\
 & 9.76\tablenotemark{$\dagger$} & 2.5-3.0 & 2.65 & $0.9^{+0.2}_{-0.3}$ & $1.0^{+0.5}_{-0.6}$ & $0.293^{+0.005}_{-0.005}$ & $6.9^{+3.3}_{-4.2}$ & $14.1^{+6.6}_{-8.6}$\\ \hline
9.9-10.2 & 10.04\tablenotemark{$\star$} & 0.2-0.4 & 0.27 & $5.1^{+0.6}_{-0.5}$ & $13.8^{+2.9}_{-2.4}$ & $0.504^{+0.014}_{-0.008}$ & $0.5^{+0.1}_{-0.1}$ & $1.2^{+0.2}_{-0.2}$\\
 & 10.05\tablenotemark{$\star$} & 0.4-0.6 & 0.49 & $5.4^{+0.5}_{-0.5}$ & $11.8^{+2.1}_{-2.2}$ & $0.472^{+0.016}_{-0.022}$ & $1.5^{+0.3}_{-0.3}$ & $3.1^{+0.6}_{-0.6}$\\
 & 10.04\tablenotemark{$\dagger$} & 0.6-0.8 & 0.68 & $5.3^{+0.2}_{-0.5}$ & $8.5^{+0.7}_{-1.7}$ & $0.360^{+0.011}_{-0.013}$ & $2.4^{+0.2}_{-0.5}$ & $5.0^{+0.4}_{-1.0}$\\
 & 10.05\tablenotemark{$\dagger$} & 0.8-1.0 & 0.89 & $3.9^{+0.2}_{-0.4}$ & $5.7^{+0.8}_{-1.5}$ & $0.288^{+0.005}_{-0.009}$ & $3.1^{+0.4}_{-0.8}$ & $6.3^{+0.9}_{-1.7}$\\
 & 10.05\tablenotemark{$\dagger$} & 1.0-1.2 & 1.09 & $4.0^{+0.3}_{-0.4}$ & $7.0^{+1.4}_{-1.6}$ & $0.364^{+0.002}_{-0.004}$ & $6.1^{+1.2}_{-1.4}$ & $12.5^{+2.5}_{-2.9}$\\
 & 10.04\tablenotemark{$\dagger$} & 1.2-1.6 & 1.39 & $3.3^{+0.2}_{-0.2}$ & $4.7^{+1.0}_{-1.0}$ & $0.260^{+0.003}_{-0.006}$ & $7.3^{+1.6}_{-1.5}$ & $14.8^{+3.2}_{-3.0}$\\
 & 10.04\tablenotemark{$\dagger$} & 1.6-2.0 & 1.87 & $2.9^{+0.2}_{-0.2}$ & $5.5^{+0.8}_{-0.8}$ & $0.274^{+0.005}_{-0.003}$ & $17.0^{+2.6}_{-2.4}$ & $34.6^{+5.2}_{-4.9}$\\
 & 10.04\tablenotemark{$\dagger$} & 2.0-2.5 & 2.28 & $2.2^{+0.2}_{-0.3}$ & $3.2^{+0.9}_{-1.1}$ & $0.277^{+0.006}_{-0.003}$ & $15.9^{+4.3}_{-5.5}$ & $32.4^{+8.8}_{-11.2}$\\
 & 10.04\tablenotemark{$\dagger$} & 2.5-3.0 & 2.73 & $1.7^{+0.3}_{-0.3}$ & $2.5^{+1.3}_{-1.2}$ & $0.308^{+0.004}_{-0.007}$ & $18.8^{+9.3}_{-8.8}$ & $38.2^{+18.9}_{-17.9}$\\ \hline
10.2-10.5 & 10.35 & 0.2-0.4 & 0.29 & $7.5^{+0.5}_{-0.4}$ & $18.1^{+2.4}_{-1.7}$ & $0.534^{+0.016}_{-0.005}$ & $0.7^{+0.1}_{-0.1}$ & $1.6^{+0.2}_{-0.1}$\\
 & 10.34 & 0.4-0.6 & 0.49 & $7.9^{+0.4}_{-0.4}$ & $16.4^{+1.9}_{-1.8}$ & $0.495^{+0.010}_{-0.018}$ & $2.1^{+0.2}_{-0.2}$ & $4.3^{+0.5}_{-0.5}$\\
 & 10.35\tablenotemark{$\star$} & 0.6-0.8 & 0.68 & $4.8^{+0.5}_{-0.5}$ & $9.8^{+2.2}_{-2.1}$ & $0.386^{+0.025}_{-0.009}$ & $2.8^{+0.6}_{-0.6}$ & $5.7^{+1.3}_{-1.2}$\\
 & 10.35\tablenotemark{$\star$} & 0.8-1.0 & 0.89 & $5.1^{+0.4}_{-0.4}$ & $9.5^{+1.7}_{-1.6}$ & $0.311^{+0.005}_{-0.006}$ & $5.1^{+0.9}_{-0.9}$ & $10.5^{+1.9}_{-1.8}$\\
 & 10.35\tablenotemark{$\star$} & 1.0-1.2 & 1.09 & $5.4^{+0.3}_{-0.3}$ & $9.2^{+1.3}_{-1.2}$ & $0.375^{+0.020}_{-0.009}$ & $8.1^{+1.1}_{-1.0}$ & $16.6^{+2.3}_{-2.1}$\\
 & 10.34\tablenotemark{$\dagger$} & 1.2-1.6 & 1.38 & $4.6^{+0.4}_{-0.3}$ & $8.3^{+1.5}_{-1.4}$ & $0.279^{+0.001}_{-0.012}$ & $12.5^{+2.3}_{-2.1}$ & $25.4^{+4.7}_{-4.2}$\\
 & 10.33\tablenotemark{$\dagger$} & 1.6-2.0 & 1.81 & $4.2^{+0.4}_{-0.2}$ & $9.3^{+1.7}_{-1.0}$ & $0.325^{+0.008}_{-0.009}$ & $26.8^{+4.8}_{-2.9}$ & $54.5^{+9.7}_{-5.9}$\\
 & 10.33\tablenotemark{$\dagger$} & 2.0-2.5 & 2.36 & $3.7^{+0.4}_{-0.4}$ & $6.2^{+1.5}_{-1.5}$ & $0.330^{+0.025}_{-0.006}$ & $33.0^{+8.1}_{-7.8}$ & $67.1^{+16.4}_{-15.9}$\\
 & 10.34\tablenotemark{$\dagger$} & 2.5-3.0 & 2.81 & $3.3^{+0.3}_{-0.6}$ & $5.8^{+1.2}_{-2.3}$ & $0.343^{+0.009}_{-0.005}$ & $45.5^{+9.4}_{-18.5}$ & $92.6^{+19.2}_{-37.6}$\\ \hline
10.5-10.8 & 10.63 & 0.2-0.4 & 0.28 & $9.8^{+0.7}_{-0.7}$ & $23.9^{+3.1}_{-2.9}$ & $0.594^{+0.027}_{-0.019}$ & $0.9^{+0.1}_{-0.1}$ & $1.8^{+0.2}_{-0.2}$\\
 & 10.64 & 0.4-0.6 & 0.48 & $8.1^{+0.8}_{-0.4}$ & $18.6^{+3.4}_{-2.0}$ & $0.555^{+0.010}_{-0.026}$ & $2.4^{+0.4}_{-0.2}$ & $4.8^{+0.9}_{-0.5}$\\
 & 10.63 & 0.6-0.8 & 0.69 & $6.4^{+0.4}_{-0.6}$ & $13.3^{+1.6}_{-2.7}$ & $0.420^{+0.006}_{-0.020}$ & $3.9^{+0.5}_{-0.8}$ & $7.9^{+0.9}_{-1.6}$\\
 & 10.64 & 0.8-1.0 & 0.89 & $6.3^{+0.4}_{-0.5}$ & $11.1^{+1.5}_{-2.0}$ & $0.331^{+0.002}_{-0.010}$ & $6.0^{+0.8}_{-1.1}$ & $12.2^{+1.6}_{-2.2}$\\
 & 10.64 & 1.0-1.2 & 1.09 & $5.6^{+0.4}_{-0.4}$ & $11.3^{+1.7}_{-1.7}$ & $0.391^{+0.013}_{-0.016}$ & $9.8^{+1.5}_{-1.5}$ & $20.0^{+3.0}_{-3.0}$\\
 & 10.64\tablenotemark{$\star$} & 1.2-1.6 & 1.37 & $6.2^{+0.5}_{-0.4}$ & $11.4^{+2.0}_{-1.9}$ & $0.308^{+0.004}_{-0.004}$ & $17.1^{+3.0}_{-2.8}$ & $34.7^{+6.1}_{-5.7}$\\
 & 10.63\tablenotemark{$\star$} & 1.6-2.0 & 1.78 & $6.9^{+0.4}_{-0.3}$ & $11.8^{+1.6}_{-1.1}$ & $0.378^{+0.017}_{-0.009}$ & $32.6^{+4.4}_{-3.0}$ & $66.3^{+9.0}_{-6.1}$\\
 & 10.64\tablenotemark{$\dagger$} & 2.0-2.5 & 2.27 & $5.2^{+0.3}_{-0.3}$ & $10.0^{+1.2}_{-1.4}$ & $0.400^{+0.008}_{-0.008}$ & $48.3^{+6.0}_{-6.6}$ & $98.3^{+12.3}_{-13.4}$\\ 
 & 10.62\tablenotemark{$\dagger$} & 2.5-3.0 & 2.76 & $4.9^{+0.5}_{-0.6}$ & $11.6^{+2.6}_{-2.6}$ & $0.437^{+0.003}_{-0.012}$ & $88.4^{+19.6}_{-20.0}$ & $179.7^{+39.8}_{-40.6}$\\ \hline
10.8-11.1 & 10.95 & 0.2-0.4 & 0.27 & $9.6^{+1.0}_{-1.0}$ & $32.7^{+5.4}_{-5.1}$ & $0.882^{+0.009}_{-0.007}$ & $1.1^{+0.2}_{-0.2}$ & $2.2^{+0.4}_{-0.4}$\\
 & 10.92 & 0.4-0.6 & 0.48 & $9.2^{+0.6}_{-0.4}$ & $22.5^{+2.6}_{-2.0}$ & $0.733^{+0.030}_{-0.009}$ & $2.8^{+0.3}_{-0.2}$ & $5.7^{+0.7}_{-0.5}$\\
 & 10.92 & 0.6-0.8 & 0.69 & $6.7^{+0.7}_{-0.7}$ & $15.5^{+3.2}_{-3.0}$ & $0.556^{+0.034}_{-0.013}$ & $4.5^{+0.9}_{-0.9}$ & $9.2^{+1.9}_{-1.8}$\\
 & 10.91 & 0.8-1.0 & 0.90 & $7.1^{+0.5}_{-0.3}$ & $12.8^{+1.8}_{-1.1}$ & $0.412^{+0.025}_{-0.014}$ & $7.0^{+1.0}_{-0.6}$ & $14.3^{+2.0}_{-1.3}$\\
 & 10.92 & 1.0-1.2 & 1.10 & $7.6^{+0.5}_{-0.4}$ & $13.6^{+1.8}_{-1.6}$ & $0.509^{+0.008}_{-0.011}$ & $12.1^{+1.6}_{-1.5}$ & $24.6^{+3.2}_{-3.0}$\\
 & 10.91 & 1.2-1.6 & 1.36 & $6.3^{+0.6}_{-0.6}$ & $13.7^{+2.5}_{-2.5}$ & $0.417^{+0.010}_{-0.008}$ & $20.2^{+3.7}_{-3.7}$ & $41.0^{+7.4}_{-7.5}$\\
 & 10.92 & 1.6-2.0 & 1.79 & $7.8^{+0.3}_{-0.3}$ & $15.9^{+1.4}_{-1.2}$ & $0.479^{+0.003}_{-0.005}$ & $44.5^{+3.9}_{-3.4}$ & $90.5^{+7.8}_{-6.8}$\\
 & 10.93\tablenotemark{$\star$} & 2.0-2.5 & 2.21 & $6.9^{+0.5}_{-0.4}$ & $13.8^{+2.0}_{-1.6}$ & $0.523^{+0.012}_{-0.009}$ & $63.1^{+8.9}_{-7.4}$ & $128.4^{+18.1}_{-15.1}$\\ 
 & 10.93\tablenotemark{$\dagger$} & 2.5-3.0 & 2.72 & $6.8^{+0.5}_{-0.7}$ & $14.0^{+2.0}_{-3.0}$ & $0.583^{+0.013}_{-0.036}$ & $102.3^{+14.6}_{-21.7}$ & $207.9^{+29.6}_{-44.1}$\\ \hline
$>11.1$ & 11.20 & 0.2-0.4 & 0.27 & $9.2^{+0.9}_{-1.9}$ & $36.0^{+5.5}_{-10.8}$ & $1.250^{+0.047}_{-0.038}$ & $1.2^{+0.2}_{-0.4}$ & $2.5^{+0.4}_{-0.8}$\\
 & 11.23 & 0.4-0.6 & 0.48 & $10.3^{+2.0}_{-3.2}$ & $22.6^{+8.4}_{-13.0}$ & $1.227^{+0.008}_{-0.025}$ & $2.8^{+1.0}_{-1.6}$ & $5.7^{+2.1}_{-3.3}$\\
 & 11.20 & 0.6-0.8 & 0.69 & $7.3^{+1.2}_{-1.2}$ & $24.4^{+6.1}_{-6.5}$ & $0.897^{+0.010}_{-0.007}$ & $7.1^{+1.8}_{-1.9}$ & $14.4^{+3.6}_{-3.8}$\\
 & 11.20 & 0.8-1.0 & 0.90 & $8.6^{+0.9}_{-1.0}$ & $16.6^{+3.6}_{-4.0}$ & $0.681^{+0.013}_{-0.009}$ & $9.2^{+2.0}_{-2.2}$ & $18.7^{+4.0}_{-4.5}$\\
 & 11.20 & 1.0-1.2 & 1.10 & $10.1^{+0.7}_{-0.8}$ & $21.6^{+2.8}_{-3.1}$ & $0.881^{+0.018}_{-0.013}$ & $19.3^{+2.5}_{-2.8}$ & $39.2^{+5.1}_{-5.7}$\\
 & 11.20 & 1.2-1.6 & 1.35 & $11.1^{+0.9}_{-1.1}$ & $19.6^{+3.5}_{-4.3}$ & $0.762^{+0.012}_{-0.022}$ & $28.2^{+5.1}_{-6.2}$ & $57.4^{+10.3}_{-12.5}$\\
 & 11.20 & 1.6-2.0 & 1.78 & $14.3^{+0.7}_{-1.0}$ & $28.7^{+3.0}_{-4.0}$ & $0.835^{+0.022}_{-0.009}$ & $79.7^{+8.4}_{-11.0}$ & $162.0^{+17.1}_{-22.4}$\\
 & 11.22 & 2.0-2.5 & 2.22 & $13.7^{+1.1}_{-0.9}$ & $25.0^{+4.1}_{-3.6}$ & $0.737^{+0.023}_{-0.025}$ & $115.3^{+18.9}_{-16.5}$ & $234.4^{+38.5}_{-33.6}$\\
 & 11.23\tablenotemark{$\star$} & 2.5-3.0 & 2.71 & $11.3^{+0.9}_{-0.8}$ & $23.2^{+3.8}_{-3.5}$ & $0.767^{+0.021}_{-0.030}$ & $169.4^{+27.8}_{-25.7}$ & $344.4^{+56.5}_{-52.3}$
\enddata
\tablecomments{Median stacking-based average 1.4~GHz radio flux densities and derived average quantities for all our bins in mass and redshift for {\emph{the entire mass-selected sample}}. A \citet{CHAB03} IMF is assumed. Radio luminosities are stated in units of $L_c$, the threshold luminosity below which \citet{BELL03A} empirically found the non-thermal radio emission to be suppressed (see Eq. \gl{eq:lum2sfr}). Resulting SFRs from bins with lower radio luminosity are hence boosted compared to e.g. the calibration of the radio-IR relation by \citet{YUNR01}. The median stellar mass and median $z$ for any given bin are also stated.}
\tablenotetext{$\dagger$}{Mass bin contains data below the limit of mass representativeness and yields an upper limit to the average SFR (see Sec. \ref{sec:comp} for further details.)}
\tablenotetext{$\star$}{First mass bin above the limit of representativeness (see Sec. \ref{sec:comp}) which contains a low fraction ($<15~\%$) of optically faint objects with $m_{AB}(i^+) \ge 25.5$ for which the photo-z accuracy is degraded (see Sec. \ref{sec:irsamp} for further details).}
\end{deluxetable*}

\begin{deluxetable*}{ccccccccc}
%\rotate
\tabletypesize{\tiny}
\tablecaption{\label{tab:fullsfg}Radio stacking results for {\emph{star forming systems}}}
%\tablenum{3}
\tablehead{\colhead{$\Delta \log(M_{*})$} & \colhead{$ \langle \log(M_{*}) \rangle$} &  \colhead{$\Delta z_{\rm{phot}}$} & \colhead{$ \langle z\st{phot} \rangle$} & \colhead{$ \langle F\st{Peak}\rangle $} & \colhead{$ \langle F\st{Total}\rangle $} & \colhead{rms} & \colhead{$  \langle L_{1.4~\rm{GHz}}\rangle / L_c$} & \colhead{  $\langle$SFR$\rangle$} \\ 
\colhead{$M_{*}~[M_{\odot}]$} & \colhead{$M_{*}~[M_{\odot}]$} & \colhead{} & \colhead{} & \colhead{$[\mu$Jy/beam]} & \colhead{$[\mu$Jy]} & \colhead{$[\mu$Jy/beam]}& \colhead{$L_c=6.4 \times 10^{21}$~W/Hz} & \colhead{$[M_{\odot}/$yr]}} 
\startdata
9.4-9.8 & 9.58 & 0.2-0.4 & 0.28 & $3.1^{+0.5}_{-0.5}$ & $8.7^{+2.3}_{-2.2}$ & $0.408^{+0.001}_{-0.002}$ & $0.3^{+0.1}_{-0.1}$ & $0.9^{+0.2}_{-0.2}$\\
 & 9.58 & 0.4-0.6 & 0.49 & $2.4^{+0.2}_{-0.4}$ & $5.1^{+1.4}_{-2.0}$ & $0.372^{+0.016}_{-0.033}$ & $0.7^{+0.2}_{-0.3}$ & $1.5^{+0.3}_{-0.5}$\\
 & 9.58 & 0.6-0.8 & 0.69 & $2.1^{+0.3}_{-0.2}$ & $3.6^{+1.2}_{-0.7}$ & $0.258^{+0.009}_{-0.002}$ & $1.0^{+0.3}_{-0.2}$ & $2.1^{+0.7}_{-0.4}$\\
 & 9.58 & 0.8-1.0 & 0.89 & $1.7^{+0.2}_{-0.2}$ & $3.2^{+0.7}_{-1.0}$ & $0.219^{+0.004}_{-0.006}$ & $1.7^{+0.4}_{-0.5}$ & $3.6^{+0.8}_{-1.1}$\\
 & 9.58\tablenotemark{$\star$} & 1.0-1.2 & 1.10 & $2.1^{+0.3}_{-0.2}$ & $3.2^{+1.0}_{-0.7}$ & $0.246^{+0.006}_{-0.009}$ & $2.9^{+0.9}_{-0.6}$ & $5.9^{+1.9}_{-1.2}$\\
 & 9.58\tablenotemark{$\dagger$} & 1.2-1.6 & 1.40 & $1.1^{+0.2}_{-0.2}$ & $1.4^{+0.5}_{-0.5}$ & $0.169^{+0.001}_{-0.004}$ & $2.2^{+0.8}_{-0.8}$ & $4.5^{+1.7}_{-1.7}$\\
 & 9.60\tablenotemark{$\dagger$} & 1.6-2.0 & 1.79 & $0.8^{+0.2}_{-0.2}$ & $1.5^{+0.9}_{-1.3}$ & $0.210^{+0.004}_{-0.004}$ & $4.2^{+2.6}_{-3.7}$ & $8.5^{+5.4}_{-7.4}$\\
 & 9.62\tablenotemark{$\dagger$} & 2.0-2.5 & 2.17 & $1.1^{+0.2}_{-0.2}$ & $1.4^{+0.6}_{-0.7}$ & $0.237^{+0.001}_{-0.000}$ & $6.3^{+2.5}_{-2.9}$ & $12.7^{+5.1}_{-5.9}$\\ \hline
9.8-10.2 & 9.99 & 0.2-0.4 & 0.28 & $6.9^{+0.4}_{-0.8}$ & $17.7^{+2.0}_{-3.5}$ & $0.523^{+0.019}_{-0.011}$ & $0.6^{+0.1}_{-0.1}$ & $1.5^{+0.1}_{-0.2}$\\
 & 9.99 & 0.4-0.6 & 0.49 & $6.2^{+0.5}_{-0.5}$ & $12.6^{+2.0}_{-1.9}$ & $0.441^{+0.011}_{-0.016}$ & $1.7^{+0.3}_{-0.3}$ & $3.4^{+0.5}_{-0.5}$\\
 & 9.98 & 0.6-0.8 & 0.68 & $5.4^{+0.3}_{-0.6}$ & $9.3^{+1.2}_{-2.2}$ & $0.336^{+0.022}_{-0.029}$ & $2.7^{+0.3}_{-0.6}$ & $5.4^{+0.7}_{-1.3}$\\
 & 9.99 & 0.8-1.0 & 0.89 & $3.9^{+0.2}_{-0.4}$ & $6.2^{+0.9}_{-1.5}$ & $0.264^{+0.002}_{-0.005}$ & $3.4^{+0.5}_{-0.8}$ & $6.8^{+1.0}_{-1.6}$\\
 & 9.99 & 1.0-1.2 & 1.09 & $3.7^{+0.3}_{-0.3}$ & $6.6^{+1.4}_{-1.3}$ & $0.305^{+0.010}_{-0.006}$ & $5.8^{+1.2}_{-1.1}$ & $11.9^{+2.4}_{-2.2}$\\
 & 9.97\tablenotemark{$\star$} & 1.2-1.6 & 1.39 & $3.2^{+0.2}_{-0.2}$ & $4.4^{+1.0}_{-1.0}$ & $0.225^{+0.014}_{-0.001}$ & $6.8^{+1.5}_{-1.5}$ & $13.9^{+3.1}_{-3.1}$\\
 & 9.98\tablenotemark{$\star$} & 1.6-2.0 & 1.85 & $2.5^{+0.2}_{-0.2}$ & $4.6^{+1.0}_{-0.8}$ & $0.228^{+0.006}_{-0.004}$ & $14.0^{+3.0}_{-2.6}$ & $28.5^{+6.2}_{-5.2}$\\
 & 9.98\tablenotemark{$\dagger$} & 2.0-2.5 & 2.25 & $2.0^{+0.2}_{-0.1}$ & $2.8^{+1.0}_{-0.5}$ & $0.248^{+0.012}_{-0.002}$ & $13.6^{+4.8}_{-2.6}$ & $27.6^{+9.8}_{-5.2}$\\
 & 9.99\tablenotemark{$\dagger$} & 2.5-3.0 & 2.71 & $1.6^{+0.2}_{-0.2}$ & $2.0^{+1.4}_{-1.5}$ & $0.256^{+0.001}_{-0.001}$ & $14.5^{+10.1}_{-11.2}$ & $29.5^{+20.6}_{-22.7}$\\ \hline
10.2-10.6 & 10.37 & 0.2-0.4 & 0.29 & $12.2^{+0.8}_{-0.3}$ & $29.0^{+3.5}_{-1.4}$ & $0.592^{+0.014}_{-0.014}$ & $1.2^{+0.1}_{-0.1}$ & $2.4^{+0.3}_{-0.1}$\\
 & 10.37 & 0.4-0.6 & 0.49 & $11.5^{+1.1}_{-0.6}$ & $23.3^{+4.4}_{-2.6}$ & $0.542^{+0.007}_{-0.011}$ & $3.0^{+0.6}_{-0.3}$ & $6.2^{+1.2}_{-0.7}$\\
 & 10.39 & 0.6-0.8 & 0.68 & $8.2^{+0.3}_{-0.3}$ & $16.0^{+1.4}_{-1.3}$ & $0.415^{+0.026}_{-0.009}$ & $4.6^{+0.4}_{-0.4}$ & $9.4^{+0.8}_{-0.7}$\\
 & 10.38 & 0.8-1.0 & 0.89 & $7.8^{+0.3}_{-0.4}$ & $14.5^{+1.4}_{-1.5}$ & $0.324^{+0.006}_{-0.002}$ & $7.9^{+0.8}_{-0.8}$ & $16.0^{+1.5}_{-1.6}$\\
 & 10.40 & 1.0-1.2 & 1.10 & $6.5^{+0.3}_{-0.3}$ & $12.0^{+1.3}_{-1.1}$ & $0.358^{+0.013}_{-0.004}$ & $10.6^{+1.1}_{-1.0}$ & $21.5^{+2.3}_{-2.0}$\\
 & 10.38 & 1.2-1.6 & 1.38 & $5.3^{+0.2}_{-0.2}$ & $9.7^{+0.9}_{-0.8}$ & $0.267^{+0.005}_{-0.004}$ & $14.7^{+1.3}_{-1.2}$ & $29.9^{+2.7}_{-2.5}$\\
 & 10.37 & 1.6-2.0 & 1.81 & $4.9^{+0.3}_{-0.3}$ & $9.8^{+1.1}_{-1.3}$ & $0.298^{+0.007}_{-0.000}$ & $28.2^{+3.2}_{-3.7}$ & $57.3^{+6.5}_{-7.5}$\\
 & 10.37\tablenotemark{$\star$} & 2.0-2.5 & 2.32 & $4.0^{+0.2}_{-0.3}$ & $7.0^{+0.7}_{-1.1}$ & $0.302^{+0.010}_{-0.001}$ & $35.6^{+3.5}_{-5.8}$ & $72.5^{+7.0}_{-11.7}$\\
 & 10.38\tablenotemark{$\star$} & 2.5-3.0 & 2.78 & $3.5^{+0.3}_{-0.3}$ & $6.6^{+1.2}_{-1.3}$ & $0.311^{+0.010}_{-0.009}$ & $50.9^{+8.9}_{-10.2}$ & $103.5^{+18.1}_{-20.8}$\\ \hline
10.6-11.0 & 10.74 & 0.2-0.4 & 0.28 & $18.8^{+1.3}_{-1.7}$ & $53.8^{+6.4}_{-8.3}$ & $0.851^{+0.017}_{-0.014}$ & $2.0^{+0.2}_{-0.3}$ & $4.0^{+0.5}_{-0.6}$\\
 & 10.75 & 0.4-0.6 & 0.48 & $13.8^{+1.2}_{-1.3}$ & $30.2^{+5.0}_{-5.4}$ & $0.707^{+0.012}_{-0.015}$ & $3.9^{+0.6}_{-0.7}$ & $7.9^{+1.3}_{-1.4}$\\
 & 10.75 & 0.6-0.8 & 0.69 & $13.3^{+1.1}_{-0.6}$ & $27.7^{+4.4}_{-2.5}$ & $0.573^{+0.035}_{-0.023}$ & $8.1^{+1.3}_{-0.7}$ & $16.4^{+2.6}_{-1.5}$\\
 & 10.75 & 0.8-1.0 & 0.89 & $10.5^{+0.6}_{-0.5}$ & $19.5^{+2.6}_{-2.2}$ & $0.433^{+0.009}_{-0.001}$ & $10.5^{+1.4}_{-1.2}$ & $21.3^{+2.8}_{-2.4}$\\
 & 10.75 & 1.0-1.2 & 1.10 & $8.1^{+0.3}_{-0.4}$ & $15.8^{+1.2}_{-1.7}$ & $0.454^{+0.005}_{-0.018}$ & $14.1^{+1.0}_{-1.5}$ & $28.7^{+2.1}_{-3.1}$\\
 & 10.75 & 1.2-1.6 & 1.37 & $8.6^{+0.4}_{-0.4}$ & $16.8^{+1.6}_{-1.8}$ & $0.333^{+0.002}_{-0.003}$ & $25.0^{+2.3}_{-2.7}$ & $50.9^{+4.7}_{-5.4}$\\
 & 10.77 & 1.6-2.0 & 1.79 & $7.9^{+0.4}_{-0.3}$ & $14.8^{+1.7}_{-1.1}$ & $0.387^{+0.011}_{-0.009}$ & $41.3^{+4.9}_{-3.0}$ & $83.9^{+9.9}_{-6.1}$\\
 & 10.77 & 2.0-2.5 & 2.22 & $6.3^{+0.7}_{-0.5}$ & $11.9^{+2.6}_{-1.9}$ & $0.397^{+0.002}_{-0.004}$ & $55.2^{+11.9}_{-8.7}$ & $112.3^{+24.2}_{-17.7}$\\
 & 10.76 & 2.5-3.0 & 2.72 & $5.6^{+0.3}_{-0.4}$ & $13.3^{+1.2}_{-1.6}$ & $0.428^{+0.005}_{-0.020}$ & $97.6^{+9.0}_{-11.9}$ & $198.4^{+18.2}_{-24.2}$\\ \hline
$> 11.0$ & 11.10 & 0.2-0.4 & 0.29 & $19.8^{+3.7}_{-4.1}$ & $75.4^{+20.8}_{-23.1}$ & $1.640^{+0.084}_{-0.100}$ & $2.9^{+0.8}_{-0.9}$ & $5.9^{+1.6}_{-1.8}$\\
 & 11.10 & 0.4-0.6 & 0.48 & $18.1^{+1.2}_{-1.8}$ & $56.3^{+6.2}_{-9.3}$ & $1.364^{+0.071}_{-0.048}$ & $6.9^{+0.8}_{-1.2}$ & $14.1^{+1.6}_{-2.3}$\\
 & 11.10 & 0.6-0.8 & 0.69 & $12.7^{+1.1}_{-1.4}$ & $32.6^{+5.1}_{-6.4}$ & $1.086^{+0.040}_{-0.011}$ & $9.6^{+1.5}_{-1.9}$ & $19.5^{+3.1}_{-3.8}$\\
 & 11.10 & 0.8-1.0 & 0.89 & $13.8^{+1.4}_{-1.7}$ & $32.0^{+5.6}_{-6.7}$ & $0.907^{+0.039}_{-0.007}$ & $17.1^{+3.0}_{-3.6}$ & $34.9^{+6.1}_{-7.3}$\\
 & 11.13 & 1.0-1.2 & 1.10 & $13.4^{+1.5}_{-1.3}$ & $26.9^{+5.6}_{-4.9}$ & $0.881^{+0.013}_{-0.009}$ & $24.2^{+5.0}_{-4.4}$ & $49.2^{+10.2}_{-9.0}$\\
 & 11.11 & 1.2-1.6 & 1.36 & $13.4^{+1.0}_{-0.8}$ & $27.7^{+4.1}_{-3.1}$ & $0.734^{+0.023}_{-0.027}$ & $40.5^{+6.0}_{-4.5}$ & $82.3^{+12.2}_{-9.2}$\\
 & 11.11 & 1.6-2.0 & 1.80 & $15.6^{+1.1}_{-1.6}$ & $30.3^{+4.3}_{-6.3}$ & $0.701^{+0.013}_{-0.014}$ & $85.9^{+12.1}_{-17.9}$ & $174.6^{+24.6}_{-36.4}$\\
 & 11.15 & 2.0-2.5 & 2.22 & $11.9^{+0.6}_{-0.5}$ & $21.8^{+2.3}_{-1.8}$ & $0.594^{+0.023}_{-0.044}$ & $100.1^{+10.6}_{-8.5}$ & $203.5^{+21.6}_{-17.2}$\\
 & 11.17 & 2.5-3.0 & 2.71 & $11.1^{+0.7}_{-1.1}$ & $22.5^{+2.9}_{-4.6}$ & $0.645^{+0.007}_{-0.004}$ & $164.5^{+20.8}_{-33.7}$ & $334.4^{+42.4}_{-68.5}$\\
\enddata
\tablecomments{Median stacking-based average 1.4~GHz radio flux densities and derived average quantities for all our bins in mass and redshift for {\emph{star forming systems within our mass-selected sample}}. For details see caption of Tab. \ref{tab:fullall}.}
\tablenotetext{$\dagger$}{Mass bin contains data below the limit of mass representativeness and yields an upper limit to the average SFR (see Sec. \ref{sec:comp} for further details.)}
\tablenotetext{$\star$}{First mass bin above the limit of representativeness (see Sec. \ref{sec:comp}) which contains a low fraction ($<15~\%$) of optically faint objects with $m_{AB}(i^+) \ge 25.5$ for which the photo-z accuracy is degraded (see Sec. \ref{sec:irsamp} for further details). The average SFR measured in this bin might be slightly overestimated towards higher values (see Sec. \ref{sec:comp}).}
\end{deluxetable*}

In order to convert the derived average 1.4~GHz luminosities into average SFRs we use the calibration of the radio-FIR correlation by \citet{BELL03A} scaled to a Chabrier IMF\footnote{\citet{BELL03A} adopts a Salpeter initial mass function with $\rm{IMF} \propto M^{-2.35}$ in the mass range from 0.1 to 100~$M_{\odot}$ so that we divide his normalization by 1.74.}:
\for{\label{eq:lum2sfr}\langle \rm{SFR} \rangle~[M_{\odot}/yr] = \left\{ \begin{array}{cl} 3.18 \times 10^{-22} \, L & ,\; L > L_c \\ \frac{3.18 \times 10^{-22} \, L}{0.1+0.9 \, (L/L_c)^{0.3}} & \; , L \le L_c \end{array} \right.} 
where $L=\langle L_{1.4~\rm{GHz}} \rangle$ is the average radio luminosity derived from the median stack according to Eq. \gl{eq:flux2lum} and $L_c = 6.4 \times 10^{21}$~W/Hz is the radio luminosity of an $L_*$-like galaxy. As \citet{BELL03A} empirically argues the low-luminosity population needs to be treated separately from higher values of radio luminosities since non-thermal radio emission might be significantly suppressed in these galaxies. Even though our work exploits the radio-faint regime our derived average 1.4~GHz luminosities lie generally above this threshold. Only at the lowest masses and $z \lesssim 0.8$ we find $\langle L_{1.4~\rm{GHz}} \rangle < L_c $ (see Tab. \ref{tab:fullall} and \ref{tab:fullsfg}). Any study relying on the calibration by \citet{YUNR01} is, consequently, directly comparable to our results as \citet{YUNR01} used a uniform normalization very similar to the case $L > L_c$ in Eq. \gl{eq:lum2sfr}.\footnote{A radio luminosity independent calibration has also been presented by \citet{COND92}. We refer to \citet{DUNN09} who present all
their results using both the \citet{BELL03A} and \citet{COND92} calibration.}
According to \citet{BELL03A} individual objects scatter about the average calibration by about a factor of two. It is not necessary to include this dispersion in the estimation of the final uncertainty on the SFR computed from the stack since the latter involves a sufficiently large number of sources to ensure that the average relation is representative. We do not attempt to take the differences of the derived SFRs caused by the discrepancy of the mentioned calibrations into consideration for the error estimates of our results. We also neglect any uncertainty on the median photo-$z$ so that all errors on the derived SFRs result from propagation of the errors derived using equation \gl{eq:errftot}.\footnote{This is justified as this error scales with the number of objects as $1/\sqrt{N}$ where $N \gg 10^2$ given our binning scheme.}

Finally, for a given sample, specific SFRs are computed as the ratio of the SFR and the median stellar mass. Based on the same arguments as before we neither take into account an uncertainty in the median mass for the error estimates of our derived SSFRs. As we exclusively deal with average quantities in this work we omit the $\langle \rangle$-notation in the following.     
 
\section{The Specific SFR (SSFR) of mass-selected galaxies over cosmic time from radio stacking}
\label{sec:ssfr}
In the remaining parts of this paper we present our measurements of the SSFR-$M_*$ relation (this Section) and discuss their implications for the evolution of the cosmic SFR density (Sec. \ref{sec:csfh}).  

\subsection{The relation between SSFR and stellar mass}
\label{sec:ssfrm}
We first consider the whole sample including all galaxies and show the redshift dependent radio-based SSFRs that are distributed in the logarithmic SSFR-$M_*$ plane as seen in the left panel of Fig. \ref{fig:ssfrvsmf}. It is clear that the SSFR for a given stellar mass increases with redshift and that it generally decreases with increasing stellar mass.

The data at the high-mass end (above $\log(M_*) \approx 10.5$) within all considered redshift slices suggest power-law relations between SSFR and stellar mass of the form
\for{\label{eq:ssfrvsm} \rm{SSFR}(M_*,z) =c(z) \times {M_*}^{\beta (z)}.} 

\begin{figure*} 
\vspace{-5cm}
\includegraphics[angle=90,width=\textwidth]{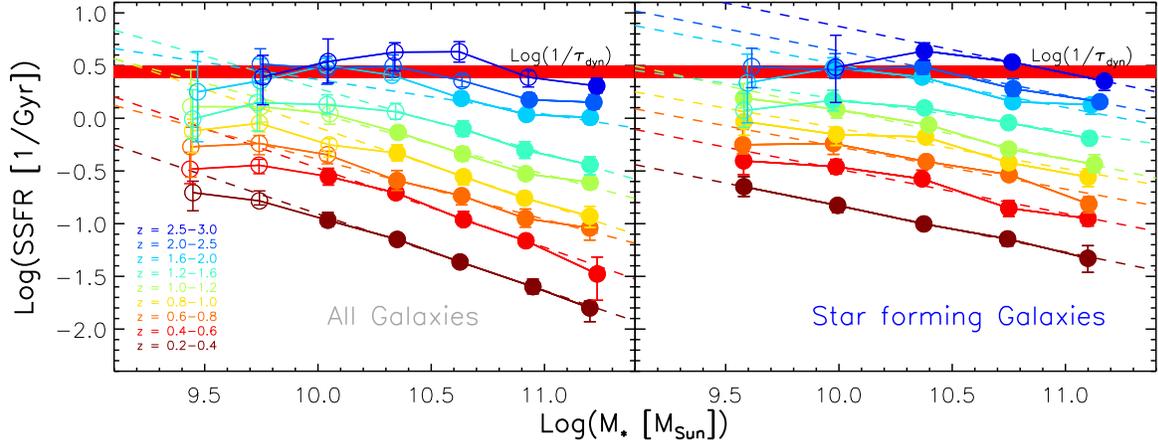}
\caption{\noindent Radio stacking based measurement of the SSFR as a function of stellar mass at $0.2<z<3.0$ for our entire galaxy sample (left) and SF systems (right). Open symbols depict samples containing galaxies less massive than the individual limits denoted in Fig. \ref{fig:compaq} and are regarded as not representative for the underlying galaxy population and rather represent upper SSFR limits. Dashed lines are two-parameter fits of the form $c \times M_*^{\beta}$ to the mass-representative data depicted by filled symbols (see Tab. \ref{tab:ssfrvsm}). The horizontal red band sketches the inverse dynamical time of $(370 \pm 50)$~Myr measured in local disk galaxies \citep{KENN98} and also found in massive disk galaxies at $z \sim 1.5$ \citep{DADD10}. Galaxies with such high levels of SSFR effectively double their mass within a dynamical time. As detailed in Sec. \ref{sec:ssfrm} this might represent an upper bound to the average SSFR. All measured data points are listed in Tab. \ref{tab:fullall} and \ref{tab:fullsfg}. The derivation of error bars involves a bootstrapping analysis combined with the uncertainties to the best-fit model of each stacking-derived average radio source (see App. \ref{sec:app_boot} and Sec. \ref{sec:im2flux}; for further details). We do not account for uncertainties associated 
with the SFR-calibration, the photometric redshift and stellar mass estimates as the large number of objects stacked for 
each data point ensures that even the joint error budget is statistically reduced to a low level that would not substantially enhance our uncertainty ranges.}
\label{fig:ssfrvsmf} 
\end{figure*}

In the following we will refer to the index $\beta$ also as slope since the relation is commonly shown in $\log$-space.
The dashed lines in Fig. \ref{fig:ssfrvsmf} depict the best fit to the data in the mass-representative regime (see Sec. \ref{sec:comp}) and indicate that only the normalization evolves while the power-law index $\beta_{\rm{ALL}}$ of the individually fitted relations shows minor fluctuations but no clear evolutionary trend. Only at $z \gtrsim 1.5$ there is tentative evidence for a somewhat shallower slope. However, at the highest redshifts probed too few mass-representative data points exist to perform the linear fit. Our evidence is solely supported by the offset between the SSFR of the most massive galaxies and those of intermediate mass remaining the same as at $z \sim 1.8$. Based on our data it therefore is justified to consider the index $\beta_{\rm{ALL}}$ in Eq. \gl{eq:ssfrvsm} a constant at least for all $z<1.5$ and $\log(M_*) \gtrsim 10.5$ .

At $\log(M_*) < 10-10.5$ we see at practically all epochs that the measured SSFRs significantly deviate from the relation fitted to the high-mass end. The extrapolation towards lower masses over-predicts the measurement. In Sec. \ref{sec:comp} we argued that all these data points -- lying below the mass representativeness limits -- likely represent upper limits. We hence believe this is a genuine deviation that is reminiscent of the bimodality (whereby quiescent galaxies preferentially populate the high-mass end) in the SSFR-$M_*$ plane confirmed at various redshifts for galaxy samples with individually measured SFRs \citep[e.g.][]{BRIN04, SALI07, ELBA07, SANT09, RODI10}.

Using our spectral classification scheme we separately study the SF galaxy population in order to break the afore mentioned bimodality.
The right panel in Fig. \ref{fig:ssfrvsmf} shows that a power-law relation according to Eq. \gl{eq:ssfrvsm} holds over the entire mass range probed, once quiescent galaxies are excluded.
Linear fits exclusively to the mass-representative regime show that, at $z \lesssim 1.5$: (i) SSFR declines towards higher mass, and that (ii) the slope $\beta_{SFG}$ is constant, as it was the case for the entire galaxy population. Compared to the entire sample, the slope is significantly shallower.\footnote{At high masses, the radio-derived SSFRs for SF galaxies lie significantly above those for all galaxies demonstrating that the SED-based pre-selection is efficient.} All theses conclusions also hold at all other epochs probed but are supported by fewer data points significantly above
the mass-representativeness limits that enter the fits. Hence we regard our conclusions as most robust at $z < 1.5$. 

Above $z \sim 1.4$ and below $\log(M_*) \approx 9.5-10$, we again find that measurements in the regime not regarded as mass-representative lie significantly below the linear fits. Since quiescent galaxies are even less frequent at these redshifts\footnote{Also at high $z$ there is evidence for the existence of quiescent systems that are predominantly massive \citep[e.g.][]{CIMA04, KRIE06, KRIE08, BRAM09}. However, as our spectral classification of SF systems is efficient to exclude passive galaxies (see I10) and as these systems are also rare we do not expect them to cause the observed trend.}, the bimodality argument is obviously insufficient to explain this observed trend. A possible explanation is that the magnitude limit of our catalog leads to 
a loss of dust-dominated systems with low masses but high star formation activity. If this were the case our previous statement that SSFRs in the under-represented mass-regime are upper limits would not necessarily hold. However, we do not expect a sufficiently high number density of low-mass dusty starbursts to make this scenario plausible.
Another explanation could lie in the dynamical considerations presented in Sec. \ref{sec:ssfrlim}.

\subsection{A potential upper limit to the average SSFR of normal galaxies}
\label{sec:ssfrlim}

\begin{deluxetable*}{cccc|ccc}
\tabletypesize{\footnotesize}
\tablecaption{\label{tab:ssfrvsm}Two parameter fits to the mass dependence of the SSFR}
%\tablenum{4}
\tablehead{\colhead{} & \colhead{All galaxies} & \colhead{} & \colhead{} & \colhead{SF systems} & \colhead{} & \colhead{} \\ \colhead{$\Delta z$} & \colhead{$\log(c\st{ALL}$~[1/Gyr])} & \colhead{$\beta\st{ALL}$} & \colhead{$\chi^2/\rm{d.o.f}$} & \colhead{$\log(c\st{SFG}$~[1/Gyr])} & \colhead{$\beta\st{SFG}$} & \colhead{$\chi^2/\rm{d.o.f}$}} 
\startdata
0.2-0.4 & $-1.63 \pm 0.04$ & $-0.73 \pm 0.03$ & 0.11 & $-1.27 \pm 0.03$ & $-0.44 \pm 0.03$ & 0.03\\
0.4-0.6 & $-1.22 \pm 0.04$ & $-0.75 \pm 0.04$ & 0.24 & $-0.90 \pm 0.03$ & $-0.42 \pm 0.03$ & 0.78\\
0.6-0.8 & $-0.96 \pm 0.08$ & $-0.57 \pm 0.08$ & 0.17 & $-0.67 \pm 0.03$ & $-0.40 \pm 0.03$ & 1.37\\
0.8-1.0 & $-0.81 \pm 0.07$ & $-0.73 \pm 0.06$ & 0.07 & $-0.48 \pm 0.03$ & $-0.38 \pm 0.03$ & 1.42\\
1.0-1.2 & $-0.53 \pm 0.05$ & $-0.58 \pm 0.05$ & 0.64 & $-0.38 \pm 0.03$ & $-0.46 \pm 0.03$ & 0.61\\
1.2-1.6 & $-0.33 \pm 0.13$ & $-0.61 \pm 0.12$ & 0.12 & $-0.12 \pm 0.03$ & $-0.30 \pm 0.03$ & 1.08\\
1.6-2.0 & $0.04 \pm 0.08$ & $-0.33 \pm 0.07$ & 1.47 & $0.10 \pm 0.07$ & $-0.41 \pm 0.07$ & 2.21\\
2.0-2.5 & & & & $0.22 \pm 0.06$ & $-0.42 \pm 0.05$ & 0.19\\
2.5-3.0 & & & & $0.43 \pm 0.16$ & $-0.44 \pm 0.15$ & 0.81\\ \hline
&$\langle \beta\st{ALL} \rangle=$  & $-0.67 \pm 0.02$ & $^{+0.34}_{-0.08}$ & $\langle \beta\st{SFG} \rangle=$ & $-0.40 \pm 0.01$ & $^{+0.10}_{-0.06}$ \\
\enddata
\tablecomments{A power-law fit of the form $c \times {(M_*/10^{11}~M_{\odot}})^{\beta}$ (Eq. \gl{eq:ssfrvsm}) was applied to the radio stacking-based SSFRs as a function of mass within any redshift slice. Fits have only been applied if more than two data points remained above the mass limit where the individual sample is regarded mass-representative. The results for all galaxies are shown in the left half of the table while those for star forming systems (see Sec. \ref{sec:class}) are given in the right half. The weighted average power-law index (over all accessible redshifts) found for each population is stated at the bottom along with the formal standard error and the scatter range yielding a more realistic uncertainty estimate.}
\end{deluxetable*}

The fact that the aforementioned deviations from the linear fits at low masses steadily grow with redshift hints at a solid upper limit to the average SSFR.
Local spiral galaxies have on average a dynamical timescale -- i.e. the rotation timescale at the outer radius of a disk galaxy -- of $\tau\st{dyn} \sim 0.37~$Gyr \citep{KENN98}. \citet{DADD10} show that this still holds at $z \sim 1.5$. The inverse of this dynamical timescale, $1/\tau\st{dyn} \sim 2.7~\rm{Gyr}^{-1}$, is similar to the threshold that seems to prevent our average SSFRs from rising continuously with decreasing mass. Note also that this dynamical timescale approximately equals the free-fall time \citep{GENZ10} which is commonly used to relate SFR volume density with gas volume density \citep[e.g.][]{SCHM59, KENN98, KRUM05, KRUM09, LERO08}.

As indicated in Fig. \ref{fig:ssfrvsmf} the population of $z > 1.5$ hence galaxies reaches average levels of star formation that enable these normal SF systems to double their mass within a dynamical time scale. Generally, star formation is thought to be limited by the rate at which cold gas is accreted onto the galaxy (\citealp[e.g.][]{DUTT09, BOUC09} and also \citealp[e.g.][where simulations actually show the cold gas inflow]{KERE05, MACC06}) while the efficiency of star formation does not appear to change out to the highest redshifts accessible to molecular gas studies in normal disk galaxies to-date \citep{DADD09, TACC10}. 
Consequently, even the highly elevated gas fractions -- i.e. the amount of gas available for star formation over the sum of gas and stellar mass -- compared to local disk systems \citep[e.g.][who find up to 60~\% at $z=1.5$]{DADD09} might not suffice to sustain a star formation activity that proceeds faster than gravity permits. As average galaxies reach inverse SSFRs comparable to their inverse dynamical -- and, most importantly, free fall -- time it is hence likely that an effective gas accretion threshold is reached. 
Hence the SSFR should stop its growth with redshift at some point. Lower mass galaxies reach this threshold at lower redshifts than the more massive systems leading to the flattening of the relation we observe at the lower mass end. We will henceforth refer to the transition from an inclined to a flat SSFR-sequence as {\emph{`crossing mass'}}.   

It is clear that carbon monoxide ALMA-studies at $z > 1.5$ of typical SF systems with $M_* \le 10^{10}~M_{\odot}$ are required to understand their molecular gas properties and to test the star formation law of this population. 

\begin{figure*} 
\vspace{-5cm}
\includegraphics[angle=90,width=\textwidth]{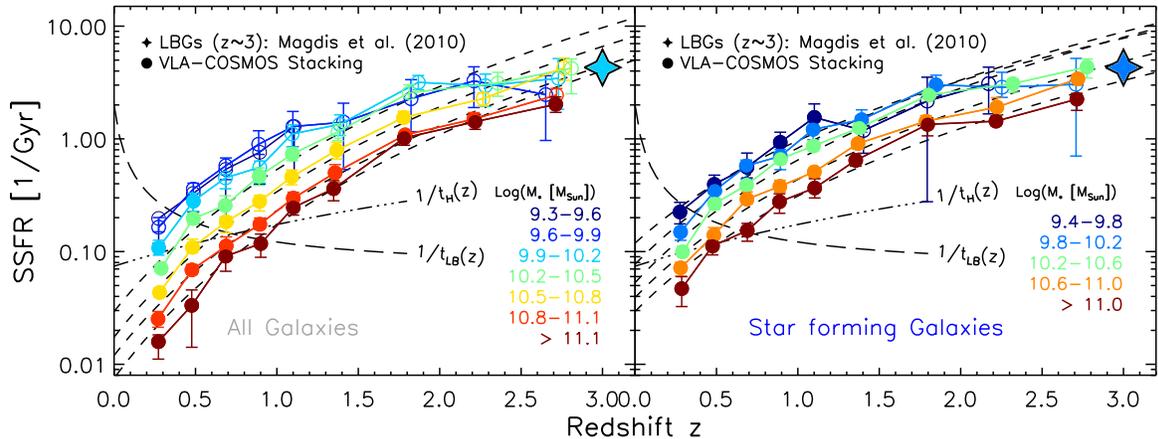}
\caption{\noindent Redshift evolution of the SSFRs for all galaxies (left) and SF systems (right) in logarithmic stellar mass bins. Two-parameter fits of the form $C \times (1+z)^n$ are applied only to data points derived from samples regarded as representative for the given underlying galaxy population which are depicted by filled symbols (see Tab. \ref{tab:ssfrvsz}). The black long-dashed line gives the mass-doubling limit above which galaxies are able to double their mass until $z=0$ assuming a constant SFR and it therefore equals the inverse lookback-time. The black dashed-dotted line depicts the inverse age of the universe at any given redshift and hence makes measured SSFRs comparable to the past average star formation activity. The SED-derived measurement of \citet{MAGD10} for LBGs at $z \sim 3$ with $\log(M_*) \sim 10$ is shown as a filled star. All data results are listed in Tab. \ref{tab:fullall} and \ref{tab:fullsfg}.}
\label{fig:ssfrvszf} 
\end{figure*}

\subsection{The redshift evolution of SSFRs as a function of mass}
\label{sec:ssfrz}
The redshift evolution of our data is shown in Fig. \ref{fig:ssfrvszf} for all galaxies and for the SF population. Both panels suggest a co-evolution of the considered mass-bins at least out to $z \sim 1.5$; while all measured SSFRs increase with redshift, the high-mass end does not evolve faster compared to lower masses and it always has the lowest SSFRs. An offset between the typical SSFRs of different mass bins is also evident for SF galaxies but it is smaller than for the entire galaxy population which shows a wider spread of SSFRs at fixed mass. Clearly, all these aspects are the direct result of our previous findings:
\begin{itemize}
\item A constant slope $\beta$ of the SSFR-$M_*$ relation is observed for all galaxies (at the high-mass end) as well as for SF galaxies alone at least out to $z \sim 1.5$.
\item The slope $\beta_{\rm{SFG}}$ is shallower for SF systems.
\end{itemize}
In Fig. \ref{fig:ssfrvszf} we also plot the mass doubling line\footnote{At a given redshift, the mass-doubling threshold is given by the inverse lookback time. A SSFR in excess of this limit is hence a mass-independent indicator for the potential of a galaxy to double its mass by $z=0$ if it were to maintain its current SFR.} (long dashed line) and the inverse age of the universe at any given redshift (dashed-dotted line). Our measurement clearly shows that virtually all SF galaxies display a higher star formation activity than is required if their entire mass had been build up at a constant rate over the whole age of the universe.\footnote{This does not necessarily imply that an individual galaxy maintains a high level of star formation activity. All statements we make here refer to an average galaxy of a given mass and cosmic epoch having a well-defined SSFR thanks to the SSFR-$M_*$ relation. Star formation might still be subsequently quenched in an individual system so that its evolutionary track does not need to coincide with those shown in Fig. \ref{fig:ssfrvszf}.} All galaxies, generally, cross the dashed-dotted line sooner or later depending on their stellar mass. The most massive systems enter the stage of sub-average star formation activity already at $z \sim 0.8$. 

At the high-mass end ($\log(M_*) \gtrsim 11$) the SSFR for SF galaxies increases by almost a factor of 50 and is about twice as much for all galaxies within $0.2 \le z \le 3$. For a given mass-bin
the redshift evolution is well described by a power law $g(z) \propto (1+z)^n$ as depicted by the dashed lines in Fig. \ref{fig:ssfrvszf}.\footnote{We fitted only data in the representative mass range.} For the most massive SF galaxies this relation holds out to the highest redshifts probed, thus no flattening is observed. However, towards lower masses and $z\gtrsim 2$ significant deviation of the data from the best-fit relation with lower SSFR towards lower masses is apparent. Again the argument of an upper SSFR limit due to dynamical reasons might explain such a deviation. For reference we also show the recent SED-based measurement by \citet{MAGD10} for a Lyman Break Galaxy (LBG) sample at $z \approx 3$.
Their study probes $M_* \approx 10^{10}~M_{\odot}$ and we see that their SSFR-measurement is significantly below the extrapolation given by our evolutionary fit in the same mass-regime even for SF galaxies. Basically, their data point is extending our measured data at the low-mass end if evolution were to stop at about $z \approx 1.5$ (\citealp[a scenario suggested by the
data of][]{STAR09} and \citealp{GONZ09}).  

\begin{deluxetable*}{cccc|cccc}
\tabletypesize{\tiny}
\tablecaption{\label{tab:ssfrvsz}Two parameter fits to the redshift evolution of the SSFR}
%\tablenum{5}
\tablehead{\colhead{} & \colhead{All galaxies} & \colhead{} & \colhead{} & \colhead{} & \colhead{SF systems} & \colhead{} & \colhead{} \\ \colhead{$\Delta \log(M_*~[M_{\odot}])$} & \colhead{$\log(C\st{ALL}$~[1/Gyr])} & \colhead{$n\st{ALL}$} & \colhead{$\chi^2/\rm{d.o.f}$} & \colhead{$\Delta \log(M_*~[M_{\odot}])$} & \colhead{$\log(C\st{SFG}$~[1/Gyr])} & \colhead{$n\st{SFG}$} & \colhead{$\chi^2/\rm{d.o.f}$}} 
\startdata
&  & & & 9.4-9.8 & $-0.92 \pm 0.45$ & $3.02 \pm 0.15$ & 0.65\\
10.2-10.5 & $-1.53 \pm 0.56$ & $4.18 \pm 0.05$ & 3.05 & 9.8-10.2 & $-0.92 \pm 0.33$ & $3.42 \pm 0.07$ & 1.63\\
10.5-10.8 & $-1.76 \pm 0.93$ & $4.28 \pm 0.05$ & 1.48 & 10.2-10.6 & $-1.11 \pm 0.28$ & $3.62 \pm 0.04$ & 5.24\\
10.8-11.1 & $-1.92 \pm 1.48$ & $4.27 \pm 0.05$ & 1.75 & 10.6-11.0 & $-1.28 \pm 0.47$ & $3.48 \pm 0.04$ & 1.73\\
$> 11.1$ & $-2.12 \pm 3.93$ & $4.53 \pm 0.07$ & 2.37 & $> 11.0$ & $-1.41 \pm 0.82$ & $3.40 \pm 0.06$ & 1.52\\ \hline
& $\langle n\st{ALL} \rangle =$ & $4.29 \pm 0.03$ & $^{+0.24}_{-0.11}$ & & $\langle n\st{SFG} \rangle =$ & $3.50 \pm 0.02$ & $^{+0.12}_{-0.48}$ \\
\enddata
\tablecomments{A power-law fit of the form $C \times (1+z)^{n}$ was applied to the radio stacking-based SSFRs as a function of redshift within any mass bin. Fits have only been performed if more than two data points remained above the mass limit where the individual sample is regarded mass-representative. The results for all galaxies are shown in the left half of the table while those for star forming systems (see Sec. \ref{sec:class}) are given in the right half. The weighted average power-law index (over all accessible masses) found for each population is stated at the bottom along with the formal standard error and the scatter range yielding a more realistic uncertainty estimate.}
\end{deluxetable*}

Summarizing our findings a separable function of the form 
\for{\label{eq:ssfrmz}\rm{SSFR}(M_*,z) \propto f(M_*) \times g(z) = M_*^{\beta} \times (1+z)^n}
describes well the mass-dependent evolution of the SSFR given our data within the restrictions discussed. For SF galaxies we find $\beta_{\rm{SFG}} \approx -0.4$ and $n_{\rm{SFG}} \approx 3.5$. We emphasize again that the dynamical arguments discussed in Sec. \ref{sec:ssfrlim} would give rise to a value of $\beta_{\rm{SFG}} = 0$ below the crossing mass of the average SSFR and 
the upper limiting SSFR. The results of the individual fits to our data yielding the parameters $\beta$ and $n$ for all and SF galaxies are presented in Tables \ref{tab:ssfrvsm} and \ref{tab:ssfrvsz}.

It is worth noting that at $z > 1$ -- where angular diameter distance is approximately constant -- the evolutionary trend we find is very close to the redshift dependence of the radio luminosity of about $(1+z)^{3.8}$ (see Sec. \ref{sec:im2flux}). As the SSFR is proportional to the radio luminosity this is yet another argument to support that our inferences are not challenged by systematic errors to the median redshift even in our broader bins at $z > 1$ (see also the corresponding discussion in Sec. \ref{sec:im2flux}).

In order to alternatively probe our inferences in the redshift range below $z=1.5$ where our data yields the most robust results we also stacked the same bins in redshift and mass into the Spitzer 24 and 70~$\mu$m COSMOS maps. We inferred SFRs from the total (8-1000~$\mu$m) IR luminosity predicted by the best-fitting IR SED \citep{CHAR01} given the joint flux density information. The results do not deviate significantly from those derived from the radio emission so that all our conclusions remain robust also when derived from the IR data. All these and further results will be presented and discussed in detail in a separate publication (Sargent et al., in prep.).    

\begin{figure*}
\vspace{-5cm} 
\includegraphics[angle=90,width=\textwidth]{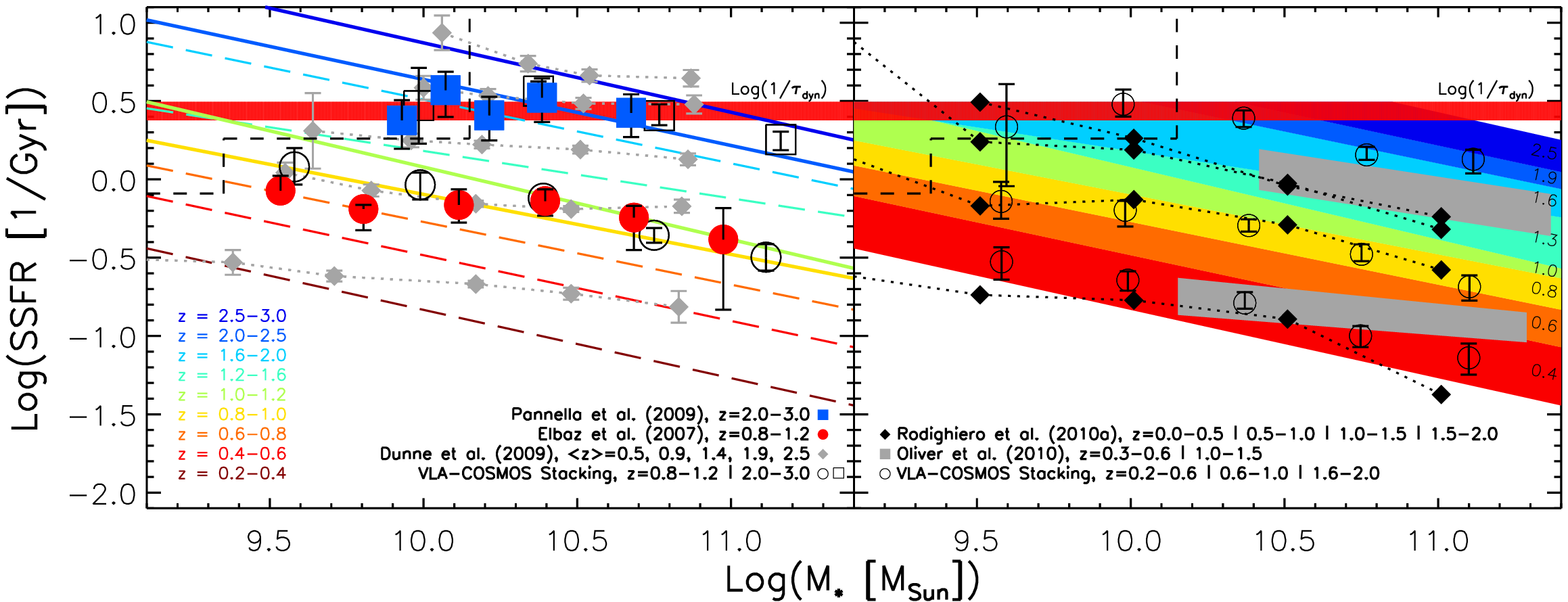}
\caption{\noindent {\textbf{Left Panel:}} Comparison of our results (dashed and solid lines of different color) of the mass dependence of the SSFR for SF systems at different redshifts to \citet{DUNN09}; grey diamonds, \citet{ELBA07}; red circles at $z \sim 1$, as well as \citet{PANN09}; blue squares at $z \sim 2$. The mass limits above which our sample is representative are denoted as black dashed lines in the upper left. {\textbf{Right Panel:}} The corresponding measurements by \citet{OLIV10}; grey shaded bands and \citet{RODI10}; black diamonds at various redshifts along with our results (color bands with mean redshift scale at the right hand side). The Herschel/PACS based SSFR-sequence \citep{RODI10} suggests a mild 'upsizing' trend and appears to stop its evolution at $z > 1.5$ where the only (apparent) deviation from our data occurs. See Sec. \ref{sec:others} and Appendix \ref{sec:app_sfg} for a discussion of effects introduced by selection biases. For immediate comparison our data are shown as open symbols in both panels, rebinned in $z$ in order to cover the same range in redshift as each referenced study. The inverse horizontal red band sketches the inverse dynamical time as detailed in Sec. \ref{sec:ssfrm}.}
\label{fig:pannelb} 
\end{figure*}

\subsection{Comparison to other studies}
\label{sec:others}
In this section we compare our findings with results in the literature, with a particular focus on those least dependent on extinction corrections because they use either radio stacking or stacking of IR imaging by Spitzer and, most recently, Herschel.
Literature data we show in the Figures belonging to this Section are based on a Salpeter IMF and have been converted to the Chabrier scale.

The evolutionary power law we derived for {\emph{all}} mass-selected galaxies is in excellent agreement with the results presented by \citet{DAME09B}, both in terms of the evolutionary exponent and the normalization of the trend. We hence concur, in particular, with those findings of \citet{DAME09B} resulting from a detailed comparison of their results with predictions from the semi-analytical model of \citet{GUOW08}. The study of \citet{DAME09B}, which is based on SFRs from 24~$\mu$m and UV detections in conjunction with deep $K$-band observation in the Chandra Deep Field South, is also in broad agreement with the 24~$\mu$m stacking analysis of \citet{ZHEN07A} at $z < 1$ in the same field. Consistent findings have also been presented recently in the Spitzer/MIPS stacking analysis at 70 and 160~$\mu$m by \citet{OLIV10} whose data covers the largest on-sky area of all aforementioned surveys, albeit at a reduced depth of $F_{3.6~\mu \rm{m}} = 10~\mu$Jy (an order of magnitude shallower compared to our sample) which prevents them from reliably constraining the evolution beyond $z \sim 1$. Based on a deep rest-frame NIR bolometric flux density selected galaxy sample in the northern Great Observatories Origins Deep Survey \citep[GOODS][]{GIAV04} field \citet{COWI08} measure extinction corrected UV-based SSFRs for all individual objects out to $z = 1.5$. Their average trends with mass agree well with our results for all galaxies in the comparable redshift ranges and also on absolute scales both studies are consistent at all masses.  

Radio-based measurements of the SSFR-$M_*$ relation have been presented by \citet{DUNN09}. In terms of the evolution of the SSFR-sequence both their and our study show a good agreement. The findings by \citet{DUNN09} differ from ours (see Fig. \ref{fig:pannelb}) as well as from most other studies not restricted to SF galaxies only in that they report an almost non-existent slope $\beta_{\rm{ALL}}$ at all reliably probed epochs. Their analysis and ours share some methodological similarities (e.g. the use of a mass- (in their case: K-) selected sample\footnote{It is, however, worth noting that 
our number statistics are larger by about a factor of four.} and a radio stacking approach) and should therefore be directly comparable. Despite the technical differences in the exact implementation of the image stacking as already discussed (see Sec. \ref{sec:im2flux}) it seems unlikely that an explanation for the different trends can be found in the radio data used. It appears more likely that the derivation of individual stellar masses causes the differences as \citet{DUNN09} use a direct conversion from the rest-frame K-band magnitudes as measured from the best-fitting SED templates to stellar mass which exclusively depends on redshift. Such a conversion should not only be different for SF and quiescent sources \citep{ARNO07} but even ceases to be applicable towards lower mass SF sources as discussed in App. D of I10. It is hence likely that low-mass SF sources with higher SSFRs have migrated to higher masses producing artificially elevated SSFRs at the high-mass end. This explanation is consistent with the generally higher deviations from our results at higher masses (see Fig. \ref{fig:pannelb}). Since neither we nor \citet{DUNN09} find a significant evolution of the slope $\beta_{\rm{ALL}}$ in the SSFR-$M_*$ plane the pure evolutionary behavior reported in both studies is largely consistent.

The SSFR-$M_*$ relation for sBzK -- and hence SF K-band selected -- galaxies in the COSMOS field was derived by \citet{PANN09} based on radio stacks from the same VLA image. Our results for SF galaxies are (necessarily) in good agreement with their findings at $z \approx 2$ (left panel of Fig. \ref{fig:pannelb}) where they do not probe the highest mass-range presented here. A main conclusion of \citet{PANN09} is, however, a mass-independent SSFR at $z>1.5$ which is mainly inferred from a measurement on their entire SF sample (not further divided by redshift) and a measurement at $z \approx 1.6$ both covering the same mass-range as considered here. A similar tendency at $z \sim 2$ has previously also been reported by \citet{DADD07} in a study carried out in the GOODS fields. Their work is based on 24~$\mu$m detected galaxies down to $\log(M_*) \approx 9.5$ but also based on radio stacks of their K-band selected sample. As galaxies at $1.3 <z<1.5$ substantially contribute to the photometric redshift distribution of the \citet{PANN09} sample, it is likely that the sBzK criterion no longer selects {\emph{all}} SF objects at these low redshifts. In this context we also refer to Appendix \ref{sec:app_sfg} where the upper left panel in Fig. \ref{fig:bzkplot} shows that the sBzK criterion by construction fails to select {\emph{all}} SF sources at $z<1.5$. As we already pointed out, our SSFR-$M_*$ relation for our SF sample tends to flatten towards lower $M_*$. When considering only low to intermediate masses, all measurements based on stacking into the VLA-COSMOS 1.4 GHz map are thus in good agreement. The steeper slope $\beta_{\rm{SFG}}$ of the SSFR-$M_*$ relation for SF galaxies found in this study is thus a consequence of the fact that we span a larger mass range at $z \approx 2$.

The left panel of Fig. \ref{fig:pannelb} also shows the results at $z \approx 1$ presented by \citet{ELBA07} based on 24~$\mu$m detection resulting from deep Spitzer/MIPS observations of the GOODS fields and UV-corrected SFRs. Although this study too infers a nearly constant relation between SSFR and mass, the figure shows that the radio-derived results agree with the mid-IR measurements remarkably well. This illustrates that measurements of the slope $\beta_{\rm{SFG}}$ of the SSFR-$M_*$ for SF galaxies are quite sensitive to deviations at the edges of the mass range even if measurements at individual masses do not significantly differ between different studies. Finally, it is also worth noting that, towards lower redshifts, our slope $\beta_{\rm{SFG}}$ agrees well with the measurements by \citet{NOES07A} which are based on SFRs from emission lines, UV as well as 24~$\mu$m imaging for a K-band selected sample of the DEEP2 spectroscopic survey. Also in the local universe GALEX/UV-based values around $\beta=-0.35$, consistent with our study, have been reported for galaxies taken from the Sloan Digital Sky Survey (SDSS) \citep[]{SALI07,SHIM07}. It should be mentioned that, based on the SDSS emission lines study of \citet{BRIN04}, \citet{ELBA07} found a slightly shallower slope $\beta_{\rm{SFG}}=-0.23$ for SF galaxies in the local universe. 

Image stacking results using Herschel/PACS data at 160~$\mu$m have been presented by \citet{RODI10}. Their GOODS-North Spitzer/IRAC data is only slightly shallower compared to our COSMOS imaging. As the right panel of Fig. \ref{fig:pannelb} shows, individual measurements by \citet{RODI10} at $z<1$ are in good agreement with our findings. (The one exception being their lowest redshift which extends to $z=0$, explaining the overall slightly lower SSFRs.) At $z > 1.5$ the \citet{RODI10} results suggest that SSFRs cease to grow further at the high-mass end. While in this redshift range our radio-derived SSFRs agree with the far-IR based ones at the lowest masses probed, the radio measurement yields about 0.4~dex (i.e. significantly) higher SSFRs at the high-mass end as they do not show a different redshift trend than at lower $z$. As our highest redshift bin is centered at a slightly higher $z$ compared to the corresponding one of \citet{RODI10} the difference might be slightly lower if the bins were perfectly matched and given the high-mass SSFR-evolution for SF galaxies is continuing at $z>1.5$. We therefore see no clear evidence for strong discrepancies of radio- and far-IR stacking derived SSFRs at high $z$ as speculated by \citet{RODI10} when comparing their results to those of \citet{PANN09} and especially those of\citet{DUNN09}. We emphasize, however, that future far-IR studies could test potential mass-dependent changes in the radio-IR correlation at $z > 1.5$ responsible for the slight differences reported here.\footnote{Herschel/PACS observations of the GOODS-North field \citep{ELBA10} revealed that in the same redshift regime the total ($8-1000~\mu$m) IR luminosity appears to be overestimated when the IR template-SED fit is constrained by a single 24~$\mu$m measurement. The deviation starts at $L\st{IR}\sim10^{12}~L_{\odot}$ and grows with increasing $L\st{IR}$, SFR and consequently mass as these quantities are correlated. It is therefore necessary to test the radio-IR correlation in the proposed way using far-IR data.} 
Based on power-law fits to their data \citet{RODI10} infer a steepening of $\beta_{\rm{SFG}}$ towards higher $z$, an effect they consequently term `upsizing'. 
Note, however, that our measurements of $\beta_{\rm{SFG}}$ agree with those of \citet{RODI10} within the uncertainties. Tentative evidence for upsizing is also reported by \citet{OLIV10} who use Spitzer/MIPS stacking of late-type galaxies at 70 and 160~$\mu$m (see right panel of Fig. \ref{fig:pannelb}). In Appendix \ref{sec:app_sfg} we show that we can mimic an upsizing trend, as well as the somewhat flatter evolution of the SSFR out to $z \approx 2$ reported by, e.g., \citet{RODI10} if we restrict our SF sample to those sources with the most active star formation. 

\section{The radio-based cosmic star formation history}
\label{sec:csfh}
Based on our measurement of radio-derived SFRs as a function of mass
we directly derive accurate SFR densities (SFRDs hereafter)
for {\emph{SF galaxies}} above the limiting mass at $z < 1.5$ and further constrain the CSFH out to $z=3$. We will also introduce two alternative extrapolations to low-mass objects that we do not directly observe. Because of our generally low mass limit, the impact of the extrapolation -- especially out to $z=1.5$ -- to these faint galaxies is small compared to most other studies. 

\subsection{The mass distribution function of the SFRD at fixed redshift}
\label{sec:sfrdfunc}
At a given redshift and mass, the SFRD is computed as the product of (i) the comoving number density as inferred directly from the number of galaxies in the relevant mass bin and (ii) their average SFR as measured in our stacking analysis. 

As already pointed out, SFRs likely represent an upper limit at the smallest masses where the sampling of the underlying population is not representative.
Consequently, SFRDs at low masses are also upper bounds, as we can correct the number counts in a given low-mass bin for the lost objects.
This is done by computing the expected number from the observed mass functions derived for the same sample
of SF galaxies that is used for this study (see I10 for further details). We account for the 
slightly smaller portion of the COSMOS-field accessible to the radio-stack compared to the area used for the derivation of the mass-functions. 
The correction for the expected number counts is always small so that corrected and uncorrected values of SFRD($M_*,z$) agree within the errors. Since it
is a systematic correction it still needs to be taken into account. 

All number-count corrected and uncorrected data points for the SFRD($M_*,z$) are shown in Fig. \ref{fig:sfrdfunc}. There appears to be a characteristic mass
of $M_*  = 10^{10.6\pm 0.4}~ M_{\odot}$ that contributes most to the total SFRD at a given redshift. Up to $z \sim 1.8$ our data points sample below this characteristic mass
and the peak is well constrained. At higher redshifts this is no longer the case.

We want to motivate now that the underlying functional form for the distribution of data points in the SFRD-$M_*$ plane is actually known
because of two facts:
\begin{enumerate}
\item There is a (possibly broken) power-law relation between (S)SFR and stellar mass for SF galaxies at all $z < 3$ as measured in this study.
\item The functional form of the mass function for SF objects in the same redshift range is well determined.
\end{enumerate} 

\begin{figure*} 
\includegraphics[width=\textwidth]{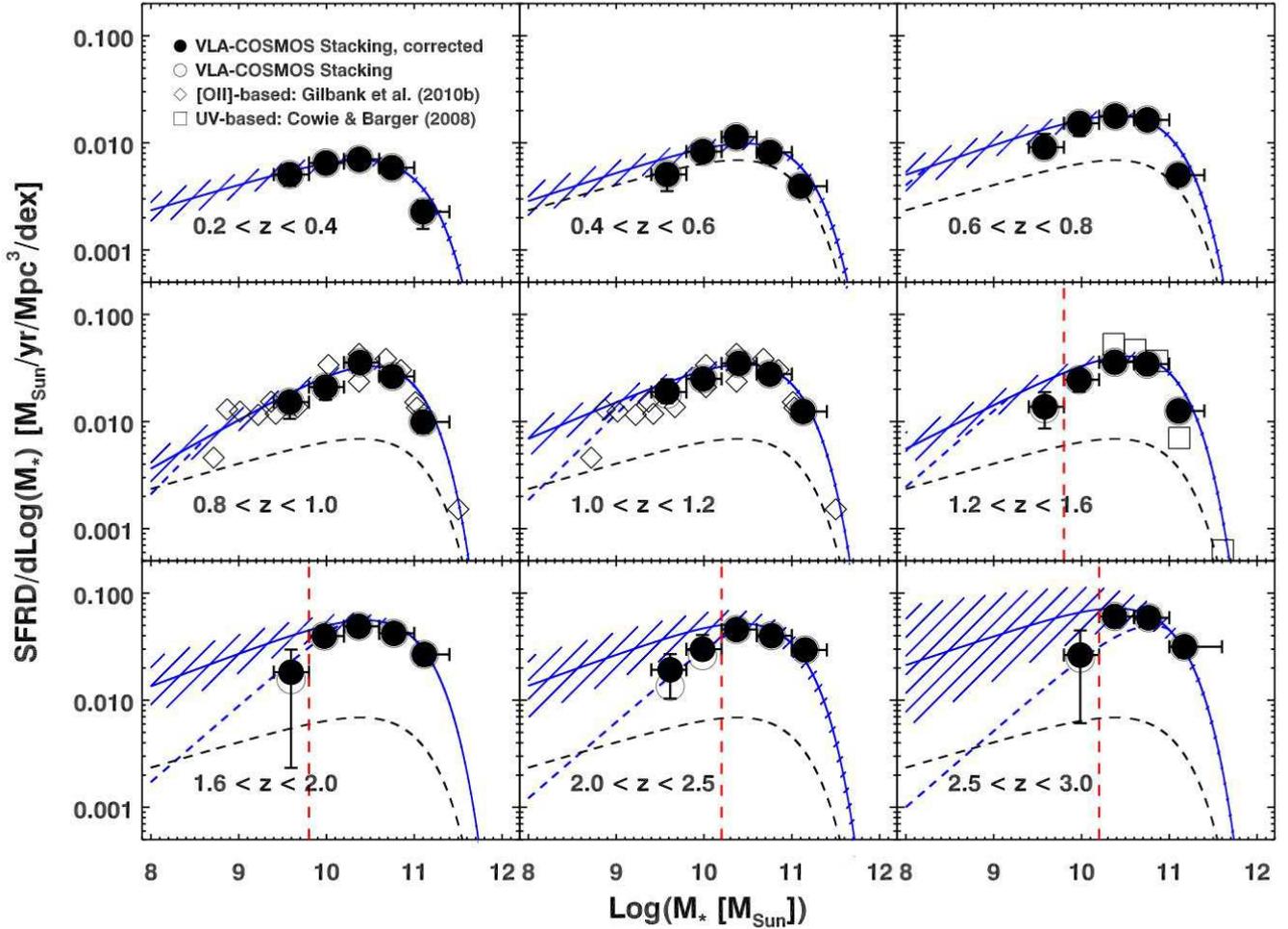}
\caption{\noindent  The distribution of the SFR density (SFRD) with respect to stellar mass, as measured at various epochs out to $z\sim 3$. In each panel the data points have been derived by
multiplying the observed number densities of SF galaxies with the average (stacking-based) radio SFRs. Below the limit where our data is regarded mass-representative -- depicted by red dashed vertical lines -- number densities have been corrected using the mass functions \citep{ILBE09B}. The uncorrected data are shown for comparison as open circles suggesting that these corrections are generally small and no corrections are needed at $z < 1.5$.  For $0.8 < z < 1.2$, the [OII]$\lambda3727$-derived SFRDs \citep[from][]{GILB10A}; open diamonds, rescaled by a constant factor of two to match our data agree well with the trends in our data. The same holds true for the UV-based results by \citet{COWI08} depicted at $1.2 < z < 1.6$ for which no rescaling was necessary. Note that their data was derived over a broader range in redshift down to $z=0.9$. As our data suggest globally only a mild evolution between $0.9 < z < 1.4$ the comparison depicted is justified. In each panel we overplot the Schechter function that results from multiplying the best-fit radio derived SSFR-sequence at a given epoch with the corresponding mass function for SF galaxies. The uncertainty range is obtained by choosing the two sets of Schechter parameters within their error margins that maximize/minimize the integral. Dashed blue lines show the distribution obtained if
an upper limit to the average SSFR at lower masses (see Sec. \ref{sec:ssfrlim} for details) is assumed (referred to as `case B' in Sec. \ref{sec:sfrdfunc}). All literature data plotted here have been converted to our Chabrier IMF.}
\label{fig:sfrdfunc} 
\end{figure*}

Regarding the second point, the mass function of SF galaxies is commonly \citep[e.g.][]{LILL95, BELL03B, BELL07, ZUCC06, ARNO07, POZZ09, ILBE05, ILBE09B} found to be
well parametrized by a power law with an exponential cutoff at a characteristic mass $M^*$ as introduced by \citet{SCHE76} of the form
\begin{eqnarray} 
\label{eq:schech}\Phi_{\rm{SFG}}(M_*) dM_* &=& \Phi^*_{\rm{SFG}} \left(M_*/M^*_{\rm{SFG}}\right)^{\alpha_{\rm{SFG}}} \\ \nonumber
&\times& \exp\left(-M_*/M^*_{\rm{SFG}}\right)d(M_*/M^*_{\rm{SFG}}).
\end{eqnarray}
This Schechter function has recently been qualitatively as well as quantitatively been modeled to be the natural consequence
of essentially two types of cessation of star formation \citep{PENG10}.\footnote{\citet{PENG10} refer to these two processes as 'environment-' and 'mass-quenching'. The former one is likely
to be explained by star formation being shut off in satellite systems as soon as galaxies fall into larger dark matter halos  
while the latter one is a continuos process stopping star formation within galaxies above the characteristic mass $M^*_{\rm{SFG}}$ at a rate proportional to their SFR. In the following we will make use of evolutionary constraints on the mass function parameters $\alpha_{\rm{SFG}}$,  $\Phi^*_{\rm{SFG}}$ and $M^*_{\rm{SFG}}$ for SF galaxies in particular.  Their trends are not only
supported by recent literature (see the further discussion in this Section) but also naturally contained in the empirical \citet{PENG10} model.}

Multiplying $\Phi_{\rm{SFG}}(M_*)$ by the SFR-sequence, i.e. another power-law in mass, again produces a Schechter function. 
Hence we can write
\for{\label{eq:sfrdfunc}\rm{SFRD}(M_*,z) dM_*=\Phi_{\rm{SFRD}}\left(\Phi^*_{\rm{SFRD}},\tilde{\alpha},M^*\right)dM_*,}
i.e. a distribution (SFRD function hereafter) of the same functional form as Eq. \gl{eq:schech} with the three parameters $\Phi^*_{\rm{SFRD}}$, $\tilde{\alpha}$ and $M^*$.  
While the exponential cutoff mass $M^*$ is the same as the one in the mass function (defined above as $M^*_{\rm{SFG}}$), $\tilde{\alpha}=\alpha_{\rm{SFG}} + \tilde{\beta}_{\rm{SFG}}$ is the sum of the low-mass slope of the mass function of SF galaxies and the slope\footnote{Please note that $\tilde{\beta}_{\rm{SFG}}$
denotes the slope of the SFR-$M_*$ relation for SF galaxies which is connected to the slope $\beta_{\rm{SFG}}$ of the SSFR-sequence (see Sec. \ref{sec:ssfrm} and Tab. \ref{tab:ssfrvsm}) by $\tilde{\beta}_{\rm{SFG}}=\beta_{\rm{SFG}}+1$.}
of the SFR-sequence \citep[see also][for a similar parameterization]{SANT09}. The parameter $\Phi^*_{\rm{SFRD}}$ acts as a normalization and its role in the global evolutionary picture
will be discussed in Sec. \ref{sec:sfrdevo}.

\begin{deluxetable}{cccc}
\hspace{-0.5cm}
\tablecaption{\label{tab:schech}Schechter parameters for the stellar mass function of {\emph{star forming galaxies}}}
%\tablenum{6}
\tablehead{\colhead{$\Delta z_{\rm{phot}}$} & \colhead{$\alpha_{\rm{SFG}}$} & \colhead{$\log(M^*_{\rm{SFG}}$)} & \colhead{$\Phi^*_{\rm{SFG}}$}\\
\colhead{} & \colhead{} & \colhead{[$M_{\odot}$]} & \colhead{[$10^{-3}$~Mpc$^{-3}$~dex$^{-1}$]}} 
\startdata
0.2-0.4 & $-1.32^{+0.01}_{-0.01}$ & $11.00^{+0.03}_{-0.03}$  &  $1.15^{+0.08}_{-0.07}$\\
0.4-0.6 & $-1.32^{+0.01}_{-0.01}$ & $11.04^{+0.03}_{-0.03}$  &  $0.70^{+0.05}_{-0.04}$\\
0.6-0.8 &  $-1.32^{+0.01}_{-0.01}$ & $10.95^{+ 0.02}_{-0.02}$  &  $0.86 ^{+0.05}_{-0.05}$\\
0.8-1.0 &  $-1.16^{+0.01}_{-0.01}$ & $10.86^{+0.02}_{-0.02}$  &  $1.38 ^{+0.06}_{-0.06}$\\
1.0-1.2 &  $-1.19^{+0.02}_{-0.02}$ & $10.92^{+0.02}_{-0.02}$  &  $0.94^{+ 0.05}_{-0.05}$\\
1.2-1.5 &  $-1.28 ^{+0.02}_{-0.02}$ & $10.91^{+0.02}_{-0.02}$  &  $0.68^{+0.03}_{-0.03}$\\
1.5-2.0 &  $-1.29^{+0.02}_{-0.02}$ & $10.96^{+0.02}_{-0.02}$  &  $0.46^{+0.02}_{-0.02}$\\
2.0-2.5 &  $-1.29^{+0.03}_{-0.03}$ & $10.95^{+0.03}_{-0.03}$  &  $0.32^{+0.06}_{-0.06}$\\
2.5-3.0 &  $-1.29^{+0.03}_{-0.03}$ & $10.95^{+0.03}_{-0.03}$  &  $0.27^{+0.05}_{-0.05}$\\
\enddata
\tablecomments{At $z<2$ all parameters have been derived by I10. At $z>2$ 
we assume a non-evolving shape so that $\alpha_{\rm{SFG}}$ and $M^*_{\rm{SFG}}$ are taken to be the average of the respective lower $z$ values. $\Phi^*_{\rm{SFG}}$ was then derived
by matching the number densities to those observed in the mass-representative regime of our data.}
\end{deluxetable}

The index $\alpha_{\rm{SFG}}$ and also the cutoff mass $M^*$ for SF galaxies are constant in the redshift regime considered \citep[e.g.][]{BELL03B, BELL07, ARNO07, PERE08, POZZ09,ILBE09B}. We assume here that $\alpha_{\rm{SFG}}$ and $M^*_{\rm{SFG}}$ stay constant also at $z > 2$. These assumptions are tentatively supported by the few observational
constraints reported for these high redshifts as detailed in Sec. \ref{sec:sfrdevo}. As detailed in Sec. \ref{sec:ssfrm}, the power-law index $\beta_{\rm{SFG}}$ -- that enters the parameter
$\tilde{\alpha} = \alpha_{\rm{SFG}} + \beta_{\rm{SFG}} + 1$ in Eq. \ref{eq:sfrdfunc} -- is also found to be a constant. However, we explained in Sec. \ref{sec:ssfrlim} that at masses lower than the crossing mass between the SSFR-sequence and a possible SSFR-threshold $\beta_{\rm{SFG}}=0$ should be assumed. In the following we will hence consider two possible scenarios below the suggested crossing mass at a given redshift:
\begin{eqnarray*}
\rm{\bf{Case \; \, A}:} &\quad& \tilde{\alpha} = \alpha_{\rm{SFG}} + \beta_{\rm{SFG}} + 1\\
\rm{\bf{Case \; \, B}:} &\quad& \tilde{\alpha} = \alpha_{\rm{SFG}} + 1.
\end{eqnarray*} 

Fig. \ref{fig:sfrdfunc} shows that at $z < 1.5$ the parameterization of the SFRD function in Eq. \ref{eq:sfrdfunc} can reproduce our data at all masses sampled and irrespective of the
exact value of $\tilde{\alpha}$. For this redshift range Fig. \ref{fig:sfrdfunc} also includes results of two other studies that rely on different SFR tracers.
At $z\sim 1$ the dependence of the SFRD on stellar mass has recently been measured using the [OII]$\lambda$3727 line
to trace star formation \citep{GILB10A}. We over-plot these data points in the corresponding redshift bins in Fig. \ref{fig:sfrdfunc} and find that
our SFRD function accurately fits these measurements as well.\footnote{These data are based on a Baldry \& Glazebrook
IMF and have been converted to the Chabrier scale. An additional rescaling by a constant factor of two was necessary in order to match our calibration. This is in agreement
with the results by \citet{GILB10A} that show SFRs based on practically all probed alternative tracers to be in excess of the [OII]-derived ones. They discuss possible explanations for this
deviation. Given the well known global uncertainty in the absolute calibration of SFR tracers this deviation is, however, not significant.} The same holds for the UV-derived results based on a Salpeter IMF by \citet{COWI08} in the GOODS-North field at $0.9 < z < 1.5$. Given our results, the global evolution of the SFRD-function between $0.9 < z < 1.4$ is mild such that we can plot these data in Fig. \ref{fig:sfrdfunc} in the bin $1.2 < z < 1.6$. It is worth noting that the \citet{COWI08} measurements at $z<0.9$ equally support our finding that the peak of the SFRD does not shift with redshift to higher values. 

Below the limiting stellar mass (dashed red lines in Fig. \ref{fig:sfrdfunc}) our data points are lower than the prediction of Eq. \ref{eq:sfrdfunc} if we assume case A for $\tilde{\alpha}$ (even though
we have applied a number density correction). Moreover, we remind the reader that -- in keeping with our previous discussion -- these data points are likely upper limits.
Given the comparatively large uncertainties of our SFRD functions at high $z$ these deviations are not highly significant but the trend is systematic and suggest a steepening of the low-mass slope of the SFRD function. It is directly related to the fact that the corresponding data points deviate from the best-fit (S)SFR-$M_*$ relation at lower mass. In Sec. \ref{sec:ssfrlim} we proposed an upper limit to the average SSFR due to dynamical reasons as a possible explanation for the trends. Taking into account this limit of $\rm{SSFR} = 1/\tau_{\rm{dyn}} \sim 2.7~$Gyr$^{-1}$ yields an index $\beta_{\rm{SFG}} = 0$ below the mass at which our fitted high-mass SSFR-$M_*$ relation crosses the supposed SSFR limit at a given epoch. We plot the SFRD function for $\tilde{\alpha}=\alpha_{\rm{SFG}} + 1$ as dashed blue lines in Fig. \ref{fig:sfrdfunc}. As the crossing mass increases with redshift and lies below the mass-representativeness threshold at $z < 1$ it has little impact on the mass-integrated SFRD. The reason is that the mass-dependent SFRD has declined already by at least an order of magnitude from the peak value before reaching the crossing mass. The changes are largest at $z > 1.8$ where the dashed SFRD function (case B) drops quickly towards lower masses right after the peak and, in doing so, traces our data points better than before for $\tilde{\alpha}=\alpha_{\rm{SFG}} + \beta_{\rm{SFG}}$ (case A). However, we emphasize that our data cannot clearly favor case A or B proposed given the large uncertainties of the Schechter parameters and the lack of representativeness of our data at such low masses at high $z$. Any scenario suggested, hence, awaits confirmation based on deeper data in the selection band once they are available. We robustly conclude that a single Schechter function is a good model for the SFRD-function over all masses at least out to $z \sim 1$ and that the distribution of SFRDs peaks in the same mass-range at all $z$ probed in both case A and B.    

\begin{figure} 
\includegraphics[width=0.5\textwidth]{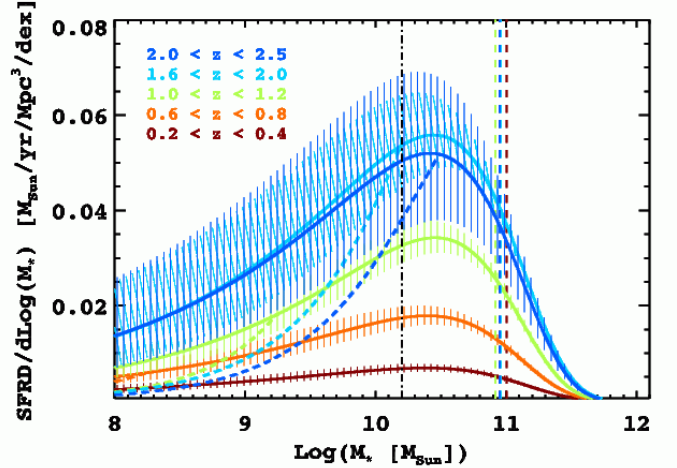}
\caption{\noindent The Schechter-function description of the SFRD (referred to as SFRD function) for $z < 2.5$, with color denoting the redshift range. This plot combines the information in the preceding Fig., now displayed with a linear scaling of the y-axis. The dashed vertical lines of different colors show the Schechter parameter $M^*_{\rm{SFG}}$ \citep[from][]{ILBE09B} for different redshifts. It is found to be nearly redshift-independent for SF galaxies. The representativeness limit of our SF sample is below $M^*_{\rm{SFG}}$ at all epochs as indicated for the highest redshift bin ($2 < z < 2.5$; dash-dotted vertical line). Thick dashed lines of different colors depict the modified low-mass end trends for our SFRD functions (referred to as `case B') that result from assuming an upper limit to the average SSFR of $\rm{SSFR} \sim 1/\tau_{\rm{dyn}}$. Here, $\tau_{\rm{dyn}} \sim 0.37~$Gyr is the dynamical timescale for normal SF galaxies (see Sec. \ref{sec:ssfrlim} for details). This plot clearly shows that the characteristic mass of SF galaxies -- which always lies below $M^*_{\rm{SFG}}$ -- does not evolve over time. This conclusion holds regardless of the exact shape of the SSFR-sequence at lower masses. Therefore, our data exclude a scenario in which the peak of the SFRD function has shifted to lower masses at later epochs.} 
\label{fig:sfrdfun} 
\end{figure}

Fig. \ref{fig:sfrdfun} shows the time evolution of our above constructed SFRD function. This non-logarithmic plot clearly illustrates
the existence of a characteristic mass of star formation at all epochs although it should be kept in mind that our results are most robust at $z < 1.5$. Our findings exclude an
evolution of this characteristic mass towards lower values with cosmic time. At $z<1.5$ this important result is supported by independent
observations at different wavelengths \citep{COWI08, GILB10B, GILB10A} and, towards $z \approx 2$, also by empirical arguments
\citep{PENG10} while the recent model of \citet{BOIS10} predicts a mild evolution of this characteristic mass by about 0.3~dex. 
In Fig. \ref{fig:sfrdfun} again dashed lines of corresponding color denote the low-mass trends of the SFRD functions once we assume
the SSFR-limit discussed. As pointed out above, within the important mass-range above $10^{10}~M_{\odot}$ -- at which the SFRD function
peaks at any epoch -- a mass-independent proportional increase is apparent even out to $z > 2$. In contrast to that, the mass-dependent SFRD appears to evolve
mildly below $10^{9}~M_{\odot}$. As a result galaxies in the stellar mass range between $10^{10}-10^{11}~M_{\odot}$ contribute most
to the global -- i.e. mass-integrated -- SFRD at any epoch but low-mass systems gain more relative importance towards low redshifts. Evidently,
low-mass systems show the same evolutionary trends as those of higher mass if case A is assumed.

Earlier observational findings appear to be at odds with the existence of a characteristic stellar mass for star formation we measure.
While they are all based on shallower survey data compared to our COSMOS imaging, the limiting magnitude of these samples
is not necessarily the reason for the different conclusions. \citet{JUNE05} -- using a mass-independent extinction correction and hence a linear
conversion from [OII] luminosity to SFR -- report that out to $z \sim 2$ the contribution to the increase of the global SFRD is more rapid,
the more massive the galaxies are. While this is a similar trend we find if we assume an upper limit in SSFR (case B), their lowest stellar mass-bin -- centered at $10^{9.6}~M_{\odot}$ -- always
appears to contribute most to the SFRD integrated over all masses. Hence, there is no clear evidence for a peak in the SFRD mass-distribution function and certainly not in the higher mass-range
our data supports. However, based on their findings in the local universe \citep{GILB10B}, \citet{GILB10A} empirically showed that a mass-dependent
calibration between [OII] luminosity and SFR is more appropriate. If true, this would modify the low the low-mass dominance in the derived SFRDs of  \citet{JUNE05} to correspond more closely to our result.\footnote{Indeed, as \citet{GILB10A} 
point out, the mass-dependence here is not only introduced by dust extinction but results from an interplay of various additional factors, e.g. the metallicity or the ionisation parameter.} 

Independently, also the radio-stacking based study by \citet{PANN09} suggests a strong dependence of the dust-attenuation correction on stellar mass at higher ($z > 1.5$) redshifts. Similarly to \citet{JUNE05}, \cite{BAUE05} derive their [OII]-based (S)SFRs from a mass-independent calibration albeit neglecting extinction corrections.\footnote{In the context of mass-dependent
evolutionary effects not considering any extinction correction is equivalent to considering a mass-uniform one.}
Finally, \citet{BUND06}, who favor a shift of the characteristic mass towards lower values with cosmic time, entirely base their conclusion on the stellar
mass functions they derive for SF galaxies (selected by rest-frame (U-B) color and [OII] equivalent line width) within the DEEP2 survey sample. 
At $z \lesssim 1.4$ these mass functions do not show an evolution of the faint-end slope, but the Schechter parameter $M^*_{\rm{SFG}}$ appears to decrease with cosmic time,
in contrast to the already discussed broad agreement in the more recent literature according to which neither $\alpha_{\rm{SFG}}$ nor $M^*_{\rm{SFG}}$ change in this redshift range.     

To conclude, we summarize our findings as follows.
\begin{enumerate}
\item Up to normalization, the mass distribution function of the SFRD is a universal Schechter function at $z < 1$ with possible deviations only below $10^{8}~M_{\odot}$. 
\item We explain this surprising constancy in shape of this SFRD-function by the non-evolving
slope of the SFR-sequence and the constant shape of the mass function for SF galaxies.
\item Our data at $z < 1.5$ clearly disfavors a strong 'downsizing' scenario in which the characteristic mass of SF galaxies that contribute most
to the overall SFRD -- integrated over all masses at a given time -- shifts towards lower values over cosmic time. The situation does not appear to change even at $z > 1.5$ where, however,
deeper data is needed to confirm our results. The characteristic mass is $M_*  = 10^{10.6\pm 0.4}~ M_{\odot}$.
\item If we assume an upper limit to the average SSFR of the order of the inverse dynamical scale ($\tau_{\rm{dyn}} \sim 0.37~$Gyr) the SFRD of galaxies less massive than $10^{9}~M_{\odot}$
evolves less rapidly than the one of the dominant higher mass range above $10^{10}~ M_{\odot}$.
\end{enumerate} 

\subsection{The evolution of the SFRD}
\label{sec:sfrdevo}
As all introduced parameters show this remarkable constancy throughout the redshift range probed (at least out to $z\approx 1$) 
the redshift dependence of Eq. \gl{eq:sfrdfunc} is entirely contained in the normalization $\Phi^*_{\rm{SFRD}}$.
Eq. \gl{eq:sfrdfunc} becomes a separable function so that we rewrite:
\for{\label{eq:sfrdfun2}\rm{SFRD}(M_*,z) dM_*=\Phi^*_{\rm{SFRD}}(z)\,\varphi \left(\tilde{\alpha},M^*\right)dM_*,}
where mass-dependence consequently is solely contained in the universal SFRD function $\varphi$.
The global SFRD at a given redshift, integrated over all masses, is simply given by
\for{\label{eq:sfrdint}\rm{SFRD}(z) =\Phi^*_{\rm{SFRD}}(z)\,\int \varphi \left(\tilde{\alpha},M^*\right)dM_*.}
In the following we will motivate that the evolution of the integrated SFRD follows a simple power-law of the form 
\for{\label{eq:phistar}\Phi^*_{\rm{SFRD}}(z)\left[M_{\odot}/\rm{yr}/\rm{Mpc}^3\right] \propto (1+z)^{n_{\Phi^*_{\rm{SFG}}} + n_{\rm{SSFR}}},}
where the two power-law indices result from the change in stellar mass density contained in SF galaxies and the increase
of the (S)SFR-sequence with redshift.
 
As detailed in I10 and previously also found by other studies \citep[e.g.][]{BELL03B, BELL07, ARNO07, POZZ09} the stellar mass density
of SF galaxies grows after the Big Bang only until $z \sim 1$. At lower redshifts it stays constant. Consequently, as the shape of
the mass functions does not evolve, also the Schechter parameter $\Phi^*_{\rm{SFG}}$ in the mass function is constant in this redshift regime, apart from fluctuations
due to large scale density fluctuations. As shown in Fig. \ref{fig:phistar}, these fluctuations are consistent with cosmic variance as estimated by \citet{SCOV07C} and detailed in I10.\footnote{For a detailed discussion on cosmic variance in the COSMOS field we refer to \citet{MENE09}.} It is clear
that cosmic variance effects are strongest at low redshifts as the effective volume sampled in a redshift bin with $\Delta z =0.2$ increases with redshift. In the interest of simplicity and to avoid systematic errors caused by cosmic sampling variance we adopt a constant $\Phi^*_{\rm{SFG}}$ at $z<1$.

If we therefore set $n_{\Phi^*_{\rm{SFG}}}(z<1)=0$ in Eq. \gl{eq:phistar} the evolution of the integrated SFRD in the range
$0<z<1$ is entirely described by the global, i.e. mass-uniform, decline of the average SFR of SF galaxies with cosmic time.
It is important to emphasize once again that this strong decline is definitely not caused by a decreasing number of SF galaxies, in particular not
at all at the high-mass end. As the efficiency of star formation within SF systems appears not to change over cosmic time \citep{DADD09, TACC10}, we conclude that below $z \sim 1$ the strongly evolving integrated SFRD must be exclusively caused by a strongly
declining mass density of cold molecular gas available for star formation towards the local universe. Recent theoretical model predictions \citep{OBRE09, DUTT09} indeed support such an evolutionary behavior of the cosmic mass density of molecular hydrogen. As we already explained in Sec. \ref{sec:ssfrz}, systematic redshift uncertainties will not influence our conclusions either as they would only be propagated along the redshift trend inferred. 

\begin{figure} 
\hspace{-1.6cm}
\includegraphics[width=0.7\textwidth]{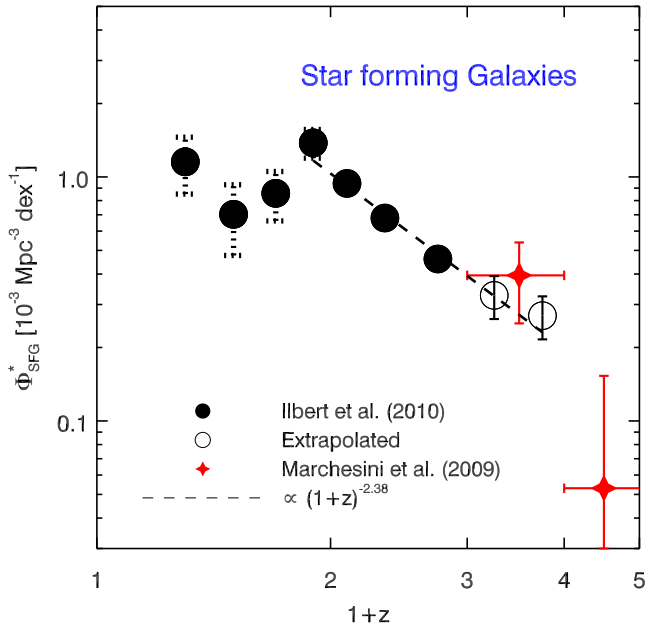}
\caption{\noindent  The redshift dependence of the Schechter parameter $\Phi^*_{\rm{SFG}}$ of the mass functions for SF galaxies as derived in the COSMOS field (I10). At $z < 1$ the normalization $\Phi^*_{\rm{SFG}}$ behaves like the other Schechter parameters and stays constant except for fluctuations very likely caused by cosmic variance. An estimate of the cosmic variance error \citep[see][for further details]{SCOV07C} is added in quadrature to the error of the Schechter fit at these redshifts, as indicated by dotted error bars. At earlier epochs the mass build-up is reasonably well characterized by a power law in $(1+z)$. This trend is consequently also seen in the evolution of the stellar mass density of SF galaxies (I10). The two highest redshift points (open symbols) were obtained under the assumption that the mass functions keep their shapes also at earlier cosmic times by matching the normalization to the numbers of galaxies observed in our mass-representative bins at the high-mass end. This assumption is supported by the direct observational constraint of $\Phi^*_{\rm{ALL}}$ from the best-fit Schechter-parameterization of the total mass-function measured by \citet{MARC09} which -- at these high redshifts -- should be representative for the SF population; red diamonds.}
\label{fig:phistar} 
\end{figure}

At earlier epochs (i.e. $z>1$) we monitor the redshift dependence of the parameter $\Phi^*_{\rm{SFG}}$ as measured by I10 (Tab. \ref{tab:schech}).  
Fig. $\ref{fig:phistar}$ shows that the above choice of a power law at $z>1$ is a reasonable assumption and we find $n_{\Phi^*_{\rm{SFG}}}(z>1) = -2.38 \pm 0.02$ from a fit to our COSMOS
data points at $1 < z < 3$.
It should be mentioned that the two highest redshift points in Fig. $\ref{fig:phistar}$ (open circles) were obtained by assuming that $\alpha_{\rm{SFG}}$ and $M^*_{\rm{SFG}}$ stay constant also at $z>2$ where I10 did not directly measure the mass function. The normalization $\Phi^*_{\rm{SFG}}(z>2)$ was therefore obtained by matching the number densities to those
observed in the mass-complete regime of our data when fixing the parameters $\alpha_{\rm{SFG}}$ and $M^*_{\rm{SFG}}$ to their average values at $z<2$. This extrapolation
is supported by the total mass function at $z>2$ measured by \citet{MARC09}. At these high redshifts we assume the total mass function to be representative for the
SF population as quiescent galaxies are not expected to significantly contribute to the number density. The study by \citet{MARC09} was carried out based on data
taken in various survey fields for which deep NIR imaging is available allowing them to estimate Schechter parameters for the total mass function also in two high-$z$
bins in the ranges $2 < z < 3$ and $3 < z < 4$. Within the errors all our extrapolated Schechter parameters at $2 < z < 3$ are in good agreement with their results.
In particular the exponential cutoff mass at $2 < z < 3$, $M^*_{\rm{ALL}} \equiv M^*_{\rm{SFG}}=10.96$, they find\footnote{As \citet{MARC09} estimate the mass function based on a Kroupa IMF,
masses are directly comparable to ours.} agrees remarkably well with our assumption (see Tab. \ref{tab:schech}). Their normalization $\Phi^*_{\rm{ALL}} \equiv \Phi^*_{\rm{SFG}}$ does not deviate
significantly from our prediction and the evolutionary trend we suggest is also supported by their data (see Fig. \ref{fig:phistar}). It is clear, however, that future measurements of the stellar mass
function at $z>1.5$ -- based on deeper NIR or mid-IR data -- are critical to confirm the validity of our assumptions.
Based on the currently available data we conclude that the stellar mass build-up in the SF galaxy population inevitably leads to a shallower decline of the integrated SFRD between $3 > z > 1$ compared to the steep decline between $1 > z > 0$.    

In order to validate our parameterization of the evolution of the integrated SFRD we proceed as follows. First, we simply add up all SFRDs measured in different mass bins 
in a given redshift bin (all data points shown in the corresponding panels in Fig. \ref{fig:sfrdfunc}) obtaining a lower limit to the integrated SFRD at that epoch. Second, in order to account for the contribution of the low-mass ($\log(M_*) < 9.5$) SF population that we cannot directly measure, we integrate Eq. \gl{eq:sfrdfunc}
from our mass-limit down to $10^5~M_{\odot}$. As we discuss in Sec. \ref{sec:sfrdfunc} a single value of the index $\tilde{\alpha}$ in Eq. \gl{eq:sfrdfunc} might not be valid over the entire low-mass range as the SSFR-$M_*$ relation might flatten as soon as an upper limiting SSFR is reached. The upper left panel in Fig. \ref{fig:sfrdall} hence shows the two alternative extrapolations and all obtained data are given in Tab. \ref{tab:madall}. The contribution of the integral generally lifts the SFRD at a given redshift by a linear factor of $\sim 1.4$ if we assume an SSFR-limit (red filled circles) and $\sim 1.7$ if a single low-mass end slope $\tilde{\alpha}$ is used (red open circles), suggesting the stacking analysis missed $\sim 30$~\% or $\sim 40$~\% respectively of the integrated SFRD. The differences between both extrapolations are largest at $z > 1$ below which either method yields practically the same results. As pointed out in Sec. \ref{sec:sfrdfunc} our data cannot clearly rule out any of the two alternative low-mass Schechter functions proposed. Consequently, our extrapolations overlap within their individual uncertainty ranges at all redshifts. In the following we favor, however, the extrapolation that includes the SSFR-limit as it assures a more conservative approach compared to the generally larger values the alternative extrapolation yields. 

It is evident that the number density corrections at our low-mass end discussed in Sec. \ref{sec:sfrdfunc} have almost no impact on our direct measurements (depicted as black open circles in the upper left panel of Fig. \ref{fig:sfrdall}) of the SFRDs. Even more, at $z < 1.5$ practically no corrections were necessary which highlights again that our inferences are most robust at these redshifts. It is therefore justified to regard the corrected values (depicted as black filled circles in the upper left and lower panel of Fig. \ref{fig:sfrdall}) as a direct and dust unaffected average measurement of the SFRD for SF galaxies to a lower mass limit of $M_* \sim 3.2 \times 10^{9}~M_{\odot}$. The evolutionary power-law, scaled to our data points and with the indices in the two redshift regimes, matches well the observed cosmic star formation history (CSFH) with respect to our measured data points presenting lower limits as well as to the integrated ones. This may seem surprising as our evolutionary model does not take into account differential effects with mass as introduced by assuming an SSFR limit (case B in Sec. \ref{sec:sfrdfunc}). However, since the bulk of the mass-integrated SFRD is contained in our direct stacking based measurements even at high $z$ -- which show a mass-independent evolution -- the model represents a good approximation.
 
In Fig. \ref{fig:sfrdall} we also compare\footnote{All literature data mentioned within the remainder of this Section are based on a Salpeter IMF and have been converted to the Chabrier scale.} our data to the CSFH derived from confirmed SF radio sources within the COSMOS field
in conjunction with extrapolations based on two distinct evolved local radio luminosity functions \citep[see][for details]{SMOL09A}. This comparison
shows how good a deep radio survey constrains the CSFH and leads us to slightly favor the extrapolations based on the \citet{COND89}
radio luminosity function in the \citet{SMOL09A} study as already our mass-limited, direct, stacking-based measurements of the SFRD on their own already reach the values which they inferred
using the \citet{SADL02} radio luminosity function at any $z < 1$. 

\begin{deluxetable}{ccc}
\tablecaption{\label{tab:madall}The total SFR density as a function of redshift (cosmic star formation history)}
\tablenum{7}
\tablehead{\colhead{$z$} & \colhead{SFRD$\st{obs}(z)$~[M$_{\odot}$/yr/Mpc$^3$]} & \colhead{SFRD$\st{int}(z)$~[M$_{\odot}$/yr/Mpc$^3$]}} 
\startdata
0.30 & 0.011 (0.011)$^{+0.001}_{-0.001}$ & $0.018^{+0.002}_{-0.006}\quad \left(0.019^{+0.004}_{-0.007}\right)$\\
0.50 & 0.015 (0.015)$^{+0.001}_{-0.001}$ & $0.023^{+0.002}_{-0.006}\quad \left(0.025^{+0.004}_{-0.008}\right)$\\
0.70 & 0.025 (0.025)$^{+0.002}_{-0.002}$ & $0.039^{+0.003}_{-0.009}\quad  \left(0.043^{+0.007}_{-0.015}\right)$\\
0.90 & 0.043 (0.043)$^{+0.003}_{-0.003}$ & $0.055^{+0.002}_{-0.008}\quad  \left(0.058^{+0.005}_{-0.010}\right)$\\
1.10 & 0.048 (0.047)$^{+0.004}_{-0.003}$ & $0.061^{+0.003}_{-0.005}\quad  \left(0.073^{+0.010}_{-0.016}\right)$\\
1.40 & 0.048 (0.048)$^{+0.004}_{-0.004}$ & $0.063^{+0.004}_{-0.007}\quad  \left(0.070^{+0.008}_{-0.018}\right)$\\
1.80 & 0.070 (0.069)$^{+0.007}_{-0.008}$ & $0.095^{+0.014}_{-0.014}\quad  \left(0.119^{+0.040}_{-0.048}\right)$\\
2.25 & 0.066 (0.062)$^{+0.007}_{-0.006}$ & $0.098^{+0.011}_{-0.009}\quad  \left(0.115^{+0.043}_{-0.046}\right)$\\
2.75 & 0.077 (0.077)$^{+0.009}_{-0.011}$ & $0.121^{+0.017}_{-0.017}\quad  \left(0.175^{+0.236}_{-0.099}\right)$\\
\enddata
\tablecomments{The central column states the number density corrected (raw) sum over all mass bins at a given redshift of the product of the average SFR and the total number of galaxies
contained in the corresponding bin down to the redshift-dependent limiting masses of this study. It is hence the sum of the data points within each panel of Fig. \ref{fig:sfrdfunc} and -- at least out to $z=1.5$ a robust direct measurement of the total dust unbiased SFRD for galaxies more massive than $\sim 3.2 \times 10^9~M_{\odot}$. The values in the right column additionally take into account the not directly measured low-mass end where we integrate over the SFRD-function at a given redshift as introduced in Sec. \ref{sec:sfrdfunc} while we assume a potential upper SSFR limit (see Sec. \ref{sec:ssfrm} for details). The values in brackets result from deriving the low-mass end contribution by integrating the single Schechter-models of the SFRD-functions and hence assuming no upper limit in SSFR. Like all other results presented in this work all values are based on a \citet{CHAB03} IMF.}
\end{deluxetable}

\begin{figure*} 
\centering
\includegraphics[width=0.7\textwidth]{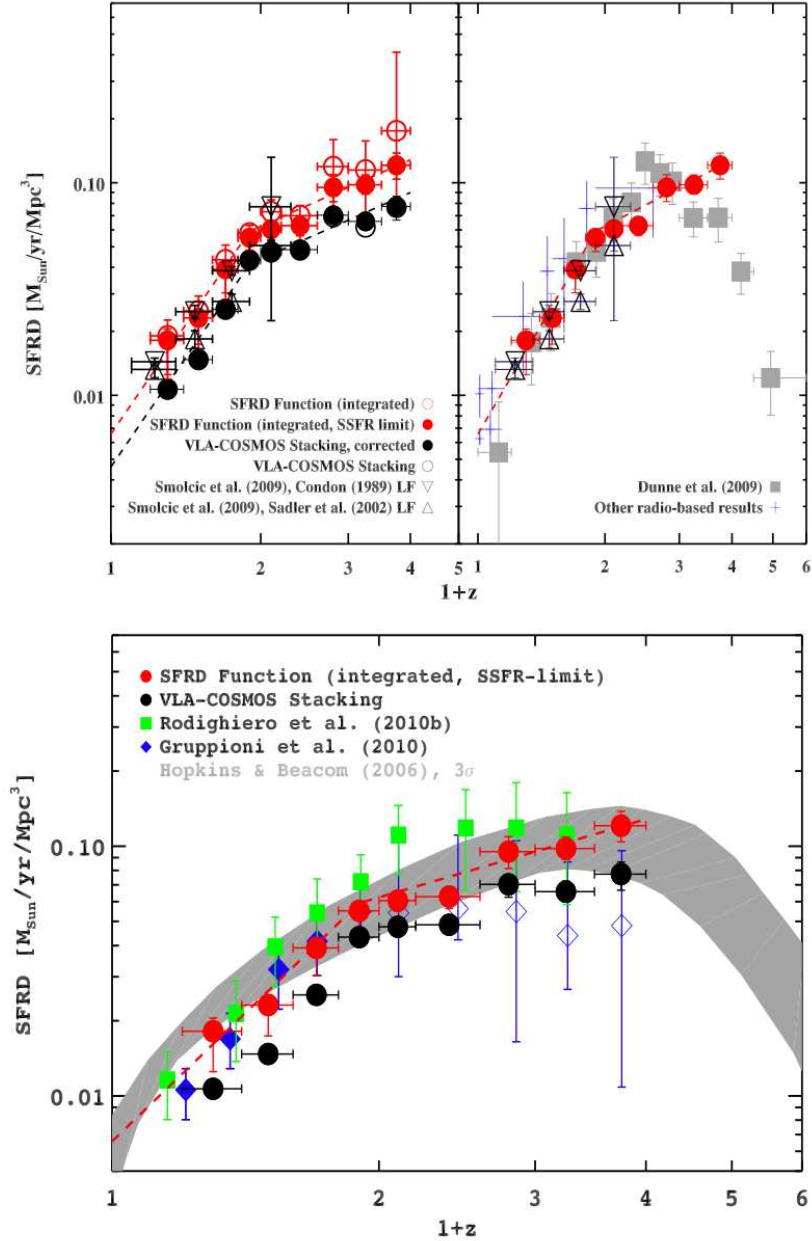}
\caption{\noindent {\textbf{Upper left:}} The cosmic star formation history (CSFH) out to $z=3$ from the VLA-COSMOS survey. Black circles (raw and number density-corrected) represent the sum of the data points in all redshift-bin panels of Fig. \ref{fig:sfrdfunc} and hence a direct -- stacking-based -- measurement of the SFRD down to the limiting mass at each epoch. Evidently, the number density corrections are always small and no corrections are necessary at $z < 1.5$ where our results are most robust. Red filled circles correspond to the 'total' SFRD at each epoch, obtained by integrating the Schechter-function fit (Fig. \ref{fig:sfrdfunc}) down to $M_* = 10^5~M_{\odot}$ and assuming an upper limit to the average SSFR (case B in Sec. \ref{sec:sfrdfunc}; see also Sec. \ref{sec:ssfrm} for details).
Red open circles are obtained by the same method assuming no upper SSFR-limit (case A in Sec.  \ref{sec:sfrdfunc}). The redshift evolution can be described by a broken power-law (dashed lines) that results from the joint (non-)evolution of the SF stellar mass density and the evolution of the (S)SFR-sequence. Down- and upward-facing triangles depict the results by \citet{SMOL09A} based on VLA-COSMOS radio detections extrapolated by two distinct radio luminosity functions (LF).
{\textbf{Upper right:}} Compilation of radio-based literature estimates of the integrated CSFH between $0< z <4$, compared to our results. The radio-stacking based results by \citet{DUNN09} are depicted in grey and suggest a clear peak of the CSFH at $z \sim 1.5$ (see Sec. \ref{sec:sfrdevo} for a full list of references and discussion).
{\textbf{Bottom:}} Mid- to far-IR measurements of SFRDs between $0< z <2.5$ along with our data and the $3\sigma$ envelope from the \citet{HOPK06} compilation. The Herschel/PACS-based results \citep{GRUP10} are lower limits at $z \gtrsim 1.2$ and should be compared to our non integrated measurements (filled black circles). Note the remarkable agreement of the IR- and radio-based data at all $z$.} 
\label{fig:sfrdall} 
\end{figure*}

The upper right panel of Fig. \ref{fig:sfrdall} shows our data along with other radio-based measurements \citep{HAAR00,MACH00,COND02,SADL02,SERJ02,SMOL09A} and the radio-stacking
derived CSFH by \citet{DUNN09}. Since the referenced measurements based on radio detections have been extensively discussed in \citet{SMOL09A}, we will focus our comparison on the study of \citet{DUNN09} as it is methodologically closest to our study. Here the extrapolations towards faint sources are based on the evolving K-band luminosity function with the fixed faint-end slope presented by \citet{CIRA10}.
As we previously pointed out, the evolutionary trends \citet{DUNN09} find are in good agreement with our results. Also on absolute scales the results of both studies are basically indistinguishable
with significant\footnote{The error bars to the data points shown here -- for which we assume an upper SSFR limit -- are smaller at high redshifts compared to those of the corresponding ones if no limit is assumed. We might have underestimated the uncertainty introduced by the extrapolations in the former case because of the assumed error to the dynamical timescale ($370 \pm 50~$Myr) which is the purported limit to the inverse SSFR. However, it should be noted that \citet{DUNN09} do not include any uncertainty caused by their extrapolation into their error budget.} deviations only at the highest redshifts probed in our study. However, at $z>1.5$ the trends observed tend to suggest different conclusions as \citet{DUNN09} find a clear peak of the CSFH around $z=1.5-2$ followed by a strong decline of the SFRD with redshift. Indeed, at the highest redshifts probed in our study the SFRD extrapolated by \citet{DUNN09} does not exceed our direct measurement (without extrapolations, see the top left panel in Fig. \ref{fig:sfrdall}). Hence, one would have to assume that galaxies below $\sim 10^{10}~M_{\odot}$ do not contribute at all to the mass-integrated SFRD at $z>2$ in order to support this peak based on our data.      

Especially when compared to dust extinction corrected UV-based studies the existence and location of the CSFH peak as measured by a dust-unbiased star formation tracer is important. Recent UV-based measurements \citep{REDD09, BOUW10} suggest a peak of the CSFH at around $z=2-2.5$ and assume that the dust obscuration for the bulk of SF galaxies does not dramatically change out to $z \sim 4$ \citep[as shown by e.g.][]{BOUW09, FINK09, MCLU10, WILK10}. Our rising SFRD at $z>2$ suggest that SF galaxies at these redshifts have a somewhat higher dust content or obey a different reddening law than the corresponding sources at lower redshifts. Whether dust-obscured sources at $z>2$ are lost in optical/UV based measurements of the CSFH cannot be definitely answered given the mentioned large error bars our data show. Future Herschel studies of the total IR luminosity evolution at $z>2$ should reveal potentially larger dust reservoirs in these systems. The even more rapid decline of the radio-based SFRD as derived by \citet{DUNN09} appears to support the picture drawn by e.g. \citet{BOUW10} that obscured star formation does not significantly contribute to the global SFRD at $z \gg 4$. Such conclusions should, however, be treated with caution as it is highly unclear at such early cosmic times what extrapolation to the directly measured radio derived SFRDs are needed given the high stellar mass limits \citep[see also][]{GALL10}. 

Our integrated CSFH is further supported by recent studies carried out at mid-IR \citep{RODI09} and far-IR \citep{GRUP10} wavelengths. The latter study is based on $\sim 210-240$ Herschel/PACS detections at 100 and 160~$\mu$m in 150~arcmin$^2$ within the GOODS-N field and provides us with a deep view of the dust-unbiased CSFH. The lower panel in Fig. \ref{fig:sfrdall} shows that the agreement of the Herschel-based results presented by \citet{GRUP10} with our radio-stacking derived mass-integrated CSFH is striking. Out to $z \sim 1$ we also find a broad agreement with the measurements of \citet{RODI09}. While \citet{GRUP10} show only lower limits at $z \gtrsim 1$ below this redshift both studies measure an evolution of $(1+z)^n$ with $n = 3.8 \pm 0.3$ ($0.4$). A recent 24~$\mu$m based study by \citet{RUJO10} confirms this result measuring $n=3.4\pm 0.2$. All these values are in remarkable agreement with our average measured evolution of the SSFR-sequence of $\langle n\st{SFG} \rangle = 3.5 \pm 0.02$ (see Sec. \ref{sec:ssfr} and Tab. \ref{tab:ssfrvsz}\footnote{Note that the scatter of the individual measurements at different redshifts of $\beta\st{SFG}$ stated in Tab. \ref{tab:ssfrvsz} is actually a more realistic uncertainty range than the formal error to their weighted mean.} and hence a strong support for both our measurement as well as our parameterization given in Eq. \gl{eq:phistar}, especially out to $z=1$. It is, however, worth mentioning that our work, compared to the other studies mentioned, draws on a far larger sample.

Finally we want to stress the fact that any shallower high redshift trend in the evolution of the SSFR-sequence also at the high-mass end would indeed lead to a decline in the evolution of the SFRD as Eq. \gl{eq:phistar} suggests. This scenario cannot be ruled out given our data as the SSFR at the low-mass end of our sample tends to flatten and the high-mass end might follow at slightly higher redshifts based on the dynamical time arguments presented in Sec. \ref{sec:ssfrlim} that result in a global upper limit to the average SSFR. Hence, again, a deeper mid-IR selected sample of SF galaxies is needed to accurately probe the regime above $z=1.5$ where the CSFH is supposed to peak.
   
\section{Summary and conclusion}
\label{sec:summ}
Based on an unprecedentedly rich sample of galaxies selected at 3.6~$\mu$m with panchromatic (FUV to mid-IR) ancillary data and mapped in 1.4~GHz radio continuum emission in the COSMOS field we have measured stellar mass-dependent average (specific) star formation rates ((S)SFR) in the redshift range $0.2 < z < 3$. These were obtained using a median image stacking
technique that is best applied in the radio regime where the angular resolution is high and the fraction of direct detections is comparatively low such that blending of sources is negligible.  

We individually measured integrated radio flux densities in each stacked image and showed that a uniform (i.e. mass-independent) correction
factor is inappropriate to convert between peak and total flux density when high angular resolution radio continuum data is used. Furthermore, we applied various criteria in order to minimize the impact of contaminating radio flux density from active galactic nuclei and discussed to which lower mass limit at a given redshift our sample remains representative with respect to the star formation properties of the underlying population. We emphasize that all our findings are to be regarded as most robust at $z < 1.5$ while our data place valuable constraints on evolutionary trends at the highest masses as far as $z=3$.

Using the template-based, rest-frame $(\rm{NUV}-r^+)\st{temp}$ color from SED-fits in the NUV-mid-IR, we separate SF galaxies from quiescent systems in order
to study the average mass dependence of their SSFR at all epochs considered. We also discussed potential effects introduced
by such a color threshold, such as mimicking a potential upturn of the SSFR-$M_*$ relation. 

Our findings are summarized as follows: 
\begin{enumerate}
\item The massive end of our global sample of mass-selected galaxies (including quiescent and SF systems) shows a power-law relation between SSFR and stellar mass ($\rm{SSFR} \propto M_*^{\beta}$) with an index of roughly $-0.7 \lesssim \beta_{\rm{ALL}} \lesssim -0.6$ and a trend towards shallower indices with increasing redshift. 
Towards lower masses the relation appears to flatten, probably because quiescent galaxies with low SSFRs preferentially occupy the massive end of the normal galaxy population.
\item For a given stellar mass we report a strong increase of the SSFR with redshift that is best parametrized by a power-law $\propto (1+z)^{4.3}$. 
\item The relation between SSFR and mass for star forming (SF) systems only (referred to as the SSFR-sequence) evolves as $(1+z)^{3.5}$ and shows a shallower power-law index of $\beta_{\rm{SFG}} \approx -0.4$ because quiescent galaxies do not lower the observed average SSFRs at the high-mass end anymore. The parameter $\beta_{\rm{SFG}}$ does not significantly change with cosmic time so that the average SSFR is best described by a separable function in mass and redshift (Eq. \ref{eq:ssfrmz} in Sec. \ref{sec:ssfrz}).  
\item Towards lower masses and $z > 1.5$ also the SSFR-sequence itself tends to flatten which might be explained by an upper limiting threshold where average SF systems already reach levels of star formation that qualify them to double their mass within a dynamical time. It is plausible that the SSFR at a given time does not continue to increase till the regime of dwarf galaxies at the rate predicted by our power-law index. We, however, cannot rule out that low-mass systems with high star formation activity but also very high dust content are missed given the limiting magnitude in our selection band.    
\end{enumerate}

We firmly conclude that, out to $z \sim 1.5$, our results hence neither support the so-called 'SSFR-downsizing' nor '-upsizing' scenarios proposed by some earlier work while they do confirm the downsizing scenario in the following take: 

{\emph{The SFR declines strongly but in a mass-independent fashion while the most massive galaxies always show the least star formation
activity and are hence the first to fall below their past-average star formation activity.}}  

By taking advantage of the simple functional form of both the (S)SFR-sequence and the mass function of SF galaxies in the redshift range we study we have shown that
\begin{itemize}
\item the mass distribution function of the comoving SFR density (SFRD) at any redshift below $z=1$ is well parameterized by a single Schechter function with a possible low-mass modification
at higher $z$.
\item the typical mass of a SF galaxy contributing most to the total (stellar mass integrated) SFRD is $10^{10.6 \pm 0.4}~M_{\odot}$, with no evidence for evolution out to $z=3$.
\end{itemize}

Out to $z \approx 1$ the evolution of the integrated SFRD, in turn, is entirely controlled by the mass-uniform evolution of the SSFR-sequence as
the number of SF galaxies in a given comoving volume does not change anymore. A strong and global decline in the mass density of molecular
gas, i.e. the reservoir out of which stars are formed, appears therefore to be the only driver of the observed decrease of the integrated SFRD with cosmic time. The rate at which the SFRD declines is in excellent agreement with the most recent other studies that use mid- to far-IR emission as an alternative dust-unbiased tracer for star formation.
Towards earlier epochs this steep trend becomes shallower as the comoving stellar mass density of SF systems decreases. In other words, there are simply less (SF) galaxies at $z>1$ while their individual SFRs further increase with redshift. This statement is certainly valid for galaxies more massive than $10^{10}~M_{\odot}$ which dominate the CSFH at all epochs out to $z=3$. Our results do not suggest any change of this trend towards the highest redshifts probed but it should be emphasized again that our data cannot constrain the situation as strongly as at $z < 1.5$. Hence, we do not rule out that the CSFH peaks in this redshift range. Indeed, the constancy of the SSFR at $z \gg 2$ suggested by other studies and motivated by the dynamical time threshold we discuss would give rise to a decline of the global SFRD at such high redshifts.   
  
\acknowledgments
The National Radio Astronomy Observatory (NRAO) is 
operated by Associated Universities, Inc., under cooperative 
agreement with the National Science Foundation.  
This work is based on observations made with the Spitzer 
Space Telescope, which is operated by the Jet Propulsion Laboratory, California Institute of Technology under NASA contract 
1407. We gratefully acknowledge the contributions of the entire COSMOS collaboration consisting of more than 100 scientists. The HST COSMOS program was supported through NASA grant HST-GO-09822.  
A.K. thanks Alvio Renzini, Simon Lilly, Yingjie Peng, David Elbaz, Simone Weinmann, Eyal Neistein, Ranga-Ram Chary, George Helou, Ryan Quadri, Maaike Damen, Andrea Maccio and David Hogg for helpful and enlightening discussions.
A.K. also thanks Henry McCracken for sharing the COSMOS K-band data with us and Loretta Dunne for interesting discussions and especially for kindly providing us with the data results from \citet{DUNN09} as well as Eric Murphy for the IR-template SED fits to our stacking derived data at Spitzer/MIPS wavelengths.
For the Figures shown in this paper we made extensive use of the Coyote Library for IDL (http://www.dfanning.com/documents/programs.html) and thank David Fanning for making it publicly available. We also made use of several routines contained in the IDL Astronomy Library (http://idlastro.gsfc.nasa.gov) and thank Wayne Landsman and the NASA Goddard Space Flight Center for making it publicly available. 
We thank the anonymous referee for very constructive suggestions that helped to improve the quality of this paper. 
C.C. thanks the Max-Planck-Gesellschaft and the Humboldt-Stiftung for support through the Max-Planck-Forschungspreis. A.M.S. is supported by a UK STFC postdoctoral fellowship. The research leading to these results has received funding from the European Union's Seventh Framework programme under grant agreement 229517. A.K. and M.T.S. acknowledge funding by DFG grant SCHI 536/3-3 as part of the Priority Programme 1177 ('Witnesses of Cosmic History:  Formation and evolution of black holes, galaxies and their environment').

{\it Facilities:} \facility{VLA}, \facility{Spitzer (IRAC)}, \facility{Spitzer (MIPS)}.

\appendix

\section{A statistical estimator for the stellar mass representativeness of a flux density-limited sample}
\label{sec:app_comp}
In principle one could roughly estimate stellar mass completeness limits by visual inspection of Fig. \ref{fig:compaq}. Given
a flux density limit (i.e. $F_{3.6~\mu \rm{m}} \approx 5~\mu$Jy for all and $F_{3.6~\mu \rm{m}} \approx 1~\mu$Jy for SF galaxies; see Sec. \ref{sec:comp}) below which no objects of a given spectral type (Sec. \ref{sec:class}) should be considered one would select by eye a stellar mass limit upward from where there are no objects below the flux density threshold.
As pointed out in Sec. \ref{sec:comp} it is, however, necessary to analytically derive these stellar mass limits in order to ensure that -- within a narrow mass-range just at any limiting mass -- we are dealing with a distribution of flux densities that can be considered representative for the one of the underlying population. A statistical estimator is needed for obtaining the actual level of representativeness we achieve at a given stellar mass. 

Our aim is to compare the properties of the exponential decline of the {\emph{observed}} distribution of $3.6~\mu$m flux densities -- i.e. the distribution of low values of flux density towards the flux density limit in our selection band -- within a narrow bin in logarithmic stellar mass to the analogously exponentially declining Gaussian distribution. As explained in the following it is sufficient for this comparison to derive the relative distance between {\emph{(a)}} the 0.95 percentile of the observed distribution of flux densities to the flux density limit and {\emph{(a)}} the 0.9 to the 0.95 percentile of the same observed distribution. The choice of the two percentiles mentioned is hereby entirely arbitrary. 

First we have to derive the corresponding ratio of distances for an arbitrary normal distribution that is cut at a given percentile.
This percentile sets the representativeness we want to achieve, i.e. 0.95 in our case. Since the width of and hence the lengthscale defined by a Gaussian is determined by a single parameter $\sigma$ any ratio of distances between given percentiles is independent of the actual value of $\sigma$ or the normalization. Given for instance a quantity $x$ defined as the distance between the 0.9025 ($=0.95 \times 0.95$) percentile of a given normal distribution and its 0.95 percentile as well as a quantity $y$ defined as the distance from the 0.855 ($=0.95 \times 0.9$) to the 0.9025 percentile of the same distribution, their ratio $(x/y)\st{Gauss}$ yields a value of 1.467 that is universal, i.e. independent of the actually chosen normal distribution. It was obtained by taking advantage of the cumulative distribution function that connects a percentile to the corresponding actual value $x_{\sigma}$ defined by the specific Gaussian of width $\sigma$ centered at $\mu$ via the error function (erf):
\begin{equation}
\Phi_{\mu,\sigma}(x_{\sigma}) = \frac{1}{2} \left[1+ \rm{erf}\left(\frac{x_{\sigma} - \mu}{\sigma \sqrt{2}}\right) \right].
\end{equation}
Differences in percentiles $\Delta \Phi_{\mu,\sigma}=\Phi_{\mu,\sigma}({x_{\sigma}}_{,j})- \Phi_{\mu,\sigma}({x_{\sigma}}_{,i})$ thus translate into physical distances $\Delta x_{\sigma} ={x_{\sigma}}_{,j}-{x_{\sigma}}_{,i}$ solely defined by the scale $\sigma$.

\begin{figure*} 
\hspace{-1.7cm}
\includegraphics[angle=90,width=1.175\textwidth]{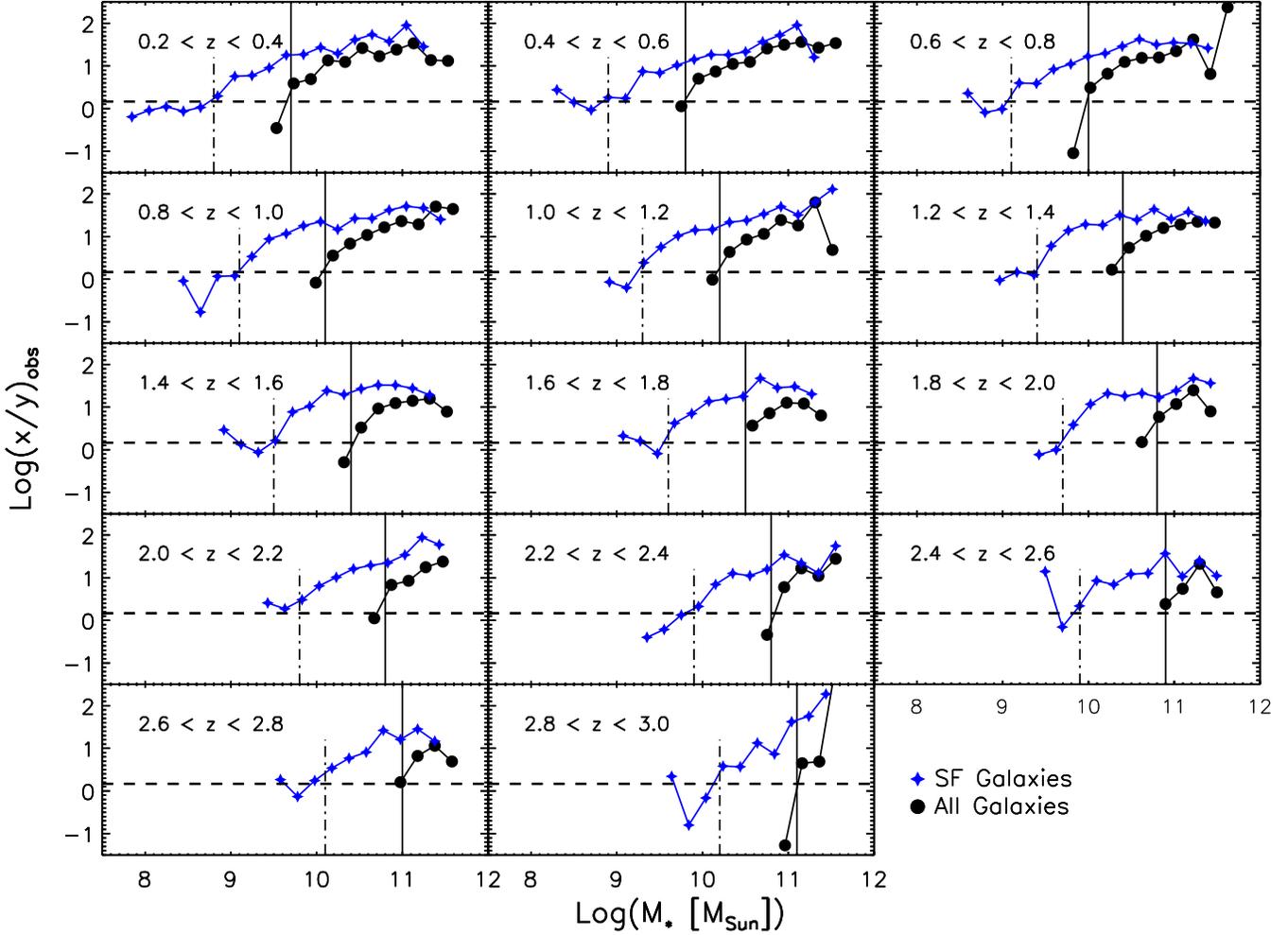}
\caption{\noindent Analytic evaluation of the statistical 95~\% statistical completeness in different redshift bins, based on a flux density threshold of $F_{3.6~\mu \rm{m}} = 5~\mu$Jy ($m\st{AB}(3.6~\mu\rm{m})= 22.15$) corresponding to the 90~\% level of intrinsic catalog completeness for all galaxies (black circles). For star forming galaxies (blue stars) the evaluation is based on a flux density threshold of $F_{3.6~\mu \rm{m}} = 1~\mu$Jy ($m\st{AB}(3.6~\mu\rm{m})= 23.9$, i.e. the magnitude limit of the catalog). For a detailed discussion of the meaning of statistical completeness, the choice of flux density thresholds and the implications on sample representativeness with respect to star formation see Sec. \ref{sec:comp}. The photometric redshift slices depicted in the individual panels are the same as in Fig. \ref{fig:compaq}. The quantity $(x/y)\st{obs}$ measures how well the the distribution of flux densities at the faint end within a given sample in a narrow mass bin ($\Delta \log (M_{\ast})=0.2$) follows the decreasing wing of a Gaussian distribution. Where it crosses the dashed horizontal line the Gaussian is cut at the 0.95 percentile and the anticipated statistical completeness limit (vertical lines) is reached. The method is described in detail in Appendix \ref{sec:app_comp}. Lines connecting the data points are meant to guide the eye.}
\label{fig:gaussall} 
\end{figure*}

Assuming that our data in narrow bins of stellar mass and redshift follows a normal distribution the distance from the 0.95 percentile of the observed distribution of $3.6~\mu$m flux densities to the flux limit of the sample yields a value $x\st{obs}$ in units of flux density. Accordingly the distance between the 0.9 and the 0.95 percentiles of the observed distribution defines a value $y\st{obs}$ and the dimensionless ratio $(x/y)\st{obs} \equiv x\st{obs}/y\st{obs}$ can be compared to the aforementioned value of $(x/y)\st{Gauss}$. Given the case that the flux density limit is located far in the tail of the observed distribution, $(x/y)\st{obs}$ will exceed $(x/y)\st{Gauss}$ and the observed distribution is statistically representative of the underlying population of objects. As the flux density limit approaches the peak of the observed distribution the observed ratio becomes lower. As soon as it overlaps with the 0.95 percentile of the (unknown) distribution of the underlying population the limiting case of 95~\% statistical completeness -- and hence the desired lower level of representativeness -- is reached so that $(x/y)\st{obs} = (x/y)\st{Gauss}$.

For our sample this effect is shown in Fig. \ref{fig:gaussall} where $(x/y)\st{Gauss}$ is indicated as a dashed horizontal line for individual ranges in photometric redshift. The finally chosen stellar mass representativeness limits are denoted by vertical lines. The data points result from an implementation of the method described in this section that additionally takes into account the detection completeness levels of the catalog as a function of $3.6~\mu$m flux density. Here we therefore obtain the mentioned percentiles using flux densities weighted by the corresponding inverse catalog detection completeness.

It is worth noting that the Gaussian distribution is just one possible parameterization and not a necessary requirement 
for the method described here. Indeed, the underlying distribution of flux densities is not even required to be symmetric.
Our method simply ensures that the observed distribution is smoothly, approximately exponentially declining to low levels of flux density,
as one may realistically expect from random processes such as photon noise and confusion. 
It is simply a practical, quantitative improvement over the alternative method of visual inspection as the latter is, essentially, assuming an unphysical
step-function rather than a continuos distribution function.

\section{Statistics}
\label{sec:app_stat}
In the following a set of $N$ pixels will always be written as $X_N$, regardless if the constituents $x_i$ ($i=1,\ldots,N$) are noise pixels or a sample of peak flux densities.
We will specify at any stage, if we are referring to background noise, in which case we will use the upper case indication 'bg'.
\subsection{Noise weighted estimators}
Due to the non-uniform noise distribution in the VLA-COSMOS map, the input samples used for stacking are ill-defined to some extent.
Solely discarding the high-noise edge regions does not remedy the fact, that there is significant variation of the rms background noise in the
cutout postage stamps originating from a broad spatial distribution across the field. 
Our aim is to find the best estimator for the representative value of the underlying population. Therefore the sample should consist of a random
and independent set of sources drawn from this population under equal circumstances. To approach the last condition, that is not achievable
in observational reality, we have to compare the outcome of the stacked sample to that of a weighted sample, in which those stamps gain
more influence, that lie in low noise regions.

Regarding the mean-stacking technique it is statistically well known, that appropriate weights are found in the reciprocal variance of each particular
stamp's noise pixel sample where the variance of the $i$th stamp is defined as $\rm{Var}_i \equiv \rm{Var}\left(X^{\rm{bg},i}_{N\sh{bg}}\right)=\sigma\st{bg}^2\left(X^{\rm{bg},i}_{N\sh{bg}}\right).$ As explained above $X^{\rm{bg},i}_{N\sh{bg}} \equiv \left\{x_{i_1}\sh{bg},\ldots,x_{i_N\sh{bg}}\sh{bg}\right\}$.\\
The noise-weighted mean of the sample $X_{N}$ of peak fluxes can thus be considered the mean of the weighted sample $\widetilde{X}_{N}$, where
the constituents $\widetilde{x}_i$ of  $\widetilde{X}_{N}$ are defined as
\for{\label{wgt}\widetilde{x}_i = \widetilde{w}_i x_i \equiv N\frac{ \rm{Var}_i^{-1}}{\sum_i \rm{Var}_i^{-1}} x_i \qquad \rm{where} \quad x_i \in X_{N}.} 
With the definitions $W_i= \rm{Var}_i^{-1}$ and $w_i = W_i/\sum_i W_i$ it is easily shown, that the mean of the $\widetilde{x}_i$ defined in \gl{wgt}
indeed equals the noise-weighted mean of the $x_i$:
\for{\langle \widetilde{X} \ran =\frac{1}{N} \sum_i \widetilde{x}_i = \frac{1}{N} \sum_i \widetilde{w}_i x_i =  \sum_i w_i x_i = \frac{\sum_i W_i x_i}{\sum_i W_i}.}
The above discussion leads to the suggestion, that in the presence of varying rms-noise in a given sample, the sample $\widetilde{X}_{N}$ is the appropriate
one to consider not only with respect to the mean value of the sample. It seems reasonable, that also its median is the best estimator for the median of the
underlying population, because both computed quantities, the median and the mean, are then referring to the same sample. We will refer to this choice of an 
estimator in the following as a noise-weighted median. 

\subsection{Bootstrapping}
\label{sec:app_boot}
In the above discussion we justified that neither the observed nor the intrinsic distribution of peak fluxes are expected to be gaussian. This needs to be taken into account no matter if we are looking for an appropriate uncertainty range to the median or mean estimator for a given sample.
In order to obtain a 68~\% confidence interval not relying on normality of the underlying parent distribution we therefore chose a bootstrapping technique for the statistical parameter of choice.

In each case we obtain the limits of the confidence interval by a bootstrapped Student's $t$-distribution. This technique is called studentizing. A $(1-\alpha)$ confidence interval for a parameter $\overline{X}$ in traditional statistics is given by 
\for{\label{ci} CI_{\alpha/2} = \overline{X} \pm t_{\alpha/2} \frac{s_{\overline{X}}}{\sqrt{N}}\, , \qquad s_{\overline{X}}: \; \mbox{standard deviation of }\overline{X}.}
Here $t_{\alpha/2}$ denotes the $\alpha/2$ percentile of the classical $t$-distribution which is equal to the $(1-\alpha/2)$ percentile due to the symmetry of Student's distribution.
Bootstrapping a $t$-distribution means to circumvent the assumption of a normally distributed population by deriving the quantity
\for{\label{deft}t_i^{\ast} = \frac{\overline{X}^{\ast}_i - \overline{X}}{s^{\ast}_{\overline{X}^{\ast}_i}/\sqrt{N}}} 
for $i=1,\ldots,N\st{bootstrap}$ samples drawn from the original sample of peak fluxes with replacement. In case of $\overline{X} \equiv \langle X \ran$ being the sample mean one thus has to compute the sample mean $\langle X\ran$ as well as all means of the bootstrapped samples $\langle X^{\ast} \ran_i$ including standard deviations $s^{\ast}_{\langle X^{\ast} \ran_i}$. The resulting distribution of $N\st{bootstrap}$ $t^{\ast}$-values is then used to compute the upper and lower confidence limits by taking its $\alpha/2$ and $(1-\alpha/2)$ percentiles: 
\begin{eqnarray}
 \label{clup}
 CI\st{up}^{1-\alpha} &=&  \overline{X} + t^{\ast}_{\alpha/2} \frac{s_{\overline{X}}}{\sqrt{N}}  \\
 \label{cllow}
 CI\st{low}^{1-\alpha} &=&  \overline{X} - t^{\ast}_{1-\alpha/2} \frac{s_{\overline{X}}}{\sqrt{N}},
\end{eqnarray}
where in general we chose $1-\alpha = 0.68$ obtaining thus a 68~\% confidence interval. Here $s_{\overline{X}} \equiv s_{\langle X\ran}$ still is just the standard deviation of the original sample.
In case of $\overline{X} \equiv \rm{Med}(X)$ denoting the sample median as the parameter of choice we have to face the problem that the denominator of \gl{deft} does not provide us with an estimator of the standard error of the median. In order to estimate this latter quantity we need to access the empirical standard deviation of a sample of medians being representative for the median of the current bootstrapped sample. Starting from this sample we thus generate a number of new bootstrapped samples hence performing a bootstrapping within the bootstrapping procedure\footnote{The number of outer bootstrapped samples is typically chosen to be an order of magnitude larger compared to the one of the inner bootstrapping.}. The standard deviation $s^{\ast}_{Med(X^{\ast \ast}_i)}$ of these subsamples' medians is then used as an estimator for the standard error of the single outer bootstrapped median as given by the denominator of Eq. \gl{deft}. In order to use Eq. (\ref{clup}, \ref{cllow}) we furthermore need the standard error of the original sample's median. This is estimated by computing the median of each outer bootstrapped sample and taking the standard deviation $s_{Med(X^{\ast})}$ of this sample of medians as the standard error.

\begin{figure*} 
\centering
\includegraphics[width=\textwidth]{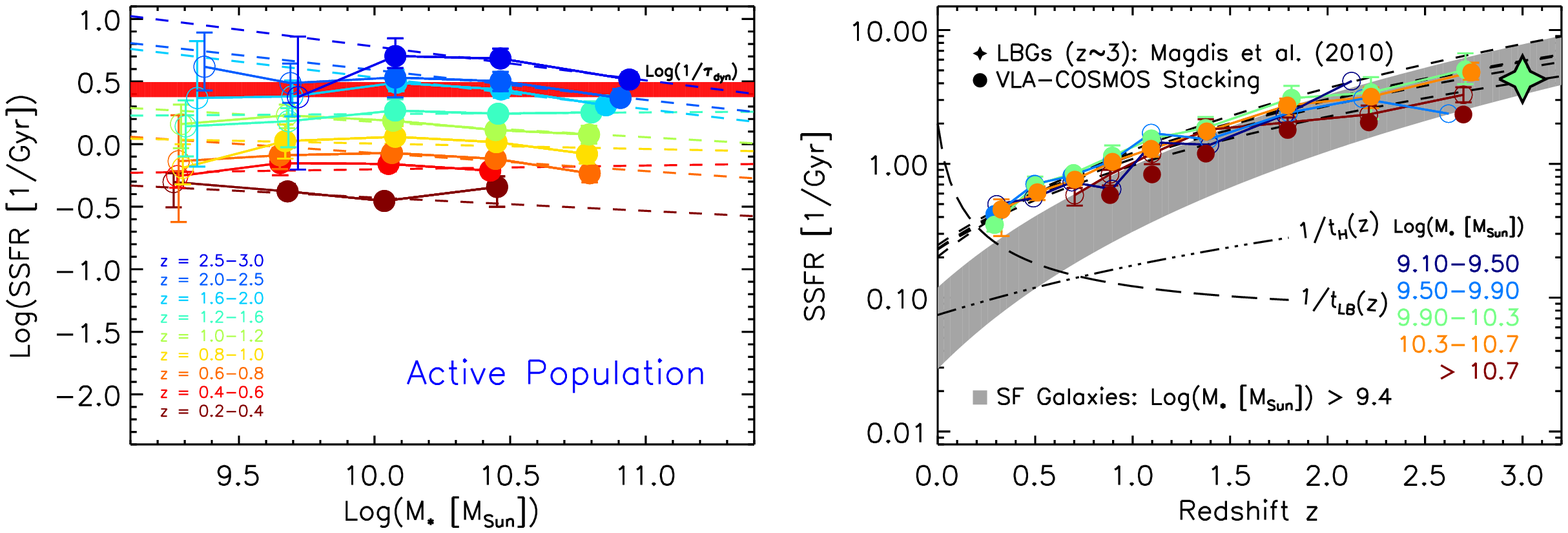}
\includegraphics[width=0.65\textwidth]{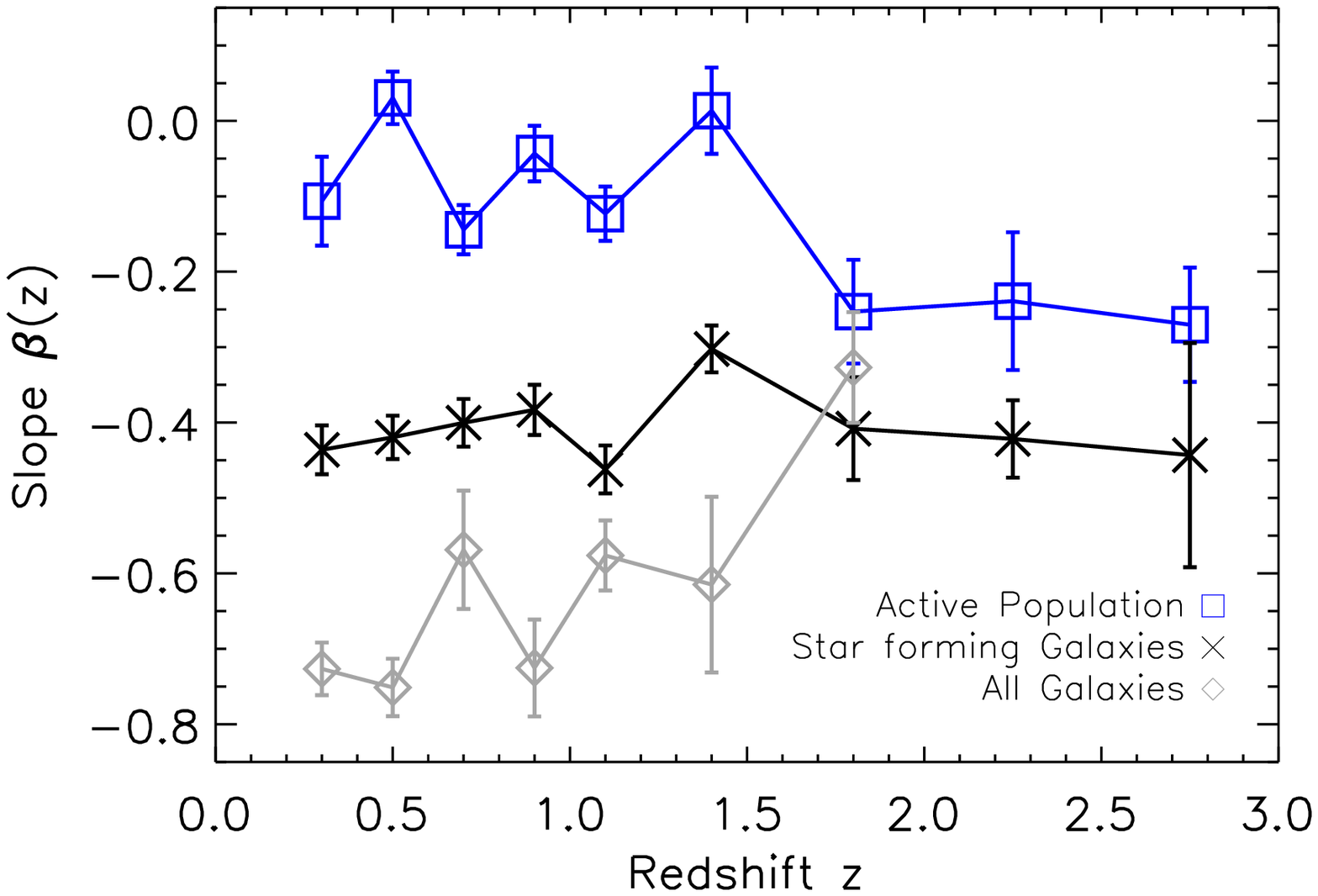}
\caption{\noindent The SSFR-$M_*$ relation (upper left) and time evolution of SSFRs in various mass-bins (upper right) for galaxies with high star formation activity ((NUV-r$^+)\st{temp} < 1.2$).
Dashed lines (color coded by redshift) denote two-parameter fits of the form $c \times M_*^{\beta}$ to the mass complete data (filled circles). The dynamical ranges are the same as in Figure \ref{fig:ssfrvsmf} and \ref{fig:ssfrvszf}, where the other quantities shown are explained. The color threshold is substantially bluer and rather arbitrary compared to the one used for the selection of SF systems. Compared to the results of the entire SF sample the radio stacks yield a flatter SSFR-sequence out to $z \sim 1.5$ ($\beta \approx -0.08 \pm 0.05$) and a mild 'upsizing' trend (lower panel) while overall a shallower evolution of the SSFR $\propto (1+z)^{2.3 \pm 0.3}$ is found for this sample of most vigorously SF galaxies (upper right, where fits to the mass-complete data in the different mass bins are depicted as dashed black lines). All these trends seen for highly active SF galaxies are hence a result of a simple selection effect. The inverse horizontal red band sketches the inverse dynamical time as detailed in Sec. \ref{sec:ssfrm}.}
\label{fig:ssfrvsma} 
\end{figure*}

\section{Selection of star forming galaxies}
\label{sec:app_sfg}
Because we cannot measure radio-based SSFRs for individual galaxies, selecting the SF population directly in the SSFR-$M_*$ plane is impossible. In Sec. \ref{sec:class} we argue that the intrinsic (dust-extinction corrected) rest-frame NUV-$r$ color is a reliable way to select SF galaxies (see also I10). In this section, we test our color selection in two ways in order to demonstrate its fidelity and to assess how our findings relate to previous measurements in the literature: we (1) choose a bluer color-cut to study the ensuing changes in the evolution of the SSFR-$M_*$ relation, and we (2) compare both color cuts to the BzK-selection of SF galaxies at high $z$ \citep{DADD04}.

\subsection{Highly active star forming galaxies}
I10 have shown that the color selection criterion ($\rm{NUV}-r^+)\st{temp} < 1.2$ leads to a morphologically clean sample of late-type spiral and irregular galaxies with template SED-based SSFRs that are clearly separated from the passive population (see Sec. \ref{sec:class}). This color threshold is somewhat arbitrary (as it is less well motivated than the cut we applied to select SF systems) but by virtue of being substantially bluer than our original choice it minimizes contamination by passive galaxies.

We derived SSFRs as a function of redshift and mass in the same way as before for galaxies with ($\rm{NUV}-r^+)\st{temp} < 1.2$. Although the exclusion of systems with intermediate star forming activity has reduced the sample size considerably, it was still possible to cover the same dynamic ranges. Only the binning scheme has been slightly modified for this strongly star formation population (see Fig. \ref{fig:ssfrvsma}). Its SSFRs usually are significantly higher compared to our original choice of SF galaxies, the slope $\beta$ of the SSFR-$M_*$ relation is shallower at low to intermediate redshifts and thus in excellent agreement with those literature results for SF galaxies discussed in Sec. \ref{sec:others} that report an almost flat SSFR-sequence. At the high-$z$ end we see a steepening of the slope $\beta$ (i.e. an 'upsizing' trend), similar or even more evident to what was found by \citet{RODI10} and, at lower significance, also by \citet{OLIV10}. The evolutionary exponent $n$ is consistent with previous measurements as well \citep[see e.g.][]{PANN09}.

The bluer color threshold hence is able to reproduce most literature findings albeit with SSFRs that tend to be comparativey high, especially at low redshift. However, it has yet to be confirmed that the galaxy population selected in this way is representative of the {\emph{entire}} SF population. 

\subsection{(s)BzK galaxies at $z\sim 2$}
We cross-matched our 3.6~$\mu$m selected catalog with the K band selected catalog for the COSMOS field \citep{MCCR10} and thus obtain a magnitude calibration in the crucial wavebands that allows us to apply the BzK selection criterion of \citet{DADD04}. Fig. \ref{fig:bzkplot} shows the BzK diagram for our sample, with galaxies in six redshift slices color coded according to their ($\rm{NUV}-r^+)\st{temp}$ color as described in Sec. \ref{sec:class}. Our `star forming' sample is the union of all galaxies plotted in blue {\emph{and}} green.

\begin{figure*} 
\hspace{-1.5cm}
\includegraphics[angle=90,width=1.15\textwidth]{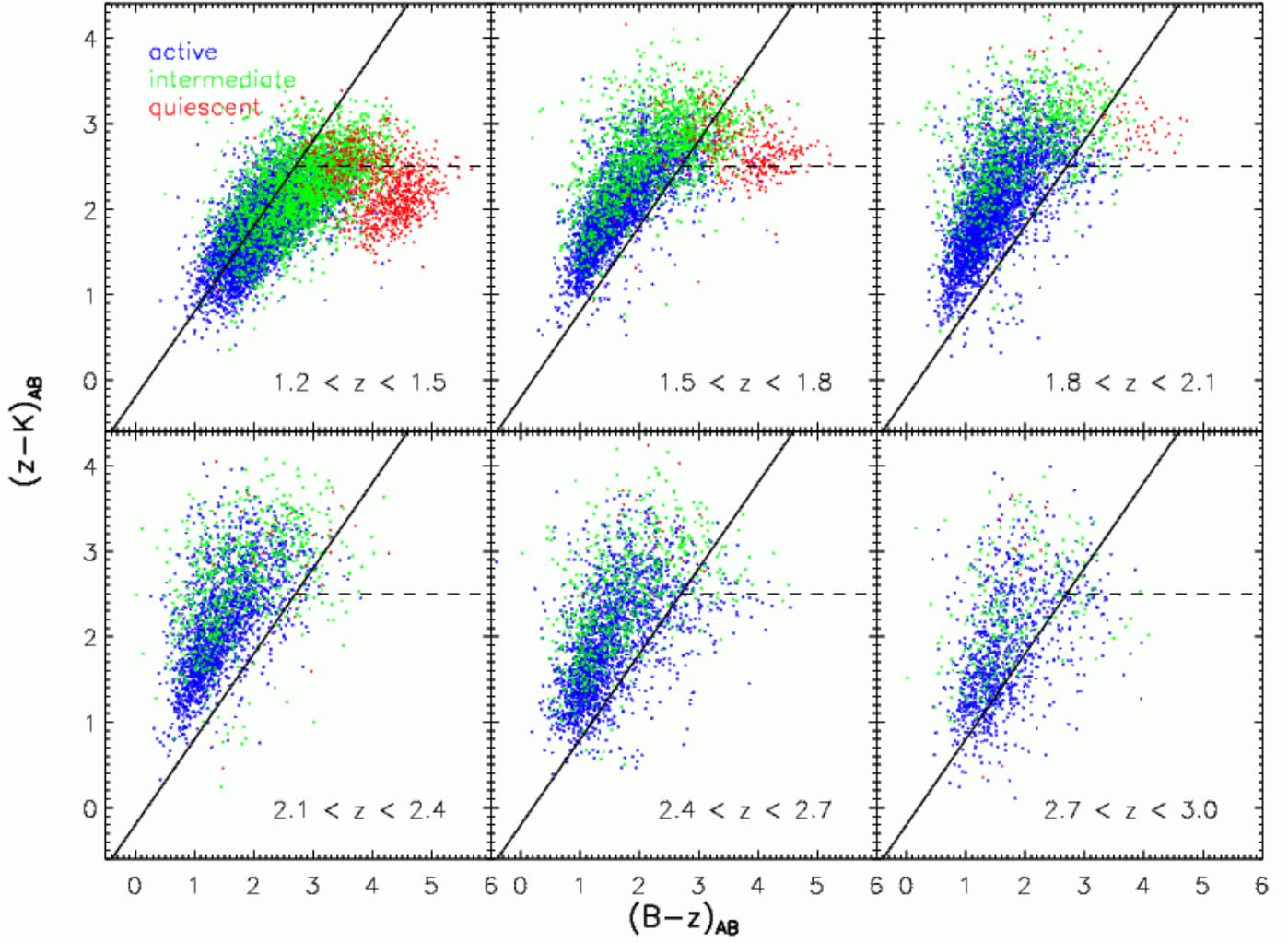}
\caption{\noindent BzK diagram of our sample in various redshift bins. The color coding refers to our choice of the $(\rm{NUV}-r^+)\st{temp}$ color threshold
in order to predefine systems with high (blue), intermediate (green) and negligible (red) star formation activity. In the lowest redshift panel the original \citep{DADD04} sBzK criterion (diagonal line) does
not appear to be efficient enough in selecting {\emph{all}} SF galaxies and is particularly missing the systems with intermediate levels of star formation. At higher redshifts our SF sample (all green and blue sources) overlaps very well with the sBzK population so that our color selection for the purpose of radio stacking is appropriate to select normal SF systems at $z>1.5$.}
\label{fig:bzkplot} 
\end{figure*}

At $z>1.5$, the sBzK criterion (all galaxies to the left of the diagonal line in each panel) is established to efficiently select normal SF systems. Fig. \ref{fig:bzkplot} illustrates that the selection window for sBzK galaxies is populated by {\emph{both}} the most actively star formation sources (blue dots) and the majority of the sources with intermediate star forming rates (plotted in green; i.e. the rest of the SF sample used throughout this article). Only a small number of objects
with moderate SF activity fall into the passive BzK region in the upper right of each panel. This is the reason for the aforementioned excellent agreement of our results with \citet{PANN09} at $z \approx 2.1$; their and our sample are virtually indistinguishable and both studies rely on the same radio data. More importantly, however, the BzK diagram strongly supports our original selection of SF objects in the crucial redshift regime $z>1.5$, where we have just shown that previously reported changes in the slope $\beta$ can be mimicked by simply selecting only very blue objects, hence, most actively forming systems.  

\clearpage

%BIBLIOGRAPHY

\bibliographystyle{apj.bst} 
\bibliography{ssfr-references.bib}

\end{document}